%% file: BUDHIES_TFR.tex
\documentclass[fleqn,usenatbib]{mnras}
\usepackage{newtxtext,newtxmath}
\usepackage[T1]{fontenc}

\usepackage{soul}
\usepackage{enumitem}
\usepackage{pgffor}
\usepackage{caption} 
\usepackage{siunitx}
\usepackage[final]{pdfpages}
\usepackage[T1]{fontenc}
\usepackage{ae,aecompl}
\usepackage{rotating}
\usepackage{xcolor}
\usepackage{ifpdf}
\usepackage[symbol]{footmisc}
\usepackage{array}
\usepackage{longtable}
\usepackage{lscape}
\usepackage{dcolumn}
\usepackage{bm}
\usepackage{ifthen}
\usepackage{booktabs}
\usepackage{ulem}
\usepackage{geometry}
\usepackage{tikz}
\usepackage{graphicx}	
\usepackage{rotating}
\usepackage{ifpdf}

\newcounter{pdfpages}

\newcommand{\kms}{km s$^{-1}$}	
\newcommand{\Msun}{M$_\odot$}

\newcommand{\wct}{W$\mathrm{_{20}^{corr}}$}
\newcommand{\wcf}{W$\mathrm{_{50}^{corr}}$}
\newcommand{\wot}{W$\mathrm{_{20}^{obs}}$}
\newcommand{\wof}{W$\mathrm{_{50}^{obs}}$}
\newcommand{\mhi}{M$\mathrm{_{HI}}$}
\newcommand{\mstel}{M$_{\star}$}
\newcommand{\mbar}{M$\mathrm{_{bar}}$}

\newcommand{\msun}{M$\mathrm{_\odot}$}
\def\H{H{\sc i}}
\def\b{BUDH{\sc i}ES}

\def\al1{ALFA100 }

\def\kms{km s$^{-1}$}
\def\mhi{$\mathrm{M_{HI}}$}

\title[The \H-based Tully-Fisher relation at z=0.2]{{\b\ V: The baryonic Tully-Fisher relation at z=0.2 based on direct \H\ detections}}

\author[Gogate et al.]{A.R. Gogate$^{1}$\thanks{E-mail: avanti@astro.rug.nl}, M.A.W. Verheijen$^{1}$\thanks{E-mail: verheyen@astro.rug.nl}, J.M. van der Hulst$^{1}$, Y.L. Jaff\'e$^{2}$ \\
$^{1}$ Kapteyn Astronomical Institute, University of Groningen, Landleven 12, 9747 AD Groningen, the Netherlands.\\
$^{2}$ Instituto de F\'isica y Astronom\'ia, Universidad de Valpara\'iso, Avda. Gran Breta\~na 1111, Casilla 5030, Valpara\'iso, Chile  \\}

\date{Accepted 2022 October 31. Received 2022 October 7; in original form 2021 September 14}

\pubyear{2022}

\begin{document}
\label{firstpage}
\pagerange{\pageref{firstpage}--\pageref{lastpage}}
\maketitle

\begin{abstract}
We present \H-based B- and R-band Tully-Fisher relations (TFRs) and the Baryonic TFR (BTFR) at z=0.2 using direct \H\ detections from the Blind Ultra-Deep \H\ Environmental Survey (\b). Deep photometry from the Isaac Newton Telescope was used for 36 out of 166 \H\ sources, matching the quality criteria required for a robust TFR analysis. Two velocity definitions at 20\% and 50\% of the peak flux were measured from the global \H\ profiles and adopted as proxies for the circular velocities. We compare our results with an identically constructed z=0 TFR from the Ursa Major (UMa) association of galaxies. To ensure an unbiased comparison of the TFR, all the samples were treated identically regarding sample selection and applied corrections. We provide catalogues and an atlas showcasing the properties of the galaxies. Our analysis is focused on the zero points of the TFR and BTFR with their slopes fixed to the z=0 relation. Our main results are: (1) The \b\ galaxies show more asymmetric \H\ profiles with shallower wings compared to the UMa galaxies, which is likely due to the environment in which they reside, (2) The luminosity-based z=0.2 TFRs are brighter and bluer than the z=0 TFRs, even when cluster galaxies are excluded from the \b\ sample, (3) The BTFR shows no evolution in its zero point over the past 2.5 billion years and does not significantly change on the inclusion of cluster galaxies, and (4) proper sample selection and consistent corrections are crucial for an unbiased analysis of the evolution of the TFR.
\end{abstract}

\begin{keywords}
galaxies: evolution – galaxies: fundamental parameters – galaxies: kinematics and dynamics – radio lines: galaxies.
\end{keywords}


\section{Introduction}

In the past few decades, several efforts have been made to improve our understanding of fundamental scaling relations between various properties of galaxies. For rotationally supported systems, i.e. for regular, late-type disc galaxies, the Tully-Fisher relation \citep[TFR,][]{Tully_Fisher77} is one such scaling relation, correlating two observed quantities of galaxies: the intrinsic luminosity, which is a proxy for the stellar mass of the galaxy, and the width of an emission line from the interstellar medium (ISM), which is directly linked to the galaxy's rotational velocity. It is now also general practice to study other manifestations of the TFR, such as the Stellar-mass TFR (STFR) and the Baryonic TFR (BTFR), by converting the luminosity and gas content of a galaxy into derived quantities such as stellar and baryonic masses. Rotational velocities can be inferred from various distance independent, kinematic measures such as the width of a global profile and the amplitude of a rotation curve \citep[see][]{Verheijen01, Ponomareva17,Lelli19}. While the TFR is standardly studied using galaxies with regular, disc-like morphologies, it is found that early-type and S0 galaxies in the Local Universe also follow a similar relation \citep[see][]{Trujillo-Gomez11,Cortese14}. In theory, however, an offset in the zero-point of the TFR for early-types can be expected due to the higher stellar Mass-to-Light ratios (M/L) of their older stellar populations. This has also been confirmed by a number of studies \citep{Bedregal06, Aragon-Salamanca06, Williams10, Davis11, denHeijer15}. Furthermore, massive, compact galaxies tend to have declining rotation curves, resulting over-estimation of the circular velocity of the dark matter halo from the measured global profile width  \citep{Casertano91, Noordermeer07}, which also may result in an offset of the zero point when using the width of the global \H\ profile.

The TFR has been used extensively for distance measurements, wherein the distance modulus to disc galaxies can be recovered from the TFR if their distance independent rotational velocities are measured properly. The observed and intrinsic scatter in the TFR, however, leads to uncertainties in the inferred distances and several studies have tried to quantify and reduce this scatter to thereby attain the tightest TFR. Using distances derived from the TFR, the Hubble constant \citep[see][]{Schombert20} as well as local cosmic flows have been studied \citep{Kashibadze08,Tully13,Boruah20,Kourkchi20}. 
Additionally, the TFR is a useful tool to provide constraints for numerical simulations of galaxy formation \citep{Navarro00,Dutton07, Vogelsberger14, Schaye15, Maccio16}, wherein the slope, scatter and zero point of the TFR need to be accurately reproduced at various cosmic epochs in order to verify the plausibility of galaxy formation scenarios. The TFR may also provide insights into internal galaxy structure and kinematics, such as the prevalence of warps and non-circular motions \citep{Franx92}.

In the Local Universe, it is general practice to use the 21-cm atomic hydrogen (\H) emission line for TFR studies. \H\ proves to be an excellent tracer of galaxy dynamics due to several factors; firstly, \H\ discs generally extend much farther out than stellar discs and in most cases, probe the outer, flat part of the rotation curve, which provides the ideal velocity measurement for a TFR study. Secondly, atomic gas has a lower velocity dispersion compared to ionised gas and hence is more directly associated with circular velocities. Thirdly, it also has a relatively constant surface density and a high area-covering factor. Notably, several \H-based TFR studies using spatially resolved rotation curves have been carried out in the recent past \citep[e.g.,][]{Verheijen01,Ponomareva16,Noordermeer07,deblok14}, providing more accurate measures of the circular velocity compared to the corrected width of the global profile.

One drawback of using \H\ measurements, however, is that \H\ emission is intrinsically weak and, therefore, its detectability is restricted to the Local Universe. At higher redshifts, \H\ becomes increasingly difficult to detect with the current generation of radio telescopes, thus requiring extremely long integration times. Consequently, \H\ surveys carried out beyond the Local Universe are limited in number \citep{Fernandez13,Hoppmann15,Gogate20,Catinella15}. A recent study by \citet{Ponomareva21} provides the deepest direct \H-based TFR study out to z$\sim$0.08 using resolved \H\ kinematics. Beyond this redshift, only one study \citep{Catinella15} has presented the BTFR at an intermediate redshift (z$\sim$0.3) using targeted \H\ observations. Their sample of extremely massive and luminous galaxies, optically selected from the Sloan Digital Sky Survey (SDSS, \citealp{York00}) seems to follow the z=0 BTFR from Several high redshift studies of the TFR exist (see Sect. \ref{redshiftevol}), though there are none that make use of \H\ as the choice of kinematic tracer. The only known \H\ study beyond the Local Universe (z$\sim$0.2) by \citet{Catinella15} shows that an optically selected sample of extremely massive, gas rich and luminous galaxies also seems to follow the local BTFR from GASS \citep{Catinella12}. Their work, however, is not a dedicated TFR study and the limitations of their data do not allow for an in-depth analysis of the evolution of the TFR.

With an increase in the amount of observational data sets over the past decades, several TFR studies of galaxy samples have been carried out using other emission line tracers of a galaxy's kinematics, such as  H\textsc{$\alpha$}, H\textsc{$\beta$},  O[\textsc{ii}] and O[\textsc{iii}] and CO \citep{Dickey92, Schoniger94,Tutui97, Lavezzi98, Tutui01, Ho07, Davis11, Tiley16}. All these tracers are usually confined to the inner, star forming regions of galaxies and typically do not accurately probe the circular velocity of the dark matter halo. While several TFR studies have been carried out beyond the Local Universe using optical emission lines \citep{Conselice05, Flores06, Kassin07, Puech08, Jaffe11a}, only one CO-based TFR study \citep{Topal18} exists to date. At intermediate redshifts, CO is a preferred kinematic tracer as compared to ionised gas due to its lower intrinsic velocity dispersion, while its emission line is relatively bright compared to \H.

TFR studies based on cosmological numerical simulations provide some insights into the \textit{expected} redshift evolution of the TFR, with a possibility to carefully construct a sample of suitable galaxies. Using semi-analytical models, \citet{Obreschkow09} concluded that their simulated TFRs in the Local Universe are in good agreement with the fiducial, observed TFR based on the HIPASS survey, but at higher redshifts, they found an increase in the scatter and a shift in the zero points towards higher velocities for a given baryonic mass (their Fig. 14).  On the other hand, \citet{Glowacki21} studied the evolution of the BTFR using the SIMBA simulation, and found that the zero point of the BTFR at higher redshifts is shifted towards lower velocities (their Fig. 3) and only detectable for disc-like galaxies with V$_{\rm flat}$ as the kinematic measure. Thus, a proper selection of kinematically regular disc galaxies is abundantly important for an apt evaluation of the redshift evolution of the TFR. 

Despite the plethora of information on the TFR at intermediate and high redshifts, there is no convergence yet on the results for the redshift evolution of the TFR (see discussion in Sect. \ref{redshiftevol}). Inconsistencies in the TFR parameters are often encountered in the literature, due to factors such as the choice of tracer and differences in photometric bands, sample selection and size, methodology adopted for the measurement of galaxy properties, corrections applied to the data etc., 
making it particularly difficult to consistently compare and study the evolution of the TFR slope, scatter and zero point over cosmic time. Consequently, these observational and selection biases often tend to introduce systematic offsets, which could be mistaken for an intrinsic evolution in the parameters of the TFR. For instance, \citet{Verheijen&Sancisi01} found that a galaxy with an uncertainty of just 1 degree on the measured inclination alone could lead to a scatter of 0.04 magnitudes (assuming a slope of -10). Another study by \citet{Bedregal06} indicates a downward shift of the Local TFR \citep[adopted in their case, from ][]{Tully00} by about 1.2 mag for a sample of lenticular galaxies, emphasising the importance of selecting proper comparison samples. For an unbiased TFR comparison, one has to ensure that the methodology adopted for measuring and correcting the observed galaxy properties used in the TFR is consistent, since it is the relative offsets between these properties that are of significance \citep{Ponomareva17}. While the observed scatter in the TFR can be minimised when using a carefully selected sample of regular, disc-like systems with extended, flat rotation curves or clear double-horned global profiles, the same cannot be expected for studies of galaxy samples at higher redshifts due to, for example, the use of a kinematically hot tracer with limited radial extent, survey limitations such as poor spatial and velocity resolution or smaller sample sizes and uncertainties in the measurement of inclinations. 

In this paper, we aim to provide, for the first time, a meaningful and detailed comparison of the \H-based TFR and BTFR at z=0 and z=0.2 in a careful and consistent manner using a volume-limited, \H-selected sample, namely the Blind Ultra-Deep \H\ Environmental Survey \citep[\b,][referred to as Paper 1 hereafter]{Gogate20}. \b\ was undertaken using the Westerbork Synthesis Radio Telescope (WSRT) and is one of the first blind \H\ imaging surveys at z > 0.1. The surveyed volume includes a range of cosmic environments, effectively encompassing a total volume of 73,400 Mpc$^3$ with a depth of 328 Mpc and covering a redshift range of 0.164 < z < 0.224. From the 166 direct \H\ detections, a subset of suitable galaxies was chosen to represent the \b\ TF sample. With this study, we present the first thorough analysis of an \H-based TFR by comparing our results to the z=0 \H-based TFR from a previous study of the Ursa Major association of galaxies \citep{Verheijen01}. The data reduction procedures and extraction of galaxy properties were carried out in an identical manner for both samples, which is crucial for a proper analysis. Our goal is to also present the effect on the observed statistical properties of the TFR due to the choice of corrections and prescriptions applied to the observables. It is our objective to provide this study as a reference for the next-generation \H\ surveys that will be able to study the \H-based TFR out to higher redshifts.

Sect. \ref{data} describes the \b\ sample and other literature samples used in this study. The rigorous sample selection process is presented in Sect. \ref{sampsel}. In Sect. \ref{corr}, we describe the corrections applied to the data, while the properties of the samples used in this paper are compared in Sect. \ref{compprop}. The \H\ and optical catalogues as well as an atlas containing the various observed properties of the \b\ TF galaxies, are presented in Sect. \ref{atlas_tf}. Sect. \ref{results_tf} provides the main results from this study. Finally, we discuss our findings in Sect. \ref{discussion} and summarise our work in Sect. \ref{summary}. Throughout this paper, we assume a $\Lambda$CDM cosmology, with $\mathrm{\Omega_{M}}$ = 0.3, $\Omega_{\Lambda}$ = 0.7 and a Hubble constant H$_0$ = 70 km s$^{-1}$ Mpc$^{-1}$. All magnitudes used in this paper are Vega magnitudes.


\section{The Data}\label{data}

\subsection{The \b\ data}

\b\ is a blind, volume-limited \H\ imaging survey undertaken with the primary aim of providing an \H\ perspective on the so-called Butcher-Oemler (BO) effect \citep{BO84} at an intermediate redshift of z $\simeq$ 0.2, corresponding to a look-back time of $\sim 2.5$ Gyr. To this effect, the survey was centred on two galaxy clusters: Abell 963 at z $= 0.206$, which is a massive, virialised, lensing BO cluster with a large fraction (19\%) of blue galaxies in its core and strong in X-ray emission from the Intra-Cluster Medium (ICM), and Abell 2192 at z $= 0.188$, which is a much smaller, non-BO cluster still in the process of forming and weak in X-rays. The two surveyed volumes also include the large-scale structure in which the clusters are embedded. The volumes within the Abell radii of these clusters occupy as little as 4 percent of the total surveyed volume, which is 73,400 Mpc$^3$ within the Full Width at Quarter Maximum (FWQM) of the primary beam. The average angular resolution of \b\ is \ang{;;23}$\times$\ang{;;38} (corresponding to 65 $\times$ 107 kpc$^2$ at z$\sim$0.164) while the rest-frame velocity resolution is 19 \kms. The achieved \H\ mass limits at the redshifts of the clusters and in the field centres is 2 $\times$ 10$^9$ \msun\ at the redshift of the two clusters, for an emission line width of 150 \kms. A total of 127 galaxies with confirmed optical counterparts were detected in \H\ in the cube containing Abell 963 (A963 hereafter), while 39 \H-detected galaxies were identified in the cube containing Abell 2192 (A2192 hereafter).

Apart from the \H\ data, a deep B- and R-band imaging survey of the two fields was carried out with the Isaac Newton Telescope (INT), which was utilised for counterpart identification, photometry, assessing optical morphologies as well as estimating inclinations. Additionally, u, g, r, i, z photometry as well as optical spectroscopy from the SDSS is available for the two fields. Other supporting data include deep NUV and FUV imaging with the GALaxy Evolution eXplorer (GALEX; \citealp{dcmartin05}), spectroscopic redshifts from the William Herschel Telescope \citep[WHT,][]{Yara1_13} and CO observations using the Large Millimeter Telescope \citep[LMT,][]{Cybulski16}. These data, however, have not been used in this paper. Details on the \b\ data processing, source finding and stellar counterpart identification can be found in Paper 1.

\subsection{Reference studies from the literature}\label{litsamp}

For comparison with the Local Universe TFR, we adopt \citet{Verheijen01}'s study of the Ursa Major association of galaxies (UMa, hereafter). In particular, we adopt the global \H\ profiles of the 22 UMa galaxies for which the amplitude of the outer flat part of the \H\ rotation curve could be measured from spatially resolved \H\ synthesis imaging data obtained with the WSRT \citep{Verheijen&Sancisi01}, and for which photometric imaging data is available in the B, R, I and K'-band \citep{Tully96}. These UMa galaxies are nearly equidistant at 18.6 Mpc \citep{Tully00}, consistent with the average of the Cosmicflows-3 distances \citep{Tully16} to the 22 individual galaxies as provided by the Extragalactic Distance Database\footnote{available at https://edd.ifa.hawaii.edu} \citep{Tully09}. The radio and photometric data reduction and analysis procedures used for the \b\ sample and those employed by \citet{Verheijen01} are essentially identical. Note that the UMa BRIK' and the INT B- and R-band images for the two \b\ fields are significantly deeper than the SDSS images. From his analysis, \citet{Verheijen01} found that the K'-band TFR using rotational velocities derived from the outer flat part of the rotation curves has the tightest correlation; however, for an unresolved \H\ study, the R-band TFR using corrected global \H\ line widths as proxies for rotational velocities is the preferred choice. While other, more recent \H-based z=0 TFR studies exist \citep[e.g.,][]{Ponomareva16,Lelli19}, we chose the UMa sample because of its many observational similarities with \b\ such as similar \H\ data sets, both obtained with the WSRT, and the availability of B- and R-band photometric images. Moreover, the UMa sample is volume-limited and complete to a limiting magnitude of m$_{zw}$=15.2 for late-type galaxies, while the data reduction procedures are identical to ours.

For a cursory high-redshift comparison we use the HIGHz sample by \citet{Catinella15}, which is a targeted \H\ survey with Arecibo and consists of 39 isolated galaxies optically selected from the SDSS, covering a redshift range of 0.17 $<$ z $<$ 0.25. These galaxies were selected to represent a sample of extremely massive, luminous, and star-forming galaxies at z > 0.16. From a preliminary analysis they found that these rare galaxies lie on the local BTFR adopted from \citet{Catinella12}, suggesting that they are scaled-up versions of local disc galaxies. To make this sample available for our comparative study, the SDSS photometry of these 39 galaxies was re-extracted from the DR7 database and transformed to Johnson-B and Cousins-R bands using the transformation equations by \citet{Cook14}. While the \H\ and photometric data acquisitions by \citet{Catinella15} differ from those for the \b\ and UMa samples, we consistently applied identical corrections (including K-corrections) to the line widths and photometry (see Sect. \ref{corr}) for all three samples. It is to be noted, however, that the comparison with the HIGHz sample is limited in scope, firstly due to the absence of direct B- and R- band photometry, which is required for a consistent analysis, secondly, since a reliable quantification of the offset from the z=0 TFR for the HIGHz sample is impossible due to the limited ranges in the luminosities and $\mathrm{W_{50}}$ line widths. In addition, the HIGHz sample also does not overlap in parameter space (see Sect. \ref{compprop}), making it unrepresentative for this analysis. Thus, it is included for illustrative purposes only in the various figures that follow.

\section{Sample selection} \label{sampsel}

For a robust TFR study, one of the fundamental requirements is to be able to accurately measure the rotational velocities of the dark matter halos of galaxies.  In the Local Universe, this is best achieved with resolved \H\ studies, which can provide rotational velocities from the \H\ rotation curves of galaxies.  However, for blind \H\ imaging at intermediate redshifts, such as \b, galaxies are only marginally resolved at best, making rotation curve measurements unattainable.  For this study, therefore, rotational velocities were inferred from \H\ global profile measurements. The corrected \H\ profile line widths at 20\% and 50\% of the peak flux are often used as proxies for the rotational velocities at the flat part of the rotation curve (further discussed in Sect. \ref{velmeasure}).  This makes it necessary to carefully select galaxies with larger inclinations and suitable \H\ profiles.  We constructed two sub-samples from the parent sample of 166 galaxies, as described below. A table containing a full break-down of the galaxies rejected at every stage of the sample selection is provided as supplementary material online.

\subsection{The Tully-Fisher Sample (\textit{TFS})} \label{TFSsample}

To construct this \b\ sub-sample, galaxies were rejected up-front due to the following observational and qualitative constraints:\\

\noindent
\textbf{A. Qualitative observational rejection criteria:} 
\begin{itemize}[align=left,
   leftmargin=2.4em,
   itemindent=0pt,
   labelsep=2pt,
   labelwidth=2em]
    \item[A1.] Galaxies with global \H\ profiles that are cut-off at the edges of the observed WSRT bandpass;
    \item[A2.] Galaxies lying outside the field-of-view of the INT mosaic;
    \item[A3.] Galaxies with corrupted or uncertain INT photometry, e.g. due to imaging artefacts from nearby bright stars, with stars superimposed on the optical image of the galaxy, or with nearby, overlapping companion(s). 
\end{itemize}

\noindent
\textbf{B. Rejection criteria based on optical morphologies or potential confusion of the stellar counterpart:}
\begin{itemize}[align=left,
   leftmargin=2.4em,
   itemindent=0pt,
   labelsep=2pt,
   labelwidth=2em]
    \item[B1.] \H\ detections with multiple nearby, UV-bright companions within the WSRT synthesised beam that lack an optical redshift.  Such cases do not allow for an unambiguous identification of the stellar counterpart of an \H\ detection. 
    \item[B2.] Galaxies with obvious disturbed optical morphologies such as tidal features or strong asymmetries.  The \H\ gas in these galaxies is likely not on circular orbits while the optical morphologies preclude an accurate measurement of the inclination.
\end{itemize}

\noindent
\textbf{C. \H\ profile shapes:}\\
Galaxies with Gaussian or strongly asymmetric \H\ profiles were rejected. An automated profile classifier was constructed which compared the maximum fluxes in three equally-spaced velocity bins of the \H\ profiles, and classified them into five categories: Double-Horned (type 1), Single-Gaussian (type 2), Boxy (type 3), Skewed Boxy (type 4) and Asymmetric (type 5). We retained types 1, 3 and 4, in an effort to ensure the inclusion of only galaxies with steep \H\ profile edges. Resolved \H\ synthesis imaging studies and simulations \citep[e.g.,][]{Verheijen01,Lelli16, El-Badry18} have shown that Gaussian profiles (type 2) are generally associated with rising rotation curves, and are thus unsuitable for a TFR analysis. In addition, they also often correspond to face-on systems with low inclinations. Asymmetric (type 5) profiles could be the result of blending of nearby, possibly interacting galaxies, given the relative large size of the synthesised beam in kpc. Since the primary aim of the \H\ data is to procure reliable measurements of the rotational velocities of the dark matter halo, such galaxies have therefore been excluded from this analysis.\\

\noindent
\textbf{D. Inclinations:} \\
Finally, the inclinations of the galaxies that were not rejected by the criteria mentioned above were computed using the available INT R-band images. Since inclinations based on our SExtractor photometry were not very robust, the galaxies were modelled with \textit{galfit} \citep{CPeng10}.  Parameters computed by SExtractor were used as the initial estimates required by \textit{galfit}. We fit S\'ersic models to all our galaxies. From the axis ratios ($b/a$) returned by \textit{galfit}, inclinations were calculated following:

\begin{equation}
     \mathrm{cos \: i = \sqrt{\frac{(\textit{b}/\textit{a})^2 - q_0^2}{1- q_0^2}}}
\end{equation}
 
\noindent
where \textit{a} and \textit{b} are the semi-major and semi-minor axes of the model ellipse while the intrinsic disc thickness (q$_0$) was chosen to be 0.2. For consistency with other comparison samples, galaxies with an inclination more face-on than 45$^\circ$ were rejected. Note the two galaxies (no. 14 and 26 in Column (1) of Table \ref{tab:A963_HI_table_tf}) which were both assigned an inclination of 90$^\circ$ since the axis ratios returned by \textit{galfit} were less than the assumed disk thickness.

These rejection criteria resulted in a sample of 36 galaxies, of which 29 belong to A963 and 7 belong to A2192 (note that A963 and A2192 refer to the entire survey volume, not just the Abell clusters themselves). 

\subsection{ The High-Quality Sample (\textit{HQS})}

Three additional quantitative criteria were applied to the global \H\ profile shapes of the 36 \textit{TFS} galaxies in order to ensure the best possible comparison with the high-quality data of the UMa sample. \\

\noindent    
\textbf{E. Quantitative rejection criteria:} \\

\noindent
\begin{itemize}[align=left,
   leftmargin=2.4em,
   itemindent=0pt,
   labelsep=2pt,
   labelwidth=2em]
    \item[E1.] \textit{Signal-to-noise of the \H\ profiles}: To ensure an accurate measurement of the widths of the global \H\ profiles, we imposed a threshold on the \H\ profile line width uncertainties and rejected galaxies with uncertainties in excess of 10\% in their measured line widths.\\
 
    \item[E2.] \textit{Symmetry of the \H\ profiles}: Galaxies with asymmetric \H\ profiles do not provide robust circular velocity measurements. One method of assessing an asymmetric profile is to assess the systemic velocities (V$_{\mathrm{sys}}$) derived from the \wot\ and \wof\ line widths. If this difference $\delta$V$_{\mathrm{sys}}$ is large, then the profile is most likely asymmetric. Based on our assessment, galaxies with absolute fractional differences in V$_{\mathrm{sys}}$ following |\;$\delta$V$_{\mathrm{sys}}$/\wot\;| > 0.05 were rejected. \\
 
    \item[E3.] \textit{Steepness of the \H\ profile edges}: To assess the steepness of the \H\ profile edges, the differences in \wot\ and \wof\ were considered. Based on our assessment, galaxies with\\ |\wot\ - \wof|\;>\; 50 \kms\ were rejected.
\end{itemize}

\noindent
These stricter, objective criteria on the quality and shapes of the \H\ global profiles resulted in the rejection of 17 more galaxies and yielded a sample of 19 galaxies, of which 12 galaxies are in the A963 volume and 7 galaxies are in the A2192 volume.

\subsection{Literature samples}

For the UMa sample we used the "RC/FD" sample of 22 galaxies from \citet{Verheijen01} for which spatially resolved \H\ rotation curves confirm that the corrected widths of the corresponding global \H\ profiles properly represent their rotational velocities. These UMa galaxies abide by the same qualitative selection criteria as the \b\ \textit{HQS} galaxies. Due to limitations of the HIGHz sample (see Sect. \ref{litsamp}), it was not used for a quantitative assessment of the TFR. All the galaxies in the HIGHz sample with inclinations above 45$^\circ$, however, are included in the illustrations of the $\mathrm{W_{50}}$ TFRs in this paper.

\section{Corrections to the data}\label{corr}

Before Tully-Fisher relations can be constructed, the observed \H\ line widths W$\mathrm{^{obs}_{\%}}$ and total apparent magnitudes m$\mathrm{^{obs}_{B,R}}$ need to be corrected for various instrumental, astrophysical and geometric effects such as finite spectral resolution, turbulent motions of the gas, Galactic and internal extinction, K-corrections and inclination. We ensure that these corrections are applied consistently to all galaxies in the \b, UMa and HIGHz samples. In this section we describe these corrections in some detail.

\subsection{Correction to the observed \H\ linewidths}\label{widthcorr}

\subsubsection{Conversion to rest-frame line widths}

For the \b\ galaxies, the observed widths of the redshifted global \H\ profiles were measured in MHz at 20\% and 50\% of the peak flux ($\Delta \nu_{\%}^{\rm obs}$) and converted to observed, uncorrected rest-frame line widths (W$\mathrm{_{\%}^{obs}}$) in \kms\ using the following equation.

\begin{equation}
    \mathrm{W_{\%}^{obs}=\frac{\Delta \nu_{\%}^{obs}}{\nu_{rest}} (1+z) c}
\end{equation}

\noindent
where $\mathrm{\nu_{rest}}$ is the rest frequency of the Hydrogen emission line (1420.4057517667 MHz).

For the HIGHz galaxies, \citet{Catinella15} provide observed line widths at 50\% of the peak flux, expressed as W$_{\rm 50} = \Delta$cz in \kms\ (column 6 in their Table 2), which we divide by (1+z) to obtain the observed rest-frame line widths in \kms\ such that we can consistently apply our correction for instrumental broadening as explained in the next subsection. For the galaxies in the UMa sample, we do not apply any correction for this redshift effect and adopt the measured line widths as the rest-frame values.




\subsubsection{Correction for instrumental broadening}

The effect of instrumental broadening on the observed line widths, caused by a finite spectral resolution R of the radio spectrometers, was corrected using the following equations, adopted from Paper 1, in which the authors studied this effect at different velocity resolutions:

\begin{equation}
    \mathrm{W_{20}^R} = \mathrm{W_{20}^{obs} - 0.36R}
\label{eq:w20corr_tf}
\end{equation}

\begin{equation}
     \mathrm{W_{50}^{R}} = \mathrm{W_{50}^{obs} - 0.29R} 
\label{eq:w50corr_tf}
\end{equation}

\noindent
For the \b\ galaxies, we measured the line widths after the data were spectrally smoothed to a Gaussian line spread function with a FWHM of four channels or 0.15625 MHz (R4, hereafter), corresponding to a rest-frame velocity resolution of R=33.0$\times$(1+z) \kms. For the UMa galaxies, the spectral resolution varied in the range R$=5.0-33.2$ \kms\ \citep{Verheijen&Sancisi01} while we adopted the rest-frame velocity resolutions for the HIGHz galaxies presented in Table 2 in \citet{Catinella15}.

\subsubsection{Correction for turbulent motion}

After correcting for instrumental velocity resolution effects, corrections for broadening due to turbulent motions of the \H\ gas were then made to the data. Following \citet{Verheijen&Sancisi01} we adopt the prescription by \citet{Tully85}, which corresponds to a linear subtraction by $w_{\rm t,\%}$ for \H\ profiles with W$_{\%}^{\rm R}$$>$$w_{\rm c,\%}$ and a quadratic subtraction if W$_{\%}^{\rm R}$$<$$w_{\rm c,\%}$ where $w_{\rm c,20}$=120 \kms\ and $w_{\rm c,50}$=100 \kms. Since all the \b\ and HIGHz galaxies in our samples have W$_{\%}^{\rm R}$$>$$w_{\rm c,\%}$, and assuming that they have monotonically rising rotation curves that properly sample the outer flat parts of the rotation curve, we adopt the values for $w_{\rm t,\%}$ from \citet{Verheijen&Sancisi01} as \\

\noindent
$w_{\rm t,20}^{\rm flat}=32$ \kms\ \;\; and \;\; $w_{\rm t,50}^{\rm flat}=15$ \kms, \\

\noindent
and thus our corrected line widths become:

\begin{equation}
     {\rm W}_{\%}^{\rm R,t} = {\rm W}_{\%}^{\rm R} - w_{\rm t,\%}^{\rm flat}
\label{eq:wturb}
\end{equation}

\noindent
Although \citet{Verheijen&Sancisi01} did not provide uncertainties related to the turbulent motion corrections, it can be noted that the values of $w_{\rm t,\%}^{\rm flat}$ in comparable studies \citep{Broeils92,Rhee96} are quite similar and hence we adopt the corresponding errors of 5 and 4 \kms\ for $w_{\rm t,20}^{\rm flat}$ and $w_{\rm t,50}^{\rm flat}$ by \citet{Broeils92} respectively. It is important to note that these corrections are based on a sample average. A few resultant non-physical corrected line widths (\wcf > \wct) are caused by the scatter around the sample, which may affect individual systems. Other statistical corrections in the literature would show similar results.

\subsubsection{Correction for inclination}

Uncertainties in corrections involving the inclination contribute significantly to the scatter in the TFR. Hence, it is crucial to determine the inclinations as accurately as possible, and to propagate the corresponding uncertainties through the relevant correction formulas. Sect. \ref{TFSsample} describes our method for inferring inclinations based on the observed ellipticity $\epsilon = 1-(b/a)$ of the optical images. For completeness, we note here that \textit{galfit} takes the smoothing of the \b\ INT images due to the seeing into account while the value of $(b/a)$ as returned by \textit{galfit} pertains to the effective radius instead of a specified isophotal contour of the outer stellar disc. In case a significant spherical bulge is present in a galaxy, this approach may result in a slight overestimate of $(b/a)$ and, consequently, an underestimate of the inclination of the disc component and thereby an overestimate of the circular velocity. Table 2 lists the $(b/a)$ values for the \b\ galaxies as returned by \textit{galfit}, along with the formal errors.

For the UMa galaxies, \citet{Tully96} measured the $(b/a)$ ratio as a function of radius and selected the value that is representative of the outer disc, taking the optical morphology of the galaxy into account, including the presence of a bar, bulge and spiral arms. 
They converted this representative $(b/a)$ into an inclination using the same equation (1) and q$_0$=0.2. They assigned an uncertainty of 3 degrees to the inferred inclinations.

For the HIGHz galaxies, \citet{Catinella15} adopted the $(b/a)$ values from an exponential fit to the r-band SDSS images ($expAB_r$ in the SDSS database) and q$_0$=0.2 while employing the same equation (1) to infer an inclination. This axis ratio is representative at the effective radius of a galaxy. They do not provide an error estimate for either the $(b/a)$ values or the inferred inclinations.

Although the measurements of the optical minor-to-major axis ratio $(b/a)$ may have been slightly different for the galaxies in the three samples, the same formula and value of the intrinsic thickness q$_0$ was used in all studies. With the inferred inclinations, we corrected the already partially corrected line width according to:

\begin{equation}
    \mathrm{W^{R,t,i}_{\%}} = \frac{\mathrm{W^{R,t}_{\%}}}{\mathrm{ sin}\: i}
    \label{eq:wcorr}
\end{equation}

\noindent
where W$_{\%}^{\mathrm{R,t}}$ is the \H\ line width corrected for instrumental broadening and turbulent motion and $i$ is the inclination of the galaxy. Hereafter, we refer to $\mathrm{W^{R,t,i}_{\%}}$ as $\mathrm{W^{corr}_{\%}}$ for the sake of simplicity.

\subsection{Photometric corrections}\label{photcor}

For the \b\ sample, deep Harris B- and R-band imaging was carried out with the INT on La Palma. Photometric calibration of these images was carried out using the photometry of selected stars from the SDSS DR7 catalogue, transformed to Johnson B and Cousins R bands using the transformation equations provided by Lupton 2005\footnote{http://classic.sdss.org/dr4/algorithms/sdssUBVRITransform.html} (see Paper 1). Subsequently, instrumental aperture B- and R-band magnitudes were derived with SExtractor by summing all the background subtracted pixels within adaptive Kron elliptical apertures defined by the R-band images and also applied to the B-band images. For our analysis, the resulting AUTO magnitudes from SExtractor needed to be converted to the equivalent of total model magnitudes for a consistent comparison with the other literature data sets, which consist of total extrapolated magnitudes for the UMa galaxies \citep{Tully96} or SDSS model magnitudes for the HIGHz galaxies. For this purpose, we extracted and analysed the luminosity profiles of several galaxies in the \textit{HQS}, measured the sky levels, identified the radial range where the exponential disc dominates the light, fitted an exponential profile to this radial range and calculated the total extrapolated magnitudes following \citet{Tully96}. From this exercise, we found that the differences between the SExtractor aperture (AUTO) magnitudes and our extrapolated total magnitudes were quite small: 0.038 for A963 and -0.014 for A2192. The INT aperture magnitudes were corrected accordingly to make up for these differences. Note that this statistical correction does not alter the results of our study.

To obtain absolute magnitudes in the B- and R-bands, the total, extrapolated model magnitudes of the galaxies require further corrections for Galactic extinction, cosmological reddening and internal extinction as described below. These corrections were applied consistently to all galaxies in the three samples under consideration.

\subsubsection{Galactic extinction}

The total apparent magnitudes were corrected for Galactic extinction (A$\mathrm{^g_{B,R}}$) following \citet{Schlegel98}. The \b\ galaxies within a WSRT pointing are all close together on the sky and received the same correction according to\\

\noindent
\begin{tabular}{ l c c c}
A963  &:\; $\mathrm{A^g_{B}}$ = 0.052 & ; & $\mathrm{A^g_{R}}$ = 0.031 \\
      &                      &   &                   \\
A2192 &:\; $\mathrm{A^g_{B}}$ = 0.039 & ; & $\mathrm{A^g_{R}}$ = 0.023 \\
\end{tabular}\\

\noindent
Galactic extinction corrections for UMa galaxies were also adopted from \citet{Schlegel98} and are provided in Table 1 of \citet{Verheijen&Sancisi01}. In the case of the HIGHz sample, de-reddened SDSS magnitudes (dered\_u, dered\_g, dered\_r, dered\_i, dered\_z) were used, since they also follow \citet{Schlegel98}.

\begin{figure*}
    \includegraphics[width=0.85\textwidth]{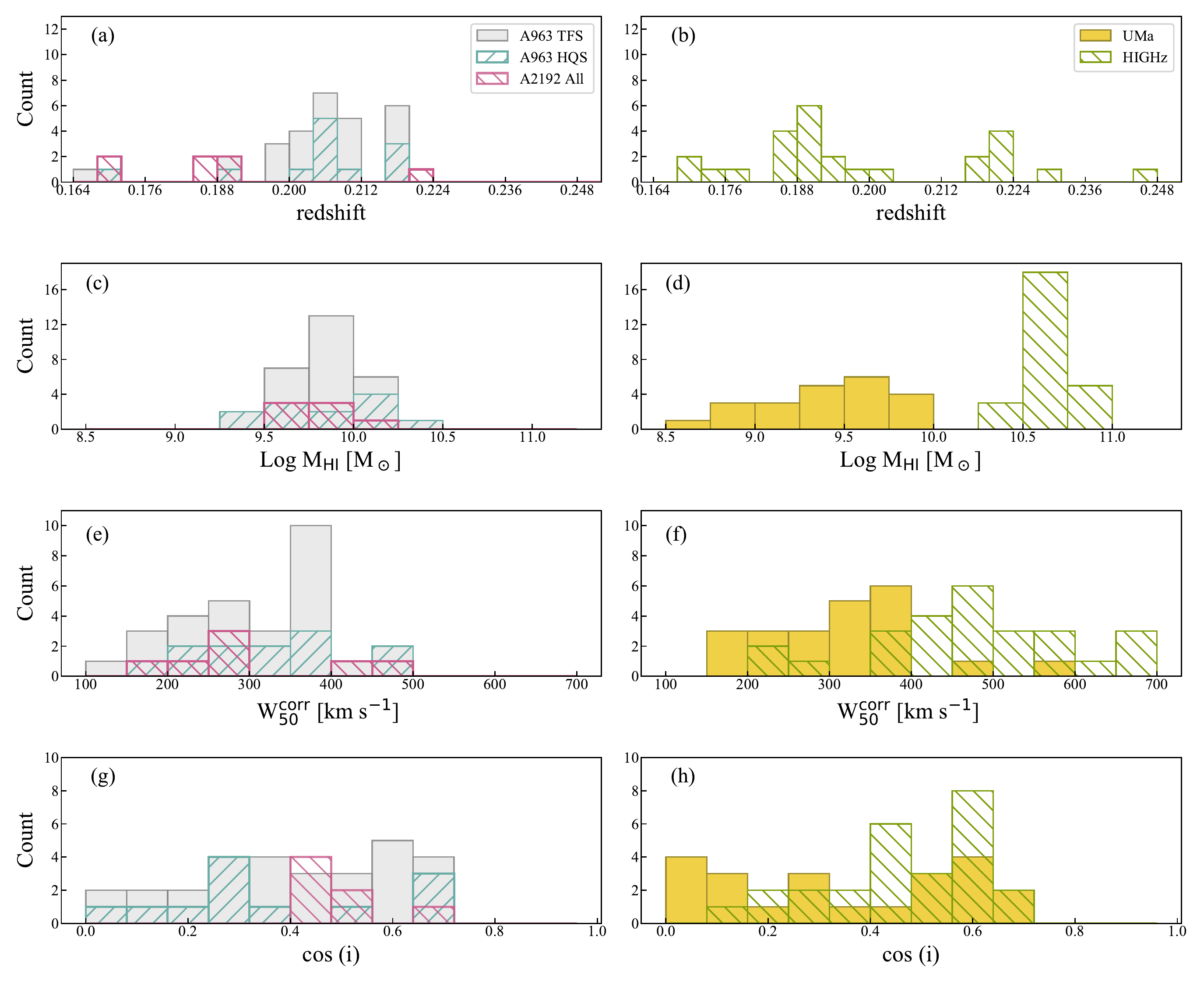}
    \caption{Histograms showing the various properties of the samples used for the TFR analysis. The \b\ samples are shown on the left. For A963, the grey histograms show the \textit{\textit{TFS}}, and the hatched cyan histograms show the \textit{HQS}. Note that the \textit{HQS} is a subset of the \textit{TFS}. For A2192, the TFS and HQS are identical, hence shown by the hatched magenta histograms. On the right, the histograms show the UMa and HIGHz samples, in orange and green (hatched) respectively. (a) and (b) illustrate the redshift distribution of the samples. The UMa sample is not a part of these histograms, since we assume an average distance of 18.6 Mpc, corresponding to z $\simeq$ 0. The \H\ mass distributions are shown in panels (c) and (d), while (e) and (f) illustrate the distribution of the \H\ line widths measured at 50\% of the peak flux. Lastly, panels (g) and (h) show the distribution of the cosine of the inclinations of galaxies in the various samples.}
    \label{fig:all_hist}
\end{figure*}

\subsubsection{K-corrections}

Corrections for cosmological reddening, known as K-corrections ($\mathrm{\kappa_{B,R}}$), were carried out with the help of the K-correction calculator by \citet{Chilingarian10}\footnote{the online K-corrections calculator can be found at http://kcor.sai.msu.ru/}. This was done for both the high redshift samples. As expected, we find that the K-corrections are larger in the B-band (average $\mathrm{\kappa_B}$ = 0.65 mag) than the R-band (average $\mathrm{\kappa_{R}}$ = 0.12 mag).

\subsubsection{Internal extinction}

Finally, the apparent magnitudes were corrected for internal extinction following \citet{Tully98}. Based on their prescription, the internal extinction correction is dependent on both the inclination and the corrected \H\ line widths of the galaxies, and is given by 

\begin{equation}
    \mathrm{A^{i}_{\mathrm{B,R}} = \gamma_{\mathrm{B,R}} \:} \mathrm{log} (a/b)
    \label{eq:intext_tully98}
\end{equation}

\noindent
where \textit{a/b} is the major-to-minor axis ratio. The $ \gamma_{\mathrm{B,R}}$ coefficient is line width dependent and calculated as

\begin{equation}
    \mathrm{\gamma_B =  1.57 + 2.75\: (log(W^{corr}_{\%}) - 2.5)} 
\end{equation} 

\begin{equation}
    \mathrm{\gamma_R = 1.15 + 1.88 \: (log(W^{corr}_{\%}) - 2.5)}
\end{equation}

\noindent
where W$^{\mathrm{corr}}_{\%}$ is the corrected \H\ line width as derived in Eq. \ref{eq:wcorr}. Note that this internal extinction correction method depends on both the inclination and the \H\ line width, recognising that galaxies of lower mass are usually less dusty. 

The final corrected magnitudes were calculated as

\begin{equation}
    \mathrm{m^{g,\kappa,i}_{\mathrm{B,R}} = m^{obs}_{\mathrm{B,R}} - A^g_{\mathrm{B,R}} - {\kappa_\mathrm{B,R}} - A_{\mathrm{B,R}}^i}
\end{equation}

\noindent
where m$^{\mathrm{obs}}_\mathrm{{B,R}}$ is the uncorrected, total apparent magnitude. The subscripts signify the choice of filter, namely B or R.

\subsubsection{Absolute magnitudes and luminosities}

Absolute B- and R-magnitudes, corrected for Galactic extinction, cosmological reddening and internal extinction, were calculated from the distance modulus equation, which takes into account the luminosity distance to each galaxy based on the adopted cosmology.

\begin{equation}
    \mathrm{M_\mathrm{B,R} = m^{g,\kappa,i}_{\mathrm{B,R}} - 5 \:log (D_{lum}/10)}
\end{equation}

\noindent
where m$\mathrm{^{g,\kappa,i}_{B,R}}$ is the corrected apparent magnitude, and the luminosity distance, D$_{\mathrm{lum}}$, is in parsecs. As mentioned before, a common distance of 18.6 Mpc was assumed for all galaxies in the UMa sample. Luminosities were computed from the corrected absolute magnitudes following the standard prescription, with adopted solar absolute magnitudes of M$_{\odot,B}$ = 5.31 and M$_{\odot,R}$ = 4.60.

\subsection{Computation of stellar and baryonic masses}\label{stelmass}
For the purpose of constructing the baryonic TFRs, stellar masses were calculated by converting the corrected, absolute B$-$ and R$-$band magnitudes using two different prescriptions that both involve a (B$-$R) colour term but with a different dependency. With the first prescription, stellar masses (M$_\star$) were determined empirically from maximum-disc rotation curve mass decompositions \citep[][ Chapter 6]{Verheijen97} of the UMa galaxies, using K$-$band luminosity profiles  and assuming a maximum-disc fit with an isothermal dark matter halo model. The stellar masses following from these rotation curve decompositions are tabulated in \citet[][ Chapter 6]{Verheijen97} and were used to calculate R$-$band stellar mass-to-light ratios \mstel/$\mathrm{L_R}$. These showed a linear correlation with the (B$-$R) colour of the UMa galaxies, expressed as: 

\begin{equation}
    \mathrm{M_{\star}^{mxd}\: [M_\odot]= L_R \times (1.35(B - R) - 0.399 })
     \label{eq:mstar1}
\end{equation}

\noindent
Maximum-disc-based stellar masses computed in this way for all the \b\ galaxies are provided in Col. (13) of Table \ref{tab:A963_opt_table_tf}. The full table is available online as supplementary material.

For comparison, we also considered an alternative stellar mass estimator, following \citet{Zibetti09}. This prescription was also adopted by \citet{Cybulski16} who calculated stellar masses for 23 \b\ galaxies using the INT B$-$ and R$-$band photometry following:

\begin{equation}
    \mathrm{M_{\star}^{zib}\: [M_\odot]= L_R \times 10^{-1.2 + 1.066 (B-R)} + 10^{0.04}}
    \label{eq:mstar2}
\end{equation}

\noindent
where the term 10$^{0.04}$ accommodates a conversion to the Kroupa initial mass function \citep{Kroupa01}. Stellar masses computed using this prescription are very similar to those based on Eq. \ref{eq:mstar1} and are therefore not tabulated. In what follows, we will use the M$_{\star}^{\rm mxd}$ prescription when referring to stellar mass M$_{\star}$.

From the inferred stellar and \H\ masses, the baryonic masses were calculated for all galaxies following:

\begin{equation}
    \mathrm{M_{bar} = M_\star + 1.4 M_{HI}}
    \label{eq:mbar}
\end{equation}

\noindent
where the factor 1.4 accounts for the contribution by Helium and metals. Molecular gas is not accounted for since its contribution to the statistical properties of the BTFR is found to be negligible \citep[e.g.,][]{Ponomareva18}.  

\section{Comparison of sample properties}\label{compprop}

As mentioned previously, all galaxies in the \b\ sample are selected to be isolated, undisturbed, \H\ rich, rotationally supported and geometrically inclined systems. In this section, some properties of the galaxy populations in the three comparison samples will be discussed. 

\subsection{Distribution of observables}


The top panels ($a$) and ($b$) of Fig. \ref{fig:all_hist} show the redshift distributions of the \b\ and HIGHz galaxies. Compared to the \b\ sample, the HIGHz sample reaches slightly further in redshift, out to z=0.245, but only 2 of the 28 galaxies are beyond the maximum redshift of the \b\ sample (z=0.224). The majority of the \textit{TFS} galaxies in the \b\ sample (29 out of 36) are located in the volume containing A963, with a significant fraction of galaxies (11 out of 29) within the redshift range that contains the large-scale structure in which A963 is embedded. The UMa sample is not shown in panel ($b$) as it is located at z=0, with an assumption that all galaxies in this sample are at a common distance of 18.6 Mpc. The HIGHz sample contains galaxies selected over a large area on the sky and, therefore, does not target a specific cosmic over-density.

\begin{figure}
\includegraphics[trim={0.2cm 0cm 0.5cm 1cm },clip,width=1\linewidth]{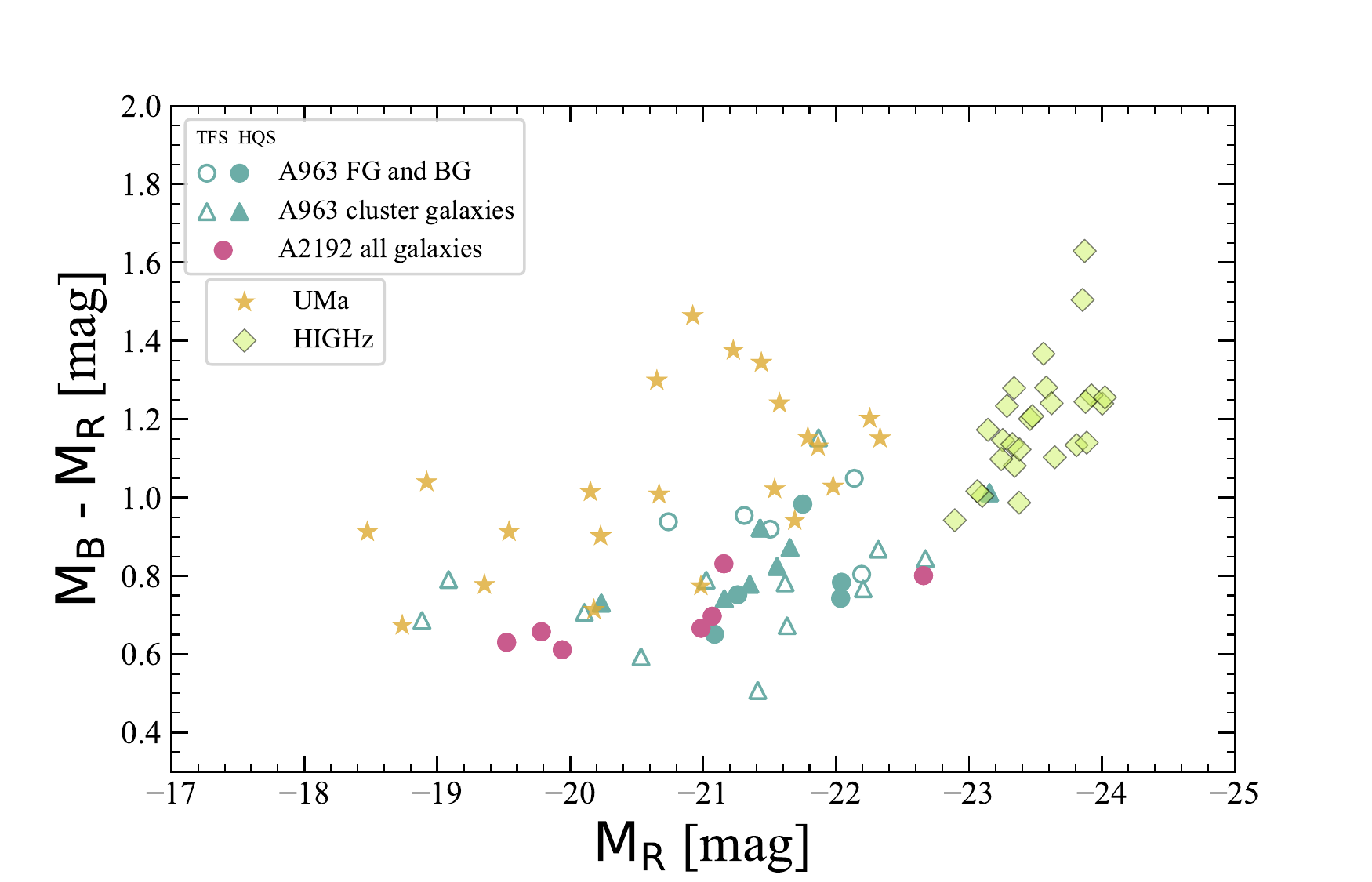}
\caption{The colour-magnitude diagram of the various samples after all photometric corrections have been applied consistently. Cyan triangles represent galaxies belonging to the cluster A963. Circles indicate galaxies from the foreground (FG) and background (BG) of the A963 volume (cyan) and all of the A2192 volume (magenta), which together make up the \b\ \textit{TFS} (open symbols), while the \textit{HQS} is shown by the solid symbols.  The UMa sample is indicated by orange stars. The HIGHz u,g,r,i magnitudes were transformed to Johnson B and Cousins R magnitudes using \citet{Cook14} and are shown by the green diamonds.}
\label{fig:cmd}
\end{figure}

\begin{figure}
\includegraphics[trim={2.5cm 0.8cm 18cm 0cm },clip,width=\linewidth]{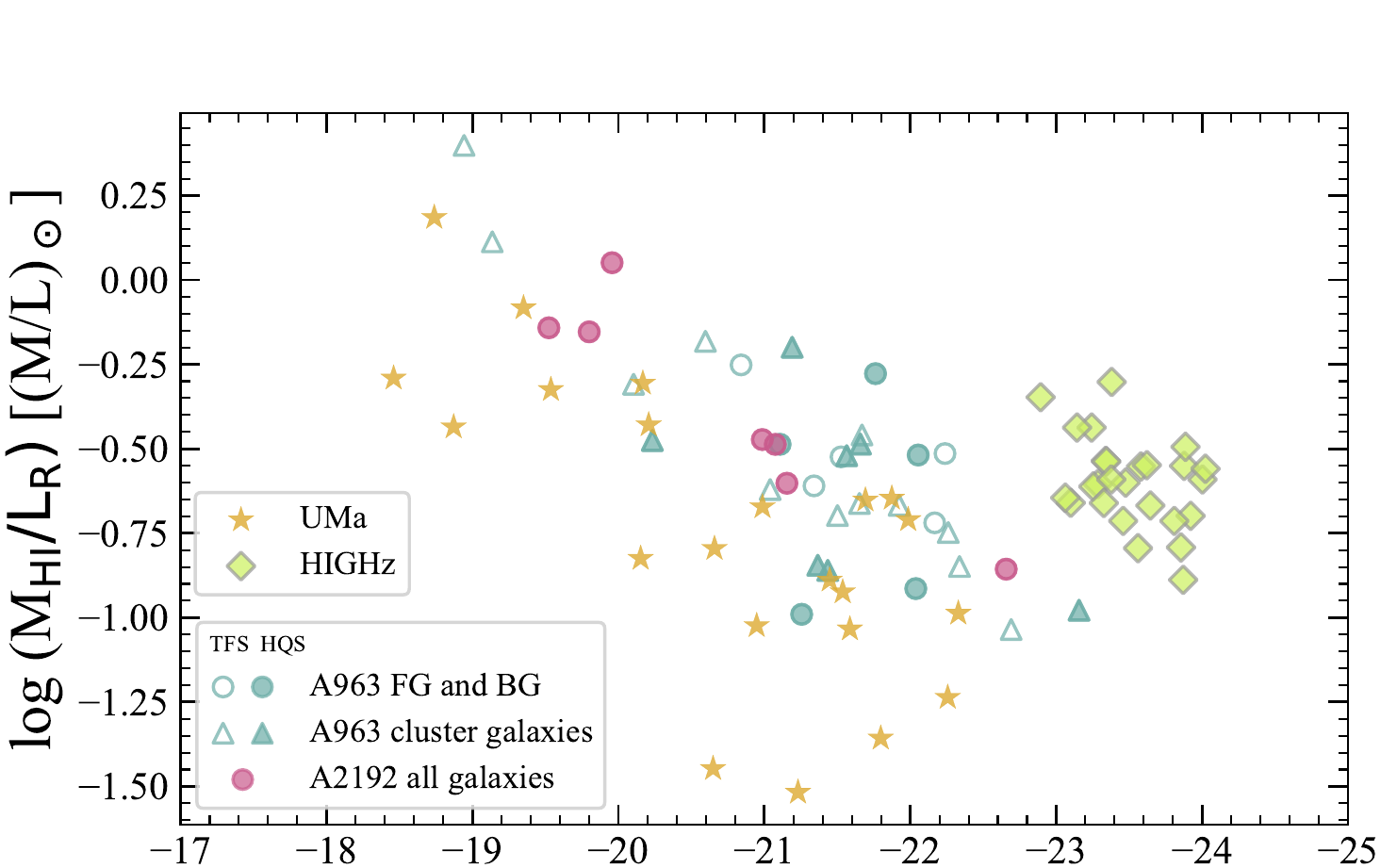}
\includegraphics[trim={17.5cm 0cm 3cm 0.6cm },clip,width=\linewidth]{MHI_L_comps.pdf}
\caption{R-band mass-to-light ratios for the various samples as a function of absolute R-band magnitude. Top: \mhi/L$_R$ as a function of absolute R-band magnitude;  Bottom: \mstel/L$_R$ as a function of absolute R-band magnitude. The colours and markers used to represent the samples are identical to Figure \ref{fig:cmd} and are also provided in the legend.}
\label{fig:mhil}
\end{figure}

\begin{figure*}
\includegraphics[trim={2.2cm 1.5cm 2.5cm 1cm },clip,width=0.95\linewidth]{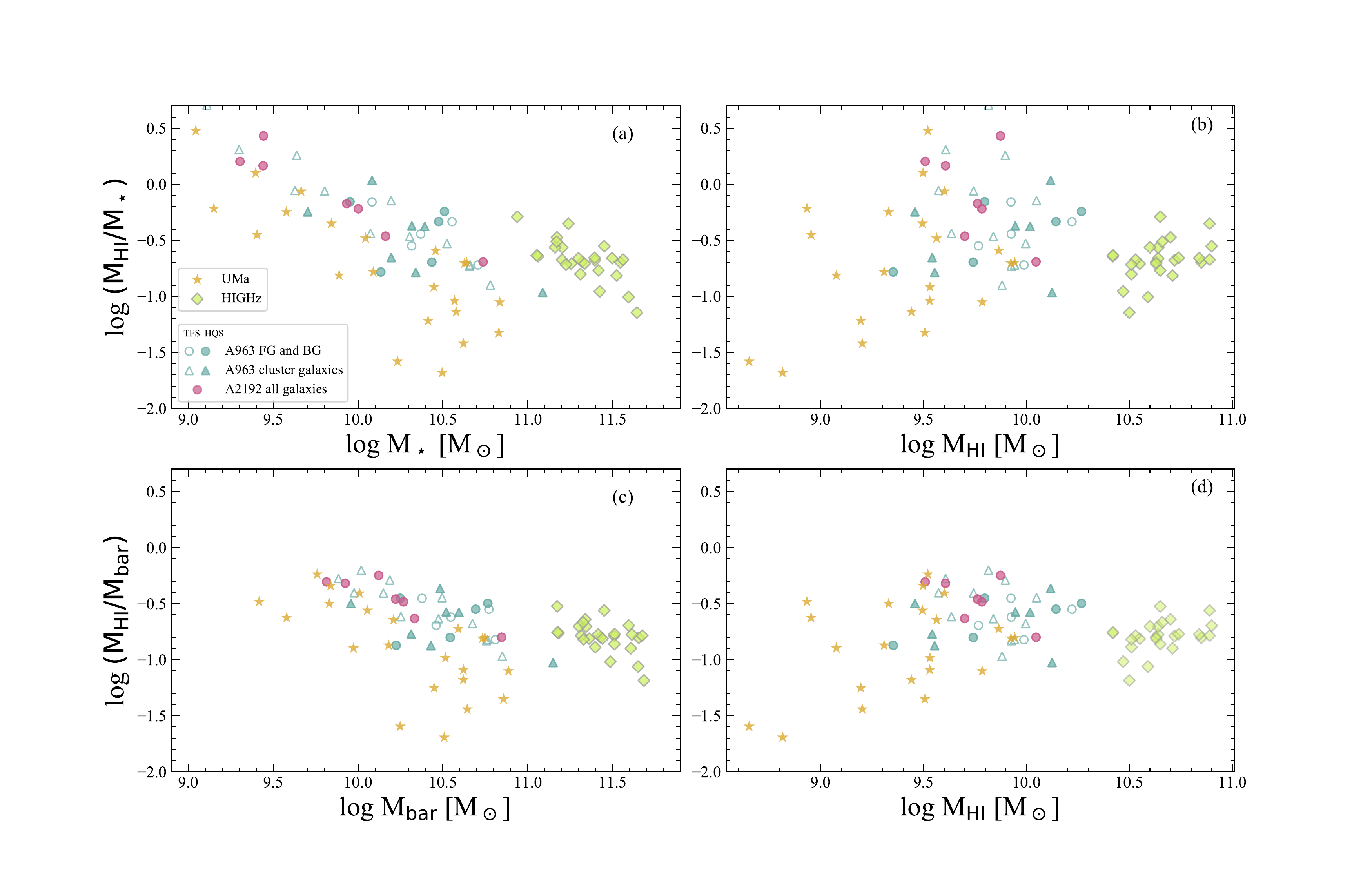}
\caption{Gas fractions of the galaxies in the various samples. Panels (a) and (b) show the \mhi/\mstel\ ratios as a function of \mstel\ and \mhi\ respectively, while panels (c) and (d) show the \mhi/\mbar\ ratios as a function of \mbar\ and \mhi\ respectively. The colours and markers are identical to Fig. \ref{fig:cmd}. Stellar and baryonic masses shown here are based on Eq. \ref{eq:mstar1} and Eq. \ref{eq:mbar}}.
\label{fig:mratios}
\end{figure*}

Panels ($c$) and ($d$) of Fig. \ref{fig:all_hist} show the distribution of \H\ gas masses of the sample galaxies, and there are some striking differences between the samples. The \b\ galaxies have \H\ masses in the range $9.3<{\rm log(M}_{\mathrm{HI}}/\rm M_\odot)<10.2$ while the UMa galaxies have notably smaller \H\ masses, covering the range $8.7<{\rm log(M}_{\mathrm{HI}}/ \rm M_\odot)<9.9$. The HIGHz galaxies, however, have significantly higher \H\ masses, covering the range $10.4<{\rm log(M}_{\mathrm{HI}}/ \rm M_\odot)<10.9$. None of the \b\ or UMa galaxies have an \H\ mass as high as the lowest \H\ mass of any HIGHz galaxy. This is not surprising as the HIGHz galaxies were selected as extremely massive, \H-rich galaxies at z$>$0.16 while the global \H\ profiles of the 'code 1' galaxies from \citet{Catinella15}, that constitute the HIGHz sample considered here, have the highest signal-to-noise and thereby a relatively high \H\ content.

Figure \ref{fig:all_hist}  ($e$) and ($f$) show histograms of the \H\ line widths measured at 50\% of the peak flux. The fastest rotators belong to the HIGHz sample ($\langle$W$_{\rm {50}}^{\rm {corr}}\rangle=$477 \kms), which is expected since this sample is selected to contain massive and luminous galaxies. On the other hand, the distributions of the \b\ (\textit{TFS}) and UMa samples are quite similar with $\langle$W$_{\rm {50}}^{\rm {corr}}\rangle=$ 313.5 \kms\ and 313.8 \kms\ respectively. The two UMa galaxies with the largest line widths are NGC 3953 and NGC 3992.

Lastly, panels ($g$) and ($h$) in Fig. \ref{fig:all_hist} illustrate the distribution of the inclinations of all our sample galaxies. Based on our sample selection criteria (see Sect. \ref{sampsel}), only galaxies with inclinations $>45^\circ$ were retained for this analysis. From the histograms, it is evident that the \b\ and UMa samples have flat distributions as expected for randomly oriented discs in a volume limited sample, whereas the HIGHz sample is biased towards more face-on systems with lower inclinations. This is an expected observational bias as galaxies with lower inclinations tend to have narrower \H\ emission lines that are easier to detect and measure.

\subsection{The colour$-$magnitude diagram}

The rest-frame M$_{\rm B}-$M$_{\rm R}$ versus M$_{\rm R}$ colour-magnitude diagram (CMD) of the three samples is shown in Fig. \ref{fig:cmd}. The magnitudes were corrected for Galactic extinction, cosmological reddening and internal extinction as described in Sect. \ref{photcor}. The three samples occupy different areas in the CMD. The \b\ and UMa samples cover a similar range in absolute magnitude ($-23<{\rm M}_{\rm R}<-19$ mag) but the \b\ galaxies ($\langle$M$_{\rm B}-$M$_{\rm R}\rangle=0.79$ mag) are on average 0.26 magnitudes bluer than the UMa galaxies ($\langle$M$_{\rm B}-$M$_{\rm R}\rangle=1.05$ mag) although there is some overlap in colour. The HIGHz galaxies ($\langle$M$_{\rm B}-$M$_{\rm R}\rangle=1.20$ mag) have similar colours as the UMa galaxies but are significantly brighter ($-24<{\rm M}_{\rm R}<-23$ mag) than the galaxies in both the \b\ and UMa samples. Only one \b\ galaxy falls in the magnitude range of the HIGHz sample. It should be recalled here that the applied correction for internal extinction not only depends on inclination but also on the corrected line width, which correlates with absolute luminosity through the TFR. This is discussed in more detail in Sect. \ref{corrdisc2} as this correction for internal extinction will eventually have some impact on the slope and zero point of the TFR. The fact that the HIGHz sample is so 'disjoint' from the \b\ and UMa samples in the CMD is another motivation to consider the HIGHz sample for illustrative purposes only and exclude it from a quantitative assessment of the cosmic evolution of the TFR zero point.

\subsection{\H\ and stellar mass-to-light ratios}

The top panel of Fig. \ref{fig:mhil} shows the \H\ mass-to-light (\mhi/L$_{\rm R}$) ratios of the galaxies in the various samples. As expected, the \b\ and UMa galaxies show a general increasing trend in the \H\ mass-to-light ratio with decreasing luminosity. For a given magnitude, however, the \b\ galaxies tend to have a slightly higher \mhi/L$_{\rm R}$ ratio. The sample averages are $\langle{\rm log}($\mhi/L$_{\rm R})\rangle=-0.52$ M$_\odot$/L$_\odot$ for the \b\ galaxies and $\langle{\rm log}($\mhi/L$_{\rm R})\rangle=-0.75$ M$_\odot$/L$_\odot$  for the UMa galaxies. The HIGHz galaxies do not follow the extrapolated trend to brighter magnitudes as they are overly gas rich with a sample average of $\langle{\rm log}($\mhi/L$_{\rm R})\rangle=-0.65$.

The bottom panel of Fig. \ref{fig:mhil} shows the stellar mass-to-light ratio in the R-band (M$_{\star}$/L$_{\rm R}$) according to Eq. \ref{eq:mstar1}. Since there is a rather strong dependence on the M$_{\rm B}-$M$_{\rm R}$ colour, the distribution of points is similar to that in the CMD, with 0.4$<$(M$_{\star}$/L$_{\rm R}$)$<$1.6 M$_\odot$/L$_\odot$. The HIGHz galaxies have similar stellar mass-to-light ratios compared to most of the UMa galaxies. The \b\ galaxies have a notably lower stellar mass-to-light ratio with a clear trend of lower M$_{\star}$/L$_{\rm R}$ values toward fainter galaxies. The sample average values $\langle$M$_{\star}$/L$_{\rm R}\rangle$ are 0.63  M$_\odot$/L$_\odot$ for the \b\ galaxies, 0.98 M$_\odot$/L$_\odot$ for the UMa galaxies and 1.2 M$_\odot$/L$_\odot$ for the HIGHz galaxies.

\subsection{\H\ mass fractions}

The \H\ mass to stellar mass ratios (\mhi/\mstel) as a function of stellar mass are shown in panel (a) of Fig. \ref{fig:mratios}. Not surprisingly, we see the same trend as in the top panel of Fig. \ref{fig:mhil} where we used the R-band luminosity instead of stellar mass. We confirm the well-known trend that lower mass galaxies tend to have a larger ratio of \H-to-stellar mass while the HIGHz galaxies seem to lie above the trend defined by the \b\ and UMa galaxies, as expected given the selection criteria for the HIGHz sample. In panel (b) of Fig. \ref{fig:mratios} we plot \mhi/\mstel\ as a function of \mhi\ and note that the correlation seen in panel (a) has disappeared. The \mhi/\mstel\ ratios for the \b\ galaxies tend to be higher than for the UMa and HIGHz galaxies with sample averages of $\langle{\rm log}($\mhi/\mstel$)\rangle=-0.74$ and $-0.73$ for the UMa and HIGHz samples respectively, while $\langle{\rm log}($\mhi/\mstel$)\rangle=-0.31$ for the \b\ sample.

In panels (c) and (d) of Fig. \ref{fig:mratios} we plot the \mhi/\mbar\ fractions as a function of \mbar\ and \mhi\ respectively. We observe in panel (c) that, compared to panel (a), the trend of \mhi/\mbar\ versus \mbar\ has become shallower as the \H\ mass is a larger fraction of \mbar\ for galaxies with a lower \mbar. Interestingly, in panel (d), plotting \mhi/\mbar\ versus \mhi\ instead of plotting \mhi/\mstel\ versus \mhi\ shows a significantly smaller scatter, while the HIGHz galaxies do not stand out significantly.

It is evident that the sample of \b\ galaxies tends to have a higher \mhi/\mbar\ ratio than the UMa and HIGHz galaxies. The sample averages for the UMa and HIGHz galaxies are $\langle{\rm log}($\mhi/\mbar$)\rangle=-0.89$ and $-0.83$ respectively, while $\langle{\rm log}($\mhi/\mbar$)\rangle=-0.58$ for the \b\ sample. From Fig. \ref{fig:mratios} we conclude that the \b\ galaxies at z=0.2 are relatively more \H-rich than the UMa galaxies at z=0, even though they have similar baryonic masses.

Finally, we remark that the larger vertical spread of the UMa sample in Figs. \ref{fig:mhil} and \ref{fig:mratios} is due to the fact that the UMa sample has a better \H\ mass sensitivity than both \b\ and HIGHz, and hence includes galaxies with significantly lower \H\ masses.

\begin{figure}
\begin{center}

\includegraphics[trim={0.69cm 5cm 1.7cm 5cm},clip, width=1\linewidth]{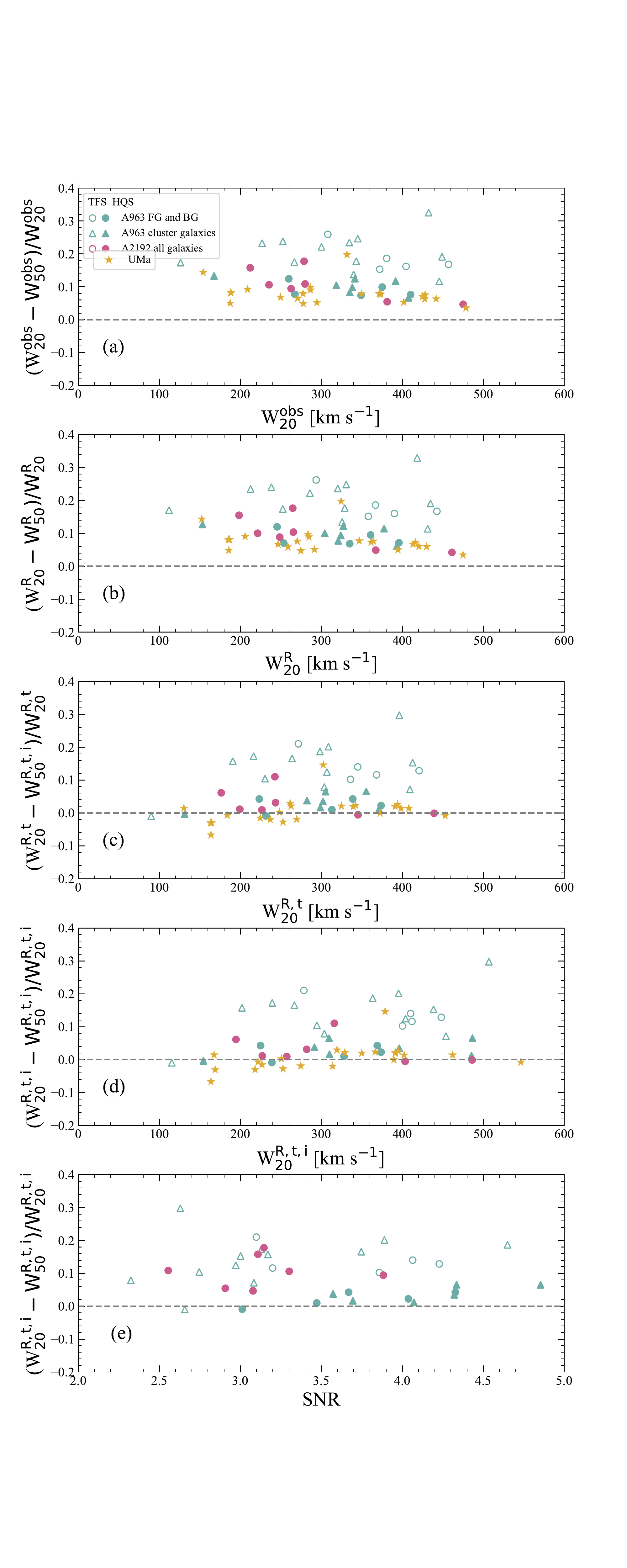}
\caption{Fractional differences in the \H\ line widths, defined as ($\mathrm{W_{20}}$-$\mathrm{W_{50}}$)/$\mathrm{W_{20}}$ for the various comparison samples.  The colours and markers used to represent the samples are identical to Figure \ref{fig:cmd} and are also provided in the legend. From top to bottom: (a) fractional differences in the observed line widths, (b) after correcting for instrumental broadening (superscript \textit{R)}, (c) after turbulent motion corrections (superscript \textit{R,t)}, (d) after inclination corrections (superscript \textit{R,t,i)}. Panel (e) illustrates the fractional differences in the corrected line widths as a function of the average SNR derived from the \H\ profiles for the \b\ galaxies.}
\label{fig:fracdiff1}
\end{center}
\end{figure}

\begin{figure*}

    \includegraphics[trim={1cm 0cm 1cm 1cm },clip,width=0.45\linewidth]{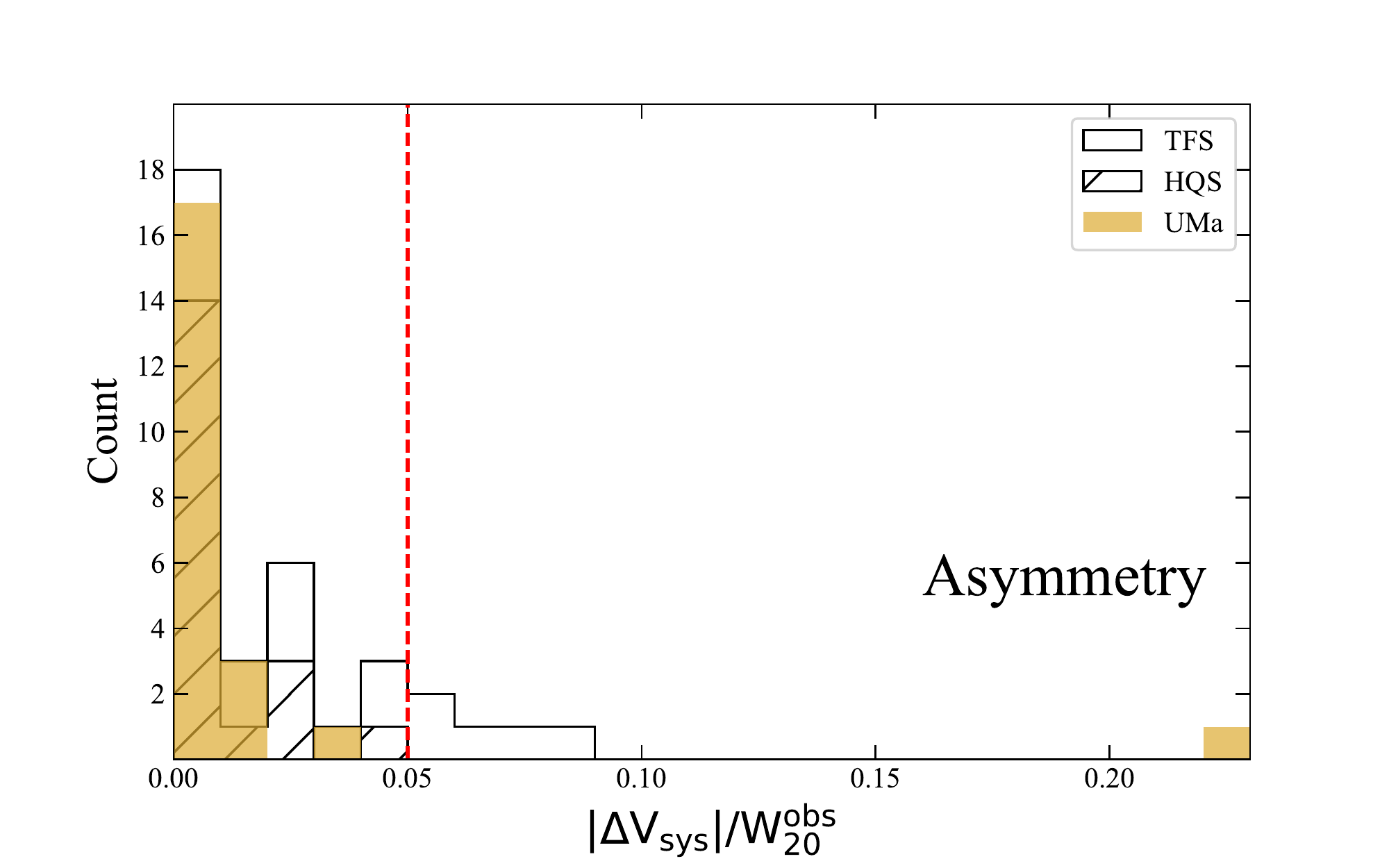} 
    \includegraphics[trim={1cm 0cm 1cm 1cm },clip,width=0.45\linewidth]{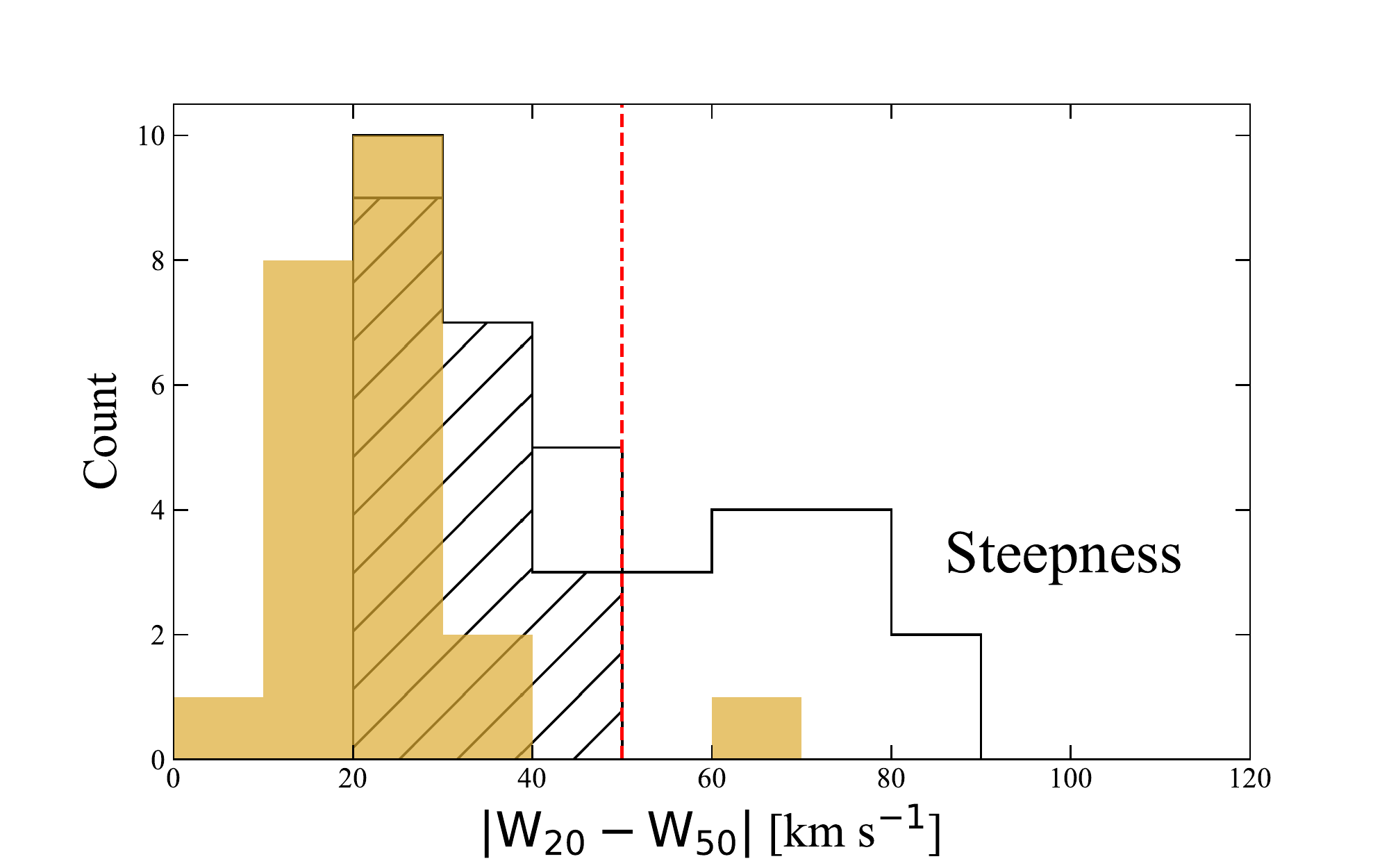}

    \caption{Histograms comparing the \H\ profile shapes of the comparison samples; the open histogram shows the \textit{TFS} while the hatched histogram shows the \textit{HQS} (which is a subset of the \textit{TFS}). The UMa sample is shown in orange. The red dashed lines indicate the thresholds applied during the sample selection process for creating the \textit{HQS}. Left: histograms showing asymmetries in the form of the absolute fractional differences in the systemic velocities derived from $\mathrm{W^{obs}_{20}}$ and $\mathrm{W^{obs}_{50}}$ respectively; Right: Histograms showing the difference in the observed line widths, which is a quantification of the steepness of the \H\ profiles. }
    \label{fig:steep}
\end{figure*}

\subsection{\H\ profile shapes}

Ideally, the \H\ profiles of isolated spiral galaxies with suitable inclinations should have steep edges, allowing the two line width measures $\mathrm{W_{20}}$ and $\mathrm{W_{50}}$, once properly corrected for instrumental spectral resolution, turbulent motion and inclination, to yield the same circular velocity. To inspect this notion is some detail, we compared the $\mathrm{W_{20}}$ and $\mathrm{W_{50}}$ profile widths of the galaxies in the various TFR samples, as illustrated in Fig \ref{fig:fracdiff1}. The figure shows the fractional differences between the observed $\mathrm{W_{20}}$ and $\mathrm{W_{50}}$ line widths (a), and the same after accumulative corrections for instrumental resolution (b), turbulent motion (c) and inclination (d). For details on the applied corrections, see Sect. \ref{widthcorr}. In panel (a), all samples deviate from the zero line, which is expected since the \H\ profile edges are not infinitely steep ($\mathrm{W^{obs}_{20}}$$>$$\mathrm{W^{obs}_{50}}$ always).

In the case of the UMa sample, this offset is corrected as we move downwards to panel (c), in which the line widths are corrected for both, instrumental spectral resolution and turbulent motion. It is also immediately evident in panel (c) that there still exists an offset and a larger scatter in the \b\ samples that could not be eliminated by applying the same corrections. The offset, however, is smaller for the \textit{HQS} (solid symbols) than for the \textit{TFS} (open symbols) by merit of the more stringent, quantitative criteria applied to the profiles of the \textit{HQS} galaxies. The average fractional difference between the $\mathrm{W^{R,t}_{20}}$ and $\mathrm{W^{R,t}_{50}}$ line widths is 0.16 for the \b\ \textit{TFS} galaxies, compared to 0.08 for the UMa galaxies. Naturally, correcting for inclination, as shown in panel (d), does not further reduce the fractional difference for any of the samples. It is important to point out in panel (e) that no trend is observed in the fractional differences as a function of the average Signal-to-Noise Ratio (SNR) of the \H\ profiles. In Sect. \ref{discussion} we address the possible reasons for this offset of the \b\ galaxies.

Figure \ref{fig:steep} shows the histograms of the asymmetries and the steepness of the \H\ profile edges of the \b\ and UMa galaxies. The red dashed lines indicate the applied cuts in the asymmetry and steepness of the profiles as quantified in Sect. \ref{sampsel} (E2 and E3) respectively. It can be noted that while these thresholds do exclude \b\ galaxies with the most asymmetric profiles or shallow profile edges from the \textit{TFS}, the profiles of the \textit{HQS} galaxies are still more asymmetric and with less steep edges than the profiles of the UMa galaxies. The origin of such profiles is discussed in Sect. \ref{velmeasure}. From the UMa sample, we note that the profile of NGC 3729 is more asymmetric than the imposed threshold while NGC 4138 has a profile with less steep edges compared to the threshold applied for the selection of the \b\ \textit{HQS} galaxies. We note that the HI rotation curve of NGC 4138 is strongly declining.

\begin{figure*}
   \includegraphics[trim={4.5cm 1cm 3.5cm 0cm},clip,width=\textwidth]{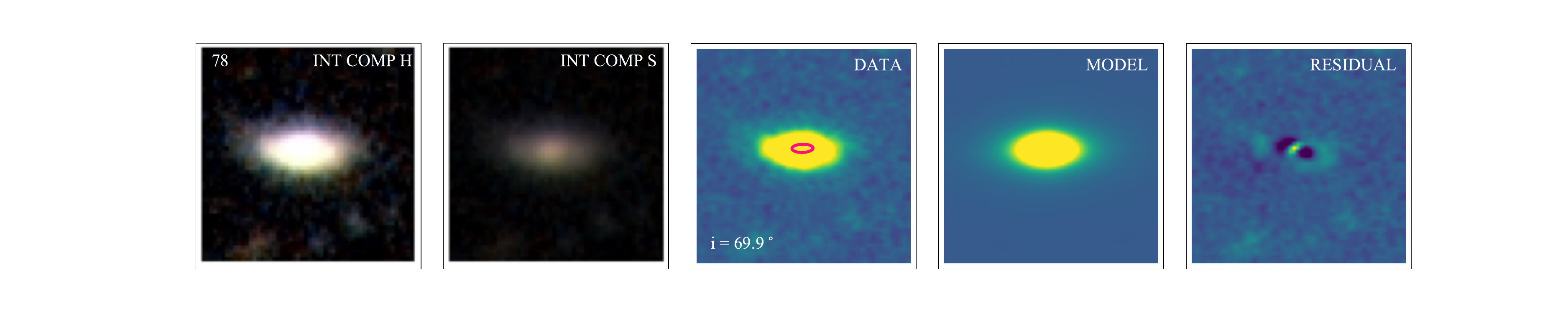}
   \includegraphics[angle =270,trim={6cm 1.5cm 10cm 3.5cm},clip,width=1\textwidth]{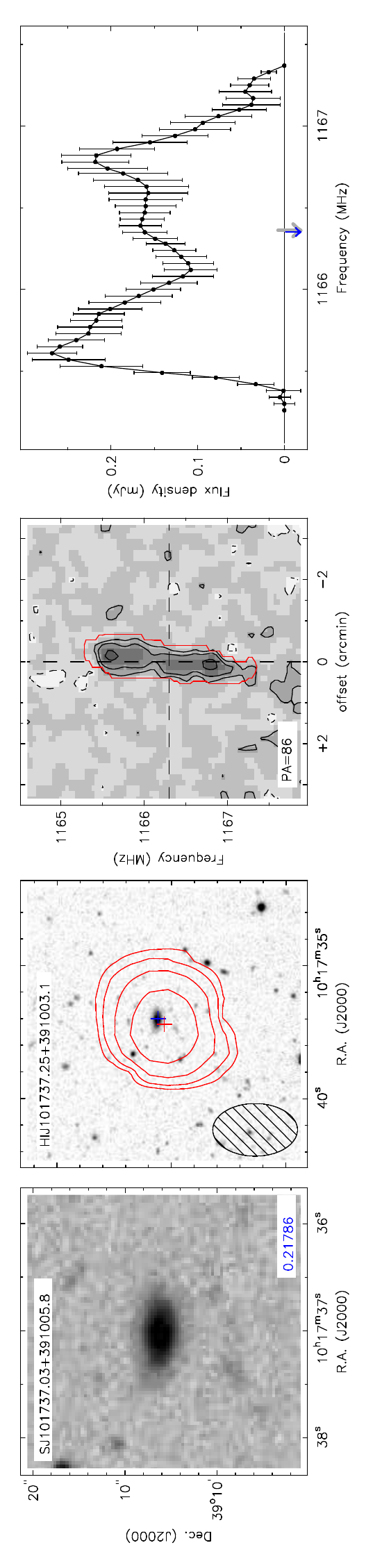}
\caption{An example of the layout of the atlas, showing a galaxy from the \textit{TFS}. The top row shows the optical properties and \textit{galfit} outputs, whereas the bottom panels mostly show the \H\ properties. The atlas layout is described in detail in Sect. \ref{atlas}. The full atlas is available online.}
\label{fig:atlaslayout}
\end{figure*}

\section{The \b\ TF catalogue and atlas}\label{atlas_tf}

We present here the tables as well as an atlas containing the \H\ and optical properties of the galaxies chosen to represent the TF sample from \b. The tables include observed, corrected and inferred properties following the methodology described in Sect. \ref{corr}.

\subsection{The catalogues}\label{catalogue}

Examples of \H\ and optical catalogues for the \b\ TF sample are provided in Tables \ref{tab:A963_HI_table_tf} and \ref{tab:A963_opt_table_tf} respectively. A description of the columns for both tables is provided in their respective table captions. The full catalogues are available as supplementary online material.

\subsection{The atlas}\label{atlas}

The catalogue presented in Tables \ref{tab:A963_HI_table_tf} and \ref{tab:A963_opt_table_tf} is accompanied by an atlas page for each galaxy. The full atlas is available online as supplementary material. Figure \ref{fig:atlaslayout} illustrates an example of the atlas layout, consisting of two rows. The top row consists of 5 panels highlighting the optical morphology and the results from the \textit{galfit} modelling. The bottom row consists of 4 panels highlighting the \H\ data, similar to the atlas pages in Paper 1. Each panel is briefly described below. \\

\noindent The top row of each atlas page displays the following from left to right:\\
\textit{Panels} (1) and (2): INT colour composite images (20$\times$20 arcsec$^2$) with a hard and soft contrast, respectively. The top-left corner of panel (1) shows the catalogue number as listed in Col. (2) of Table \ref{tab:A963_HI_table_tf}.\\
\textit{Panel} (3): The optical R-band image of the galaxy. The red ellipse depicts the fitting result returned by \textit{galfit} as derived at the effective radius and deconvolved for the seeing.\\
\textit{Panel} (4): The resulting model as returned by \textit{galfit}.\\
\textit{Panel} (5): The residual image as returned by \textit{galfit}. \\

\noindent The bottom row of each atlas page displays the following from left to right:\\
\textit{Panel} (1): INT R-band image (30$\times$30 arcsec$^2$) of the optical counterpart. The SDSS ID is given in the top-left corner. The optical redshift is printed in blue in the bottom-right corner of the image for those sources with optical spectroscopy. \\
\textit{Panel} (2): A zoomed-out INT R-band image (2$\times$2 arcmin$^2$) with \H\ contours from the total \H\ map overlaid in red. The \H\ ID is provided in the top-left corner. The optical centre is indicated by the blue cross while the \H\ centre is indicated by the red cross. The contours are set at \H\ column density levels of 1, 2, 4, 8, 16, and 32 $\times$10$^{19}$ cm$^{-2}$.  \\
\textit{Panel} (3): The position–velocity diagram along the optical major axis, extracted from the R4 ($\sim$38 \kms) cube (see Paper I). The contour levels correspond to –2 (dashed), 2, 4, 6, 9, 12, 15, 20, and 25 times the local rms noise level. The mask within which the \H\ flux was determined is outlined in red. The position angle is given in the bottom-left corner of the diagram. The central frequency and the \H\ centre are indicated by the vertical and horizontal dashed lines, respectively. \\
\textit{Panel} (4): The global \H\ profile as extracted from the R4 cube within the mask indicated in panel (3). The \H\ and optical redshifts are indicated by the grey and blue arrows respectively. Further details about the data processing can be found in Paper 1.\\

\begin{figure*}
\includegraphics[trim={0.3cm 0 0cm 0.2cm},clip, width=0.45\linewidth]{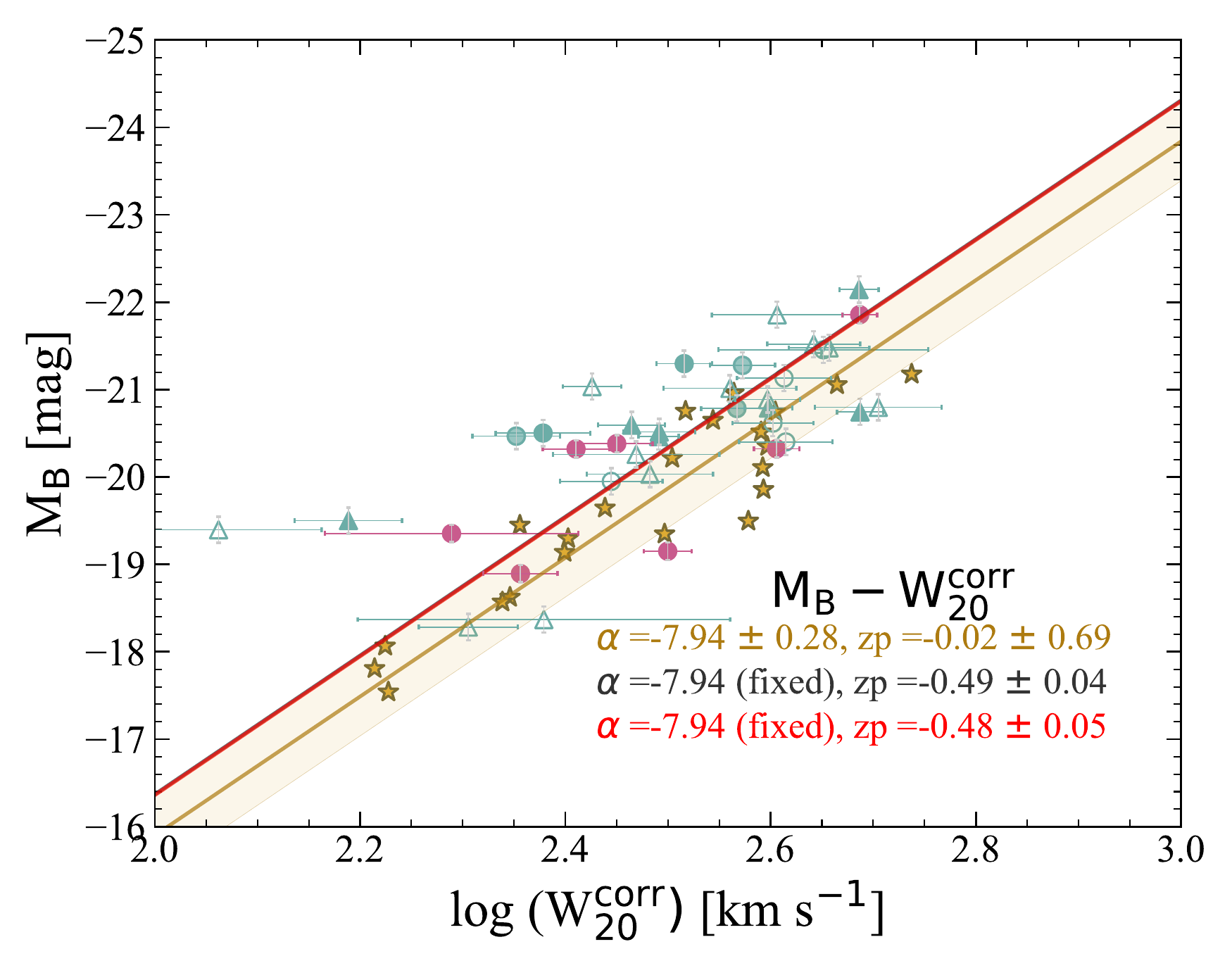}
\includegraphics[trim={0.3cm 0 0cm 0.2cm},clip, width=0.45\linewidth]{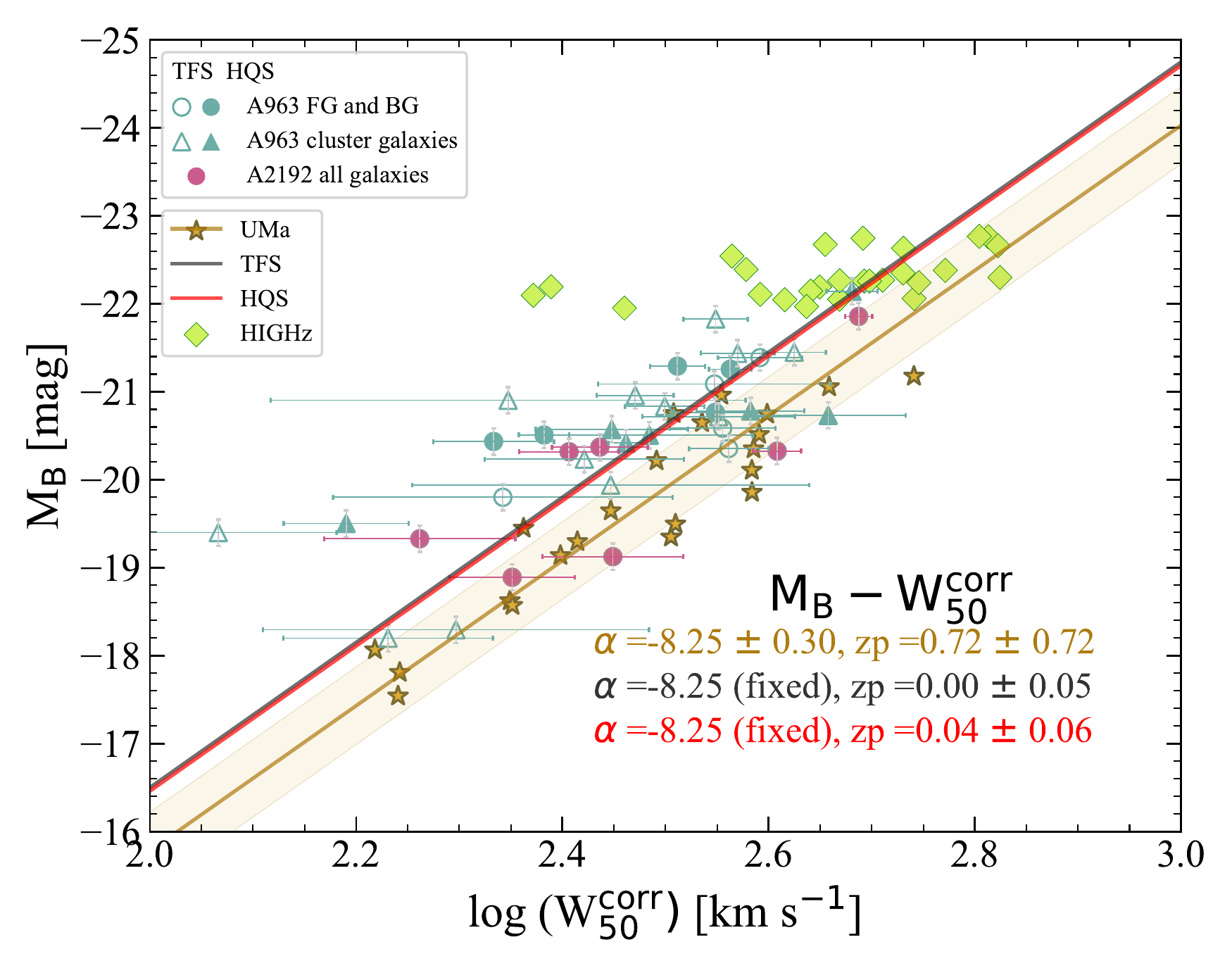}
\includegraphics[trim={0.3cm 0 0cm 0.2cm},clip, width=0.45\linewidth]{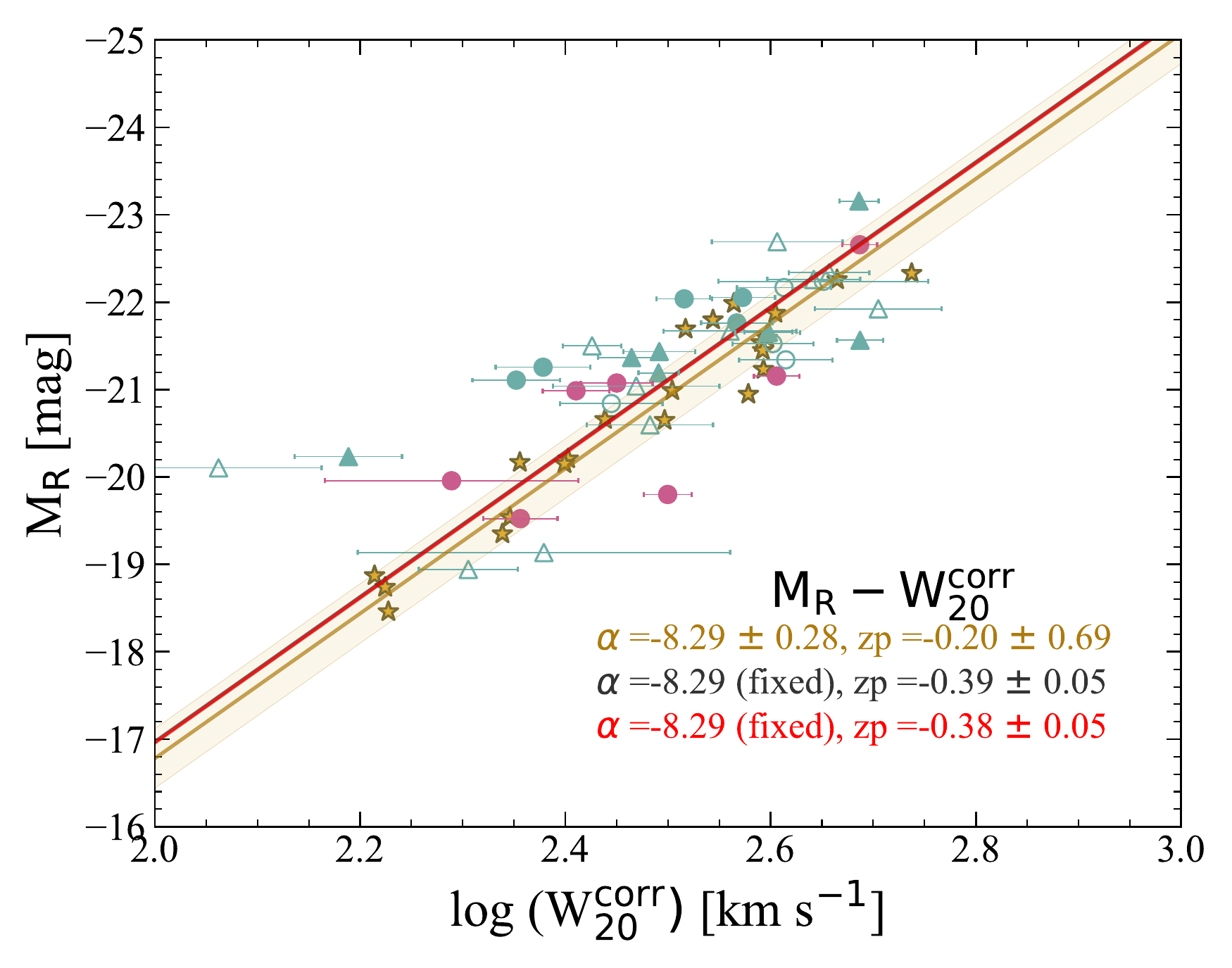}
\includegraphics[trim={0.3cm 0 0cm 0.2cm},clip, width=0.45\linewidth]{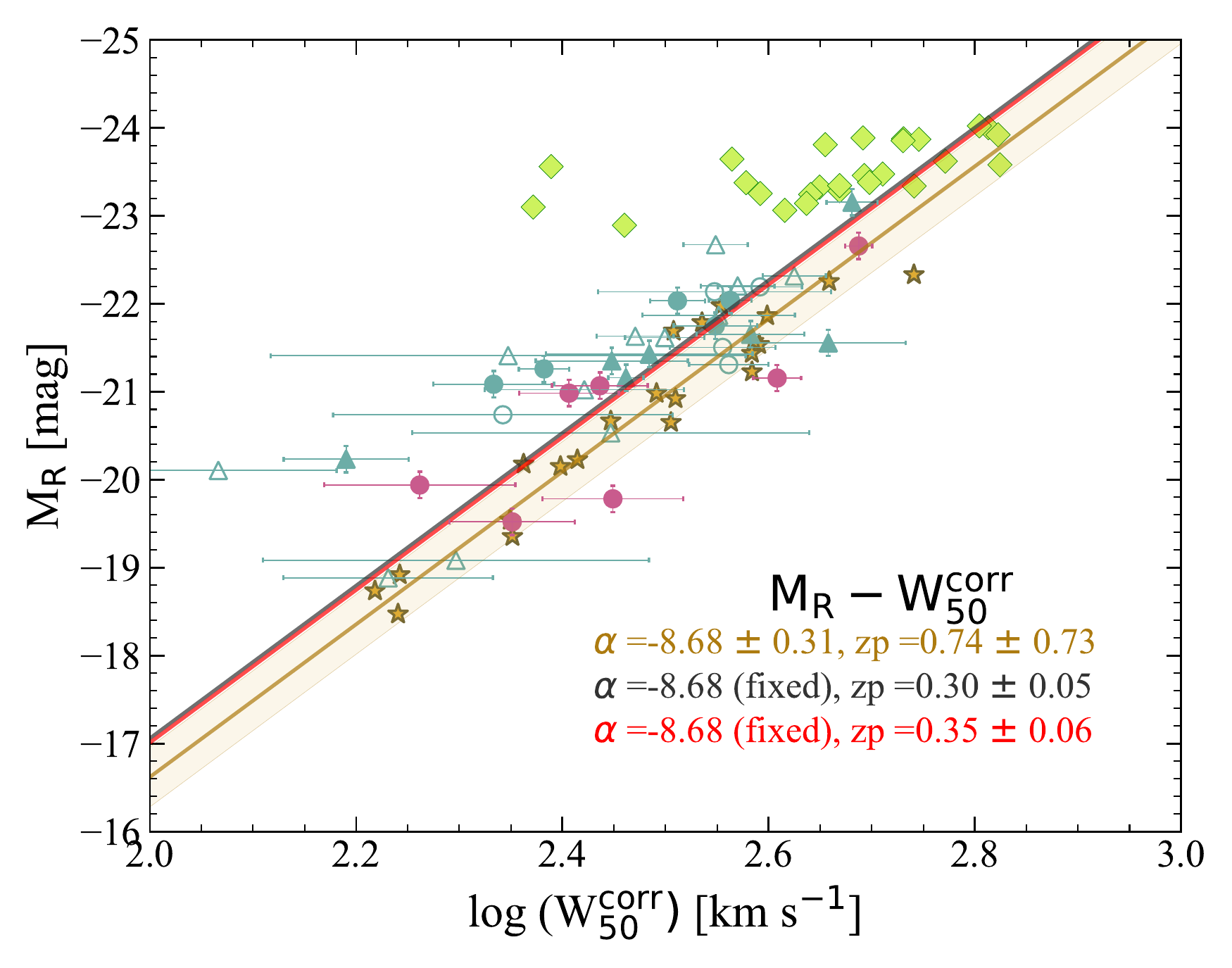}
\caption{Top panels: The B-band TFRs using the \wct\ (left) and \wcf\ (right) line widths as velocity tracers. Bottom panels: Similar to the top panels but for the R-band instead. The z=0 UMa TFR \citep{Verheijen01} is shown by the orange markers and a corresponding orange best-fit line, while the orange band represents the total observed rms scatter in the UMa data points. Overlays of the HIGHz galaxies are shown in the \wcf\ TFRs for illustrative purposes. The colours and symbols used are identical to Fig. \ref{fig:cmd}. The best fit \b\ TFRs, with slopes fixed to the UMa TFR, are shown by the grey (\textit{TFS}) and red (\textit{HQS}) lines, which are indistinguishable in most cases. The slopes ($\alpha$) and zero points (zp) of the various TFRs are printed in the bottom-right corner of the panels. Note that the large error on the zero point of the UMa fit is a consequence of the fact that the slope was a free parameter. The fit results are tabulated in Table \ref{tab:fitparams}}.
\label{fig:tfr}
\end{figure*}

\begin{figure*}
\includegraphics[trim={0cm 0 0cm 0.5cm},clip, width=0.45\linewidth]{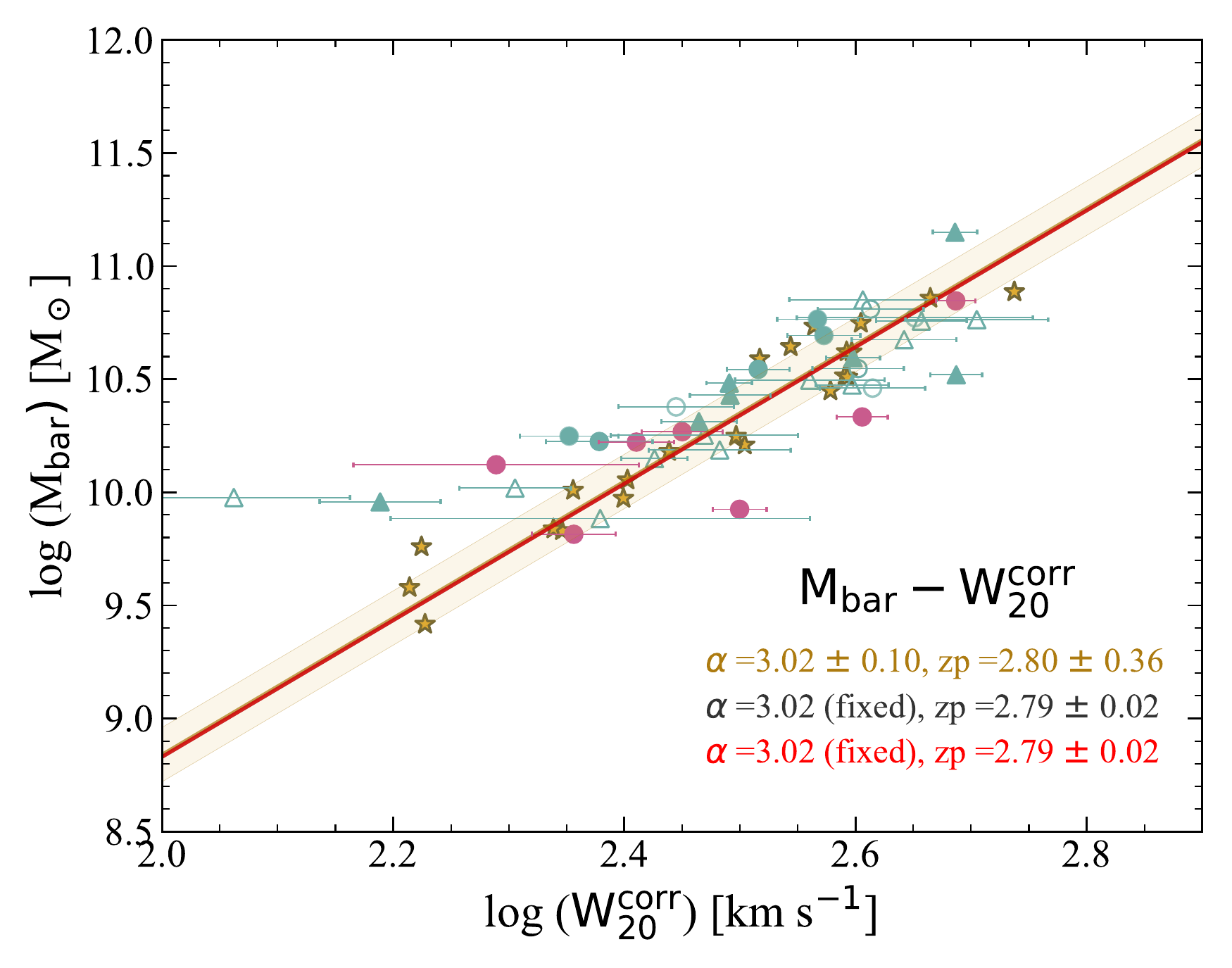}
\includegraphics[trim={0cm 0 0cm 0.5cm},clip, width=0.45\linewidth]{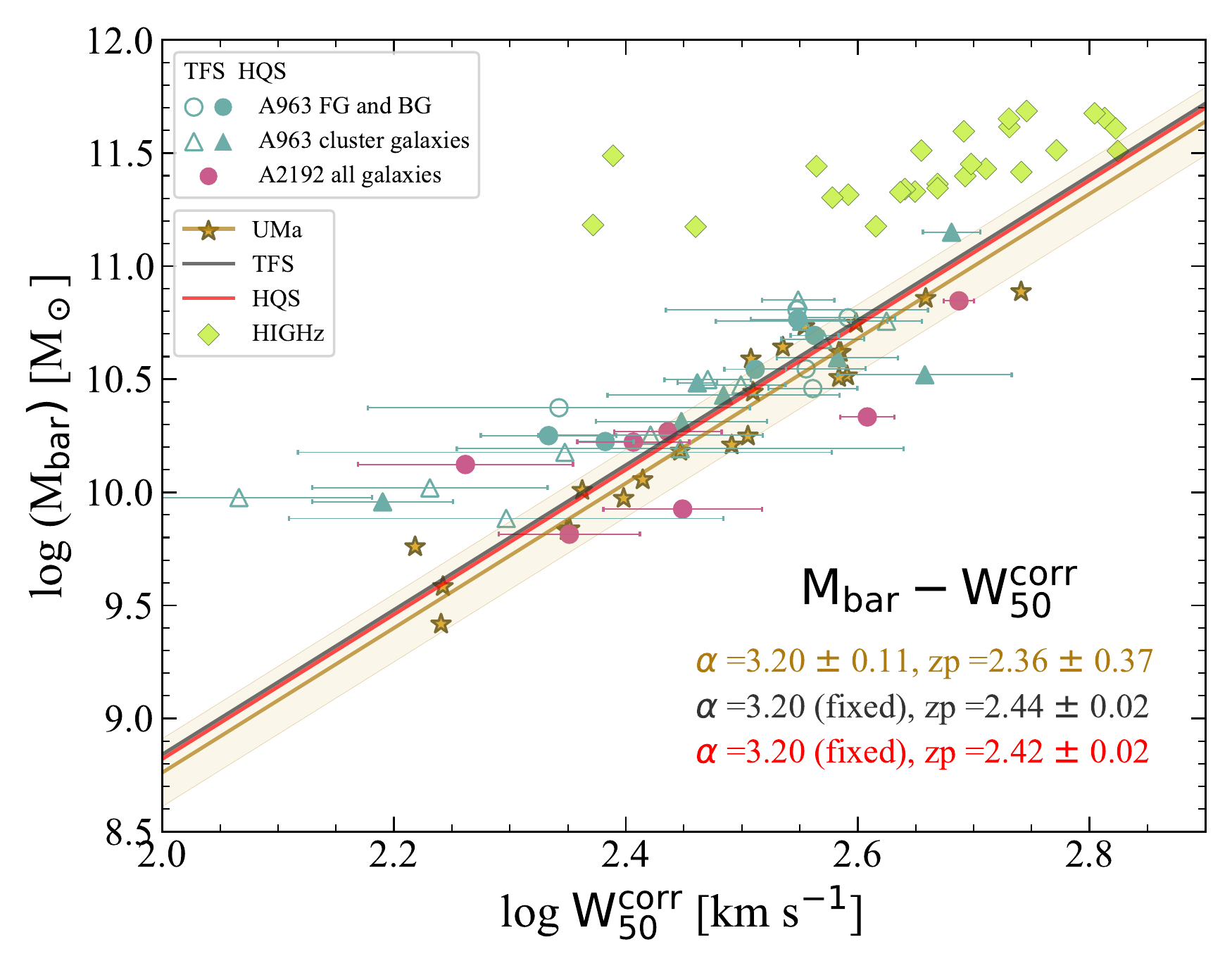}
\caption{BTFRs based on the two velocity measures, \wct\ (left) and \wcf\ (right) with baryonic masses (1.4\mhi + \mstel). Note that the large error on the zero point of the UMa fit is a consequence of the fact that the slope was a free parameter. The symbols, colours and layout are identical to Fig. \ref{fig:tfr}. The fit results are tabulated in Table \ref{tab:fitparams}. Again, note that the fits to the \textit{TFS} and \textit{HQS} are indistinguishable.}
\label{fig:btfr}
\end{figure*}

\section{The Tully-Fisher Relations}\label{results_tf}

Presented in this section are the TFRs obtained using the corrected \H\ line widths as tracers of the rotational velocities, and different photometric bands as well as derived quantities such as baryonic masses. We begin with explaining the fitting methods applied to the various samples, followed by a presentation of the luminosity-based TFRs and the baryonic TFRs using rotational velocities derived from the $\mathrm{W^{corr}_{20}}$ and $\mathrm{W^{corr}_{50}}$ \H\ line widths.

    

\begin{figure}
\includegraphics[trim={0.2cm 0cm 0.3cm 0.2cm},clip, width=0.92\linewidth]{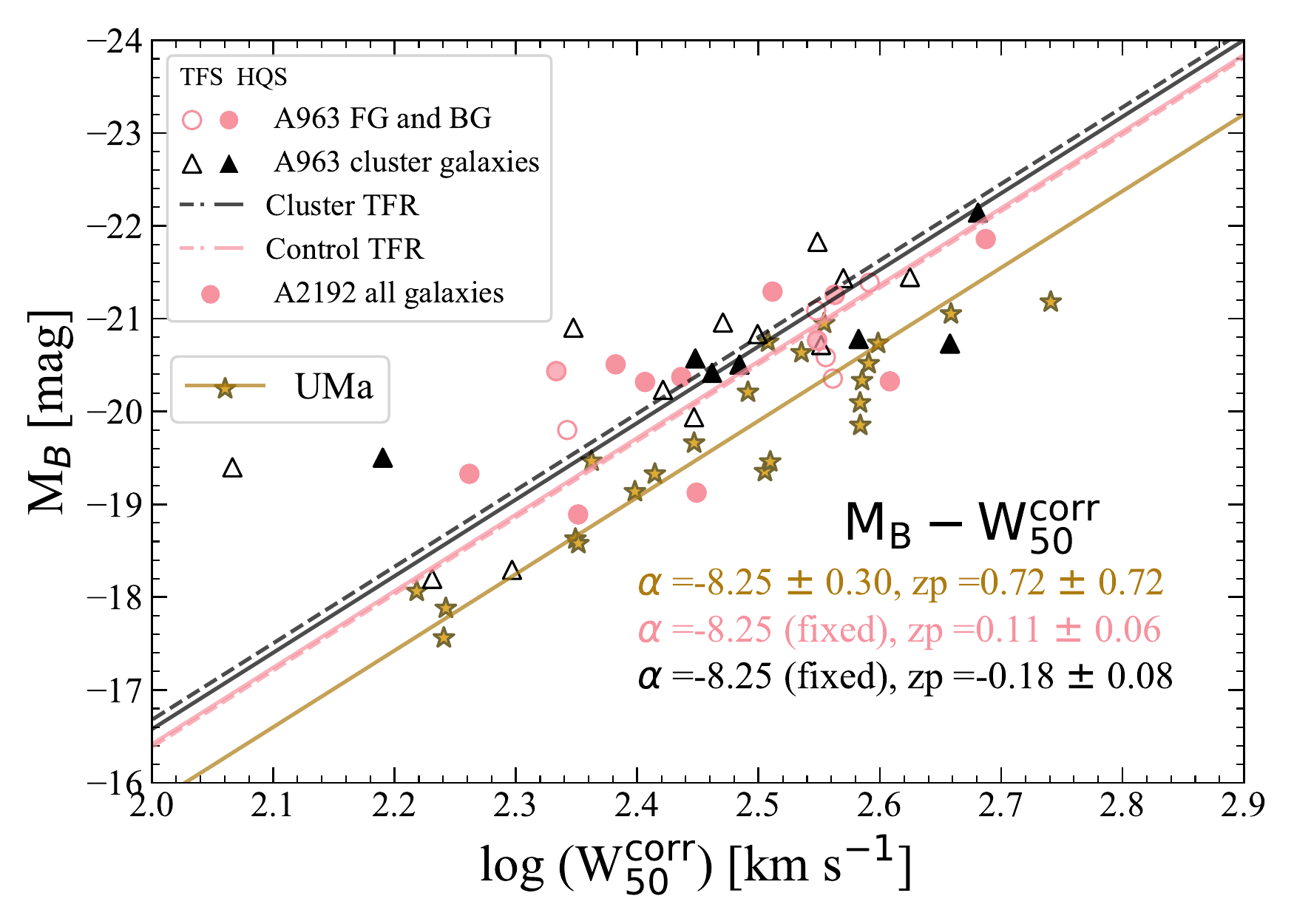}
\includegraphics[trim={0.2cm 0cm 0.3cm 0.2cm},clip, width=0.92\linewidth]{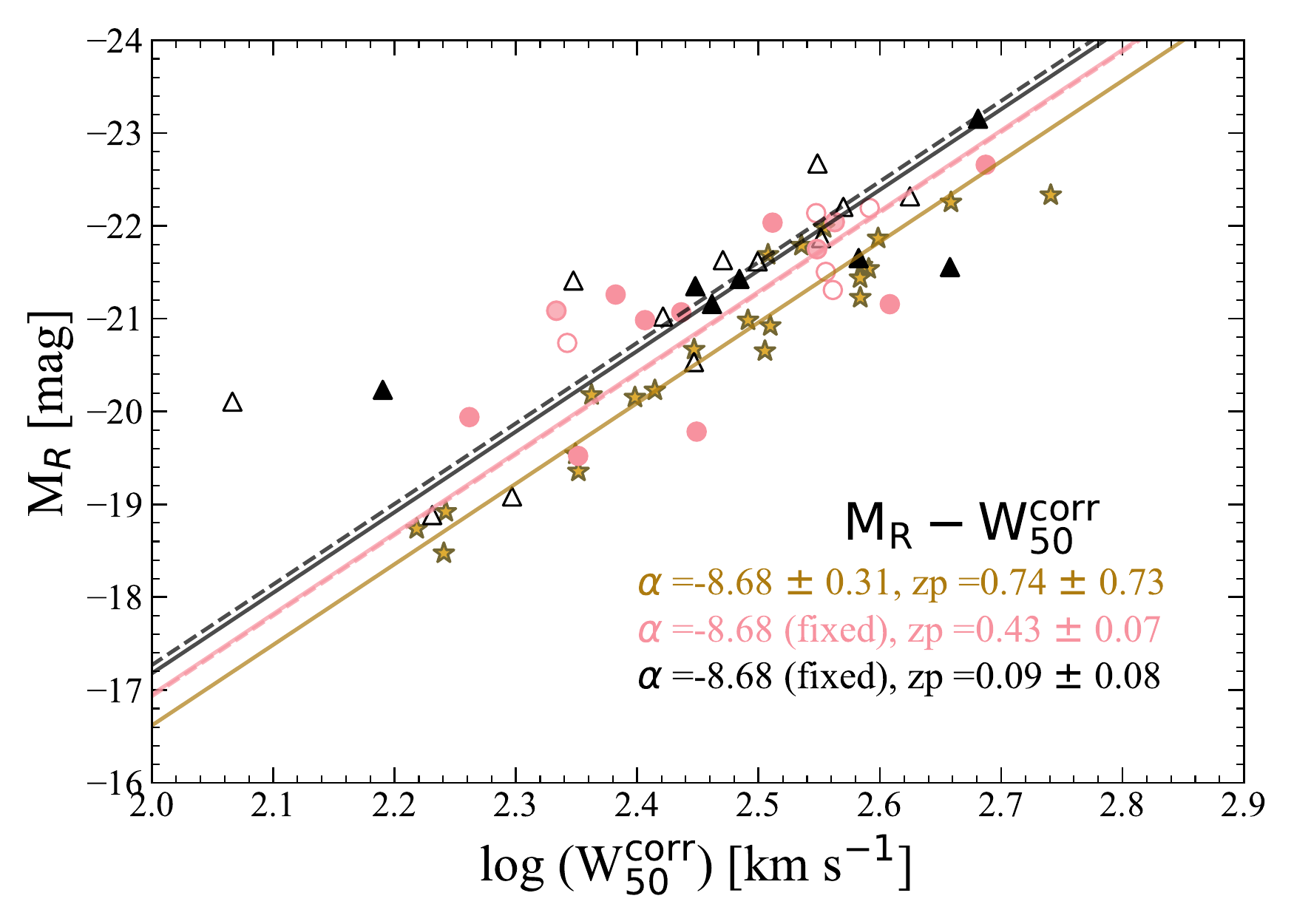}
\includegraphics[trim={0.2cm 0cm 0.3cm 0.2cm},clip, width=0.92\linewidth]{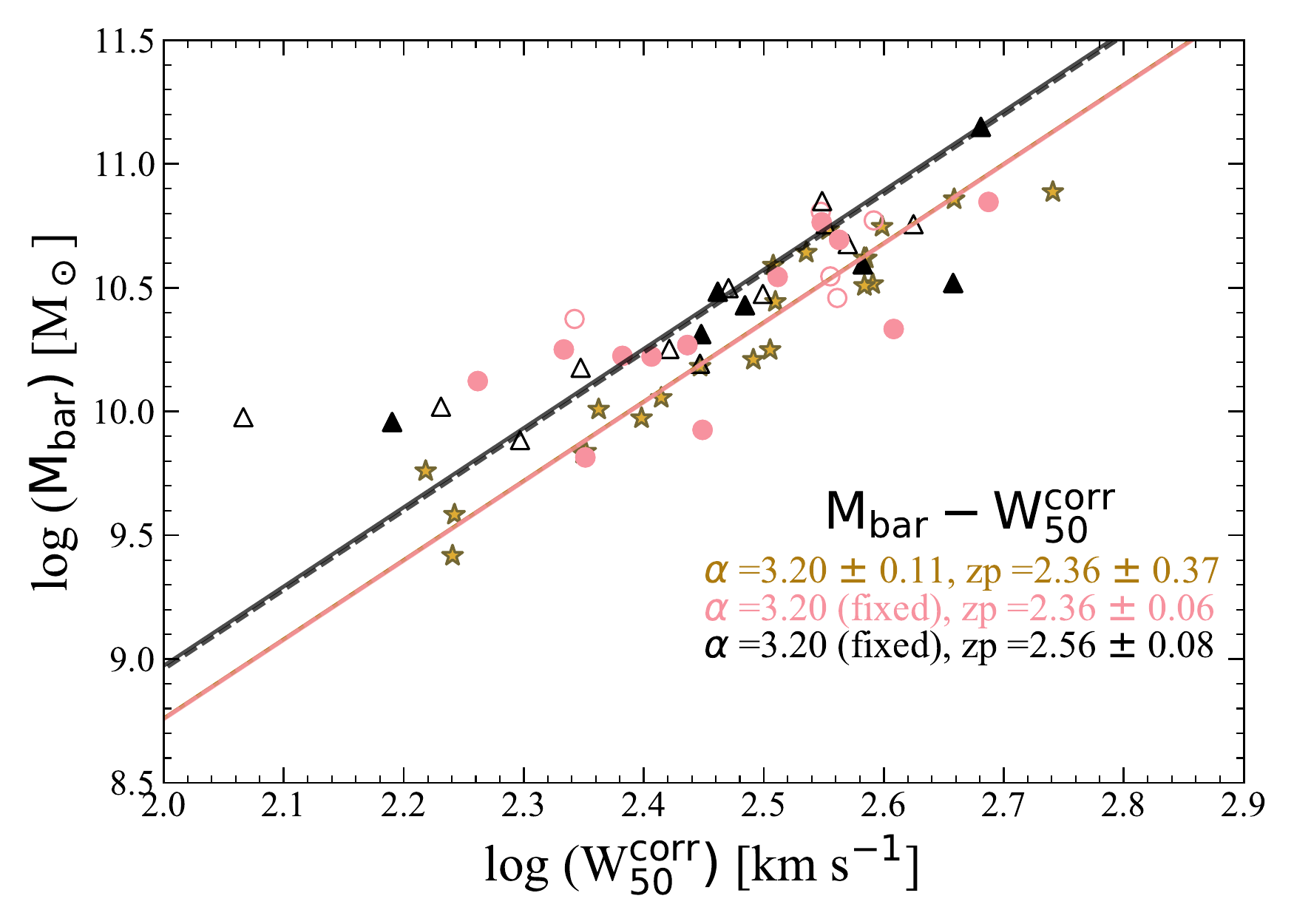}

\caption{Top and middle: The luminosity-based TFRs for the \textit{Cluster} and \textit{Control} samples in B-band and R-band respectively, using \wcf\ as the velocity measure. Bottom: BTFRs using the same velocity measure as the luminosity based TFRs. The \textit{Control} sample and its corresponding TFR is shown by the pink circles and lines while the \textit{Cluster} sample and TFR is shown by the black triangles and lines. The open and solid symbols as well as the dashed and solid lines represent the \textit{TFS} and \textit{HQS} respectively. The UMa galaxies and TFR are given in orange. The fit parameters (slope $\alpha$ and zero point zp of the UMa and \textit{TFS} TFRs are printed in the bottom-right of each panel. Other fit results can be found in Table \ref{tab:fitparamsbo}.}
\label{fig:bo}
\end{figure}

\subsection{Fitting method}







In order to minimise the Malmquist bias, inverse, weighted-least-square fits were made to the data, identical to the procedure followed by \citet{Verheijen01}, and our choice of the fiducial TFR in the Local Universe is from that study as well. A custom-made fitting algorithm using the python package \textit{scipy.optimize.curve\_fit} was implemented for the fitting. Uncertainties in the corrected line widths were estimated using standard error propagation of the errors on the observed line widths. Uncertainties in the model parameters featuring in the various correction formulas were not considered. Only errors in the corrected line widths were taken into account during the fitting, and the small inclination related co-variance between the errors in the corrected luminosities and in the line widths were ignored. The parameters of the inverse TFR and BTFR fits are provided in Tables \ref{tab:fitparams} and \ref{tab:fitparamsbo}.
Table \ref{tab:fitparams} provides the fit parameters for the luminosity-based TFRs and BTFRs for the three samples in consideration: UMa, \textit{TFS} and \textit{HQS}. Table \ref{tab:fitparamsbo} is similar, but provides the fit parameters for the \textit{Cluster} and \textit{Control} samples (see Sect. \ref{bo}). In addition, the tables also provide the offset differences with respect to each other as well as with UMa. Offsets greater than 5$\sigma$ are highlighted.

\subsection{The Luminosity-based TFR}\label{tfr}

The best-fit TFRs for the full \textit{TFS} with both the slope and the zero point left free, and using both velocity measures \wct\ and \wcf, combined with the two photometric bands B and R are given by Eqs. \ref{eq:tfr20Bfree}, \ref{eq:tfr50Bfree}, \ref{eq:tfr20Rfree} and \ref{eq:tfr50Rfree}. 

\begin{equation}
     \mathrm{M_B [mag] = (-10.6 \pm 0.7) \:log (W_{20}^{corr}) \:[km \:s^{-1}] + (6.4 \pm 1.5)}
     \label{eq:tfr20Bfree}
\end{equation}

\begin{equation}
     \mathrm{M_B [mag] = (-8.2 \pm 0.5) \:log (W_{50}^{corr}) [km\: s^{-1}] + (0.1 \pm 1.4)}
     \label{eq:tfr50Bfree}
\end{equation}

\begin{equation}
    \mathrm{M_R [mag] = (-10.9 \pm 0.7)\: log (W_{20}^{corr}) [km\: s^{-1}] + (6.3 \pm 1.5)}
    \label{eq:tfr20Rfree}
\end{equation}

\begin{equation}
    \mathrm{M_R [mag] = (-8.4 \pm 0.5) \:log (W_{50}^{corr}) [km\: s^{-1}] + (-0.3 \pm 1.4)}
    \label{eq:tfr50Rfree}
\end{equation}

\medskip

\noindent
We note that the TFRs based on the smaller \textit{HQS} are statistically indistinguishable from the TFRs based on the larger \textit{TFS}, and hence those fits to the \textit{HQS} are not presented here.


Due to the rather limited range (2.25 < log W$^{\mathrm{corr}}_\%$ [\kms] < 2.75) and relatively larger errors on the \H\ line widths for the \b\ galaxies, however, we limit our analysis to fitting and comparing the TFR zero points only. For this purpose, fits to the \textit{TFS} and \textit{HQS} were made with the slopes fixed to the corresponding TFR of the UMa sample, which displays a significantly smaller scatter.  

Fig. \ref{fig:tfr} presents the luminosity-based TFR fits with fixed slopes, based on the two velocity measures \wct\ (left) and \wcf\ (right), and the absolute B-band (top) and R-band (bottom) magnitudes. In all cases we find an offset of the \b\ TFR zero point towards brighter luminosities compared to the z=0 UMa TFRs. These offsets are smallest when using the \wct\ line widths (left panels) with 0.47$\pm$0.06 mag in the B-band and 0.19$\pm$0.06 mag in the R-band. The offsets in the zero point are significantly larger when using \wcf\ (right panels) with 0.72$\pm$0.06 mag and 0.44$\pm$0.06 mag in the B and R bands respectively. In all four cases, the vertical scatters are comparable, between 0.56 and 0.69 magnitudes (see Table \ref{tab:fitparams}). The numbers quoted above are for the \textit{TFS}, but there are no significant differences in the zero point offsets and scatters when fitting to the more restrictive \textit{HQS} (again, see Table \ref{tab:fitparams}).

\subsection{The Baryonic TFR}\label{btfr}

\noindent
The best fit inverse BTFRs with both the slope and the zero point left free are described by:

\begin{align}
    \begin{split}\label{eq:btfrfree20}
        \mathrm{log(M_{bar}/M_\odot) = (3.7 \pm 0.2) \: log (W_{20}^{corr}) \: [km\: s^{-1}] + (1.1\pm0.7)}
    \end{split}\\
    \begin{split}\label{eq:btfrfree50}
            \mathrm{log(M_{bar}/M_\odot) = (3.4 \pm 0.2) \: log (W_{50}^{corr}) \: [km\: s^{-1}] + (1.9\pm0.6)}
    \end{split}
\end{align}
\medskip

\noindent
Again, the corresponding figures for these BTFRs are not shown, since our focus is on the zero point of the BTFRs with their slopes fixed to the BTFR of UMa. 

Instead, we show in Fig. \ref{fig:btfr} the BTFRs obtained when making inverse fits with the slope fixed to the UMa value, using W$_{20}^{\rm corr}$ (left) and W$_{50}^{\rm corr}$ (right). The zero points of these \b\ BTFRs are indistinguishable from the zero points of the UMa BTFRs, with a maximum zero point offset of 0.08 dex for the \textit{TFS} when using W$_{50}^{\rm corr}$.  These results will be discussed in detail in Sect. \ref{discussion}.


\subsection{The TFRs from an environmental perspective}\label{bo}

The final versions of the TFR presented in this paper investigate the effect of the environment on the zero points, taking advantage of the fact that the \b\ samples also include galaxies in a cluster environment. In particular, we are interested in the influence of the environment from a "Butcher-Oemler" (BO) perspective. The BO effect \citep{BO84} manifests itself as a higher fraction of blue galaxies in the cores of clusters at higher redshifts. A963 is one of the nearest BO clusters and part of the seminal BO study by \cite{BO84}. In contrast, A2192, is a non-BO cluster with no blue galaxies associated with its core, weak in X-rays and still in the process of forming \citep[for more information on the two clusters, see][]{Yara0_12, Yara1_13,Yara2_15,Yara3_16}. Several studies of the BO effect, using optical and other bands, have been carried out with varying results \citep[e.g.,][]{Couch94, Lavery86, Tran03, DePropris04,Andreon04,Andreon06,Urquhart10, Lerchester11}. While some confirmed the presence of this BO effect, others claimed it was a selection bias by preferential inclusion of brighter, bluer galaxies at higher redshifts. 

Several environment- and morphology-specific studies of the TFR, both at low \citep[e.g.,][]{Vogt04,Mocz12} and high redshifts, have been carried out. In dense environments, for instance, kinematically disturbed galaxies are expected to be more common than in the field due to mechanisms such as ram-pressure stripping, tidal interactions, mergers, harassment and strangulation \citep[e.g.,][]{Oosterloo05, Poggianti17, Jaffe18, Toomre72, White78, Smith10, Kawata08, Maier16}. Inclusion of such galaxies in a TFR analysis is expected to lead to a larger scatter and possibly zero point offsets in the TFR (see Sect. \ref{sampseldiscussion} on sample selection). At higher redshifts, several environment-specific TFR studies reported no significant differences in the TFRs in cluster and field populations \citep{Ziegler03,Nakamura06,Jaffe11b,Perez21}, while other studies such as \citet{bamford05}, \citet{Milvang-Jensen03} and \citet[][for z$\sim$ 1.5]{Perez21} found an overall brightening of cluster galaxies at a fixed rotational velocity. Other morphology-specific studies such as \citet{Bedregal06,Yara14} reported fainter magnitudes at a given rotational velocity for lenticular and early-type galaxies compared to spiral galaxies.

Returning to the BO effect, the cluster substructure and the \H\ content of the blue galaxies in A963 have been investigated in a series of papers \citep{Verheijen07, Yara1_13,Yara2_15, Yara3_16}. These studies showed that none of the blue galaxies within the central 1 Mpc region of A963 were detected in \H , neither individually nor with a \H\ spectral stacking technique.
Other blue galaxies in the direct vicinity of A963, however, might still be gas bearing and experiencing an enhanced star formation activity. Our topic of interest is to investigate whether the \H -detected galaxies outside the cluster core of A963 are responsible for the blueing and brightening of the TFR zero points as observed in Fig. \ref{fig:tfr}, and whether the BO effect manifests itself in the zero points of the TFR and BTFR. Our thorough selection criteria ensured the identification of isolated, inclined \H\ discs with symmetric \H\ profiles with steep edges, thereby including only galaxies that are kinematically rather undisturbed and for which the corrected width of the global \H\ profile is a reliable proxy for the circular velocity of the halo. It might still be possible, however, that the immediate environment around the massive A963 cluster has an impact on the luminosity or the dark matter content of the blue, gas-rich galaxies, and is thereby partly responsible for the scatter and zero point offsets observed in Figs. \ref{fig:tfr} and \ref{fig:btfr}.

For both the \textit{TFS} and \textit{HQS}, we constructed two sub-samples, distinguished by their environment. Galaxies within $\pm$2.5$\mathrm{\sigma_{cl}}$ from the systemic velocity of A963, with $\mathrm{\sigma_{cl}}$ = 933 \kms\ , constitute the \textit{Cluster} sub-sample. The remainder of the galaxies in both survey volumes, consisting of those in the foreground and background of A963 as well as those in the entire A2192 survey volume, constitute the \textit{Control} sub-sample. The single galaxy detected in \H\ within a 1 Mpc projected distance from the core of A2192 did not pass our selection criteria for the \textit{TFS}. Thus none of the galaxies within the A2192 survey volume are located within a cluster environment. Moreover, A2192 has a much smaller velocity dispersion ($\mathrm{\sigma_{cl}}$ = 645 \kms) and a negligible ICM compared to A963. A total of 19 galaxies from the \textit{TFS} belong to the \textit{Cluster} sub-sample, while 17 belong to the \textit{Control} sub-sample. From the \textit{HQS}, 7 and 12 galaxies belong to the \textit{Cluster} and \textit{Control} sub-samples respectively. The environment-based TFRs and BTFRs are illustrated in Fig. \ref{fig:bo}. 

The zero points of the luminosity-based W$_{50}$ TFRs for the \textit{TFS} \textit{Control} sample are still somewhat brighter and bluer ($\Delta$M$_B$ = 0.61 $\pm$ 0.07 mag and $\Delta$M$_R$ = 0.31 $\pm$ 0.08 mag) than those of the z=0 UMa reference TFRs. The zero points of the luminosity based W$_{50}$ TFRs for the \textit{TFS} \textit{Cluster} sample, however, are significantly brighter ($\Delta$M$_B$ = 0.90 $\pm$ 0.09 mag and $\Delta$M$_R$ = 0.65 $\pm$ 0.09 mag) than those of the z=0 UMa reference TFRs but similarly blue compared to the \textit{TFS} \textit{Control} sample. The
zero point of the W$_{50}$ BTFR for the \textit{TFS} \textit{Cluster} sample is offset by 0.20$\pm$0.08 dex from the zero point of the UMa BTFR while the zero point of the W$_{50}$ BTFR for the \textit{TFS} \textit{Control} sample is indistinguishable (0.00$\pm$0.06 dex) from the zero point of the UMa BTFR. Effectively, the zero points of the \b\ BTFRs show no significant offset from the z=0 UMa BTFR. These results will be discussed further in Sect. \ref{envir}.

\section{Discussion}\label{discussion}


As discussed in the introduction, many studies of the Tully-Fisher relation exist but often with conflicting results regarding the slope, scatter and zero point of the relation, and their evolution with redshift, due to systematic differences in the target selection, observables, methodologies and corrections applied to the various samples. This section discusses our observational findings after a uniform analysis of the \b\ and UMa samples, and presents caveats that one needs to keep in mind to ensure that possible systematic differences in the comparison samples are not mistaken for an evolution in the TFR parameters.

\subsection{Impact of sample properties, observables and corrections on TFR scatter and zero points} \label{scatter}

The differences in the scatter and zero points of the luminosity-based TFRs as presented in Fig. \ref{fig:tfr} may be the result of a number of factors, such as the choice of photometric band (B versus R), the velocity measure (\wct\ versus \wcf), inaccuracies in the measurement of inclinations, intrinsic differences in the kinematics and morphologies of galaxies at higher redshifts, and the effect of environment on these galaxies. Interestingly, the Baryonic TFRs (see Fig. \ref{fig:btfr}) show a notably smaller scatter and zero point offsets than the luminosity based TFRs, as also reported in the literature \citep[e.g.,][]{Lelli16}. In all cases of the TFR, however, the z=0.2 TFRs show a larger vertical scatter and zero point compared to the z=0 TFR. Below, we explore some of the factors mentioned above in more detail.


\subsubsection{Sample selection} \label{sampseldiscussion}

Galaxies at higher redshifts are more often found to have kinematical and morphological anomalies than local galaxies \citep{Kannappan02, Flores06, Kassin07} and inclusion of these disturbed galaxies introduces a large scatter in the TFR. \citet{Weiner06} and \citet{Kassin07} showed in their studies that the scatter in the stellar mass TFR was greatly reduced by adopting a kinematic estimator $S_{0.5}$, which adds a measure of disordered, non-circular motions to the rotational velocity, effectively accounting for pressure support. This empirical approach to reduce the scatter was also confirmed by simulations \citep[see, for instance, ][]{Covington09}. Furthermore, including different galaxy types in the sample may also result in different slopes and systematic offsets in the TFR zero points. For instance, galaxies with rising or declining rotation curves are systematically offset from the TFR defined by regular spirals with flat rotation curves \citep[e.g.,][]{Verheijen01,Bedregal06, denHeijer15}. 

In our extensive sample selection procedure (see Sect. \ref{sampsel}), we identified and excluded optically disturbed and interacting galaxies since our goal is to construct a robust TFR using galaxies with reliable photometry and \H\ line widths that reflect the circular velocities of the dark matter halos. Due to the limitations in the resolution and quality of both the \H\ and the optical data, however, it is likely that some kinematically disturbed systems may still have been included in our samples, despite our best efforts. 

It is noteworthy that there is almost no difference in the TFRs of the two \b\ sub-samples (\textit{HQS} and \textit{TFS}), implying that an even stricter control over the \H\ properties of galaxies, in particular the symmetry of the \H\ global profile, does not significantly alter the TFR zero points or reduce the scatter. 
In all cases, the differences in the zero points are less than 0.1 magnitude. For simplicity, we will limit the rest of the discussion based on the results of the larger \textit{TFS}.


\subsubsection{Inclinations}

Improper inclination measurements are a dominant source of scatter in the TFR, since inclination corrections are applied not just to the kinematic measures but also to the magnitudes, and thus deserve special attention. Inclinations and their uncertainties can be estimated in a number of ways, such as making tilted-ring fits to the \H\ velocity fields, measuring \H\ disc ellipticities or optical axis ratios. With spatially unresolved \H\ data, inclinations are based on the ellipticity of the optical images. For the \b\ data, the optical axis ratios were computed from S\'ersic models fit to our deep INT R-band images \citep[][]{Gogate20} using \textit{galfit}, which also corrects for the seeing. These axis ratios were determined at the effective radius and converted to inclinations adopting a disc thickness (q$_0$) of 0.2. We found that adopting a different value of q$_0$ would not significantly impact the zero point offsets. For instance, a q$_0$ of 0.1 would result in a difference of $\sim$1 per cent in the corrected line widths for an observed axis ratio ($b/a$) of 0.7 ($i$=46.5$^\circ$), and $\sim$1.6 per cent for an observed $(b/a)$ of 0.3 ($i$=76.8$^\circ$).  Several other studies with spatially unresolved \H\ data used a similar approach \citep[e.g.,][]{Tully_Fisher77, Topal18}.

\begin{figure}
\includegraphics[width=\linewidth]{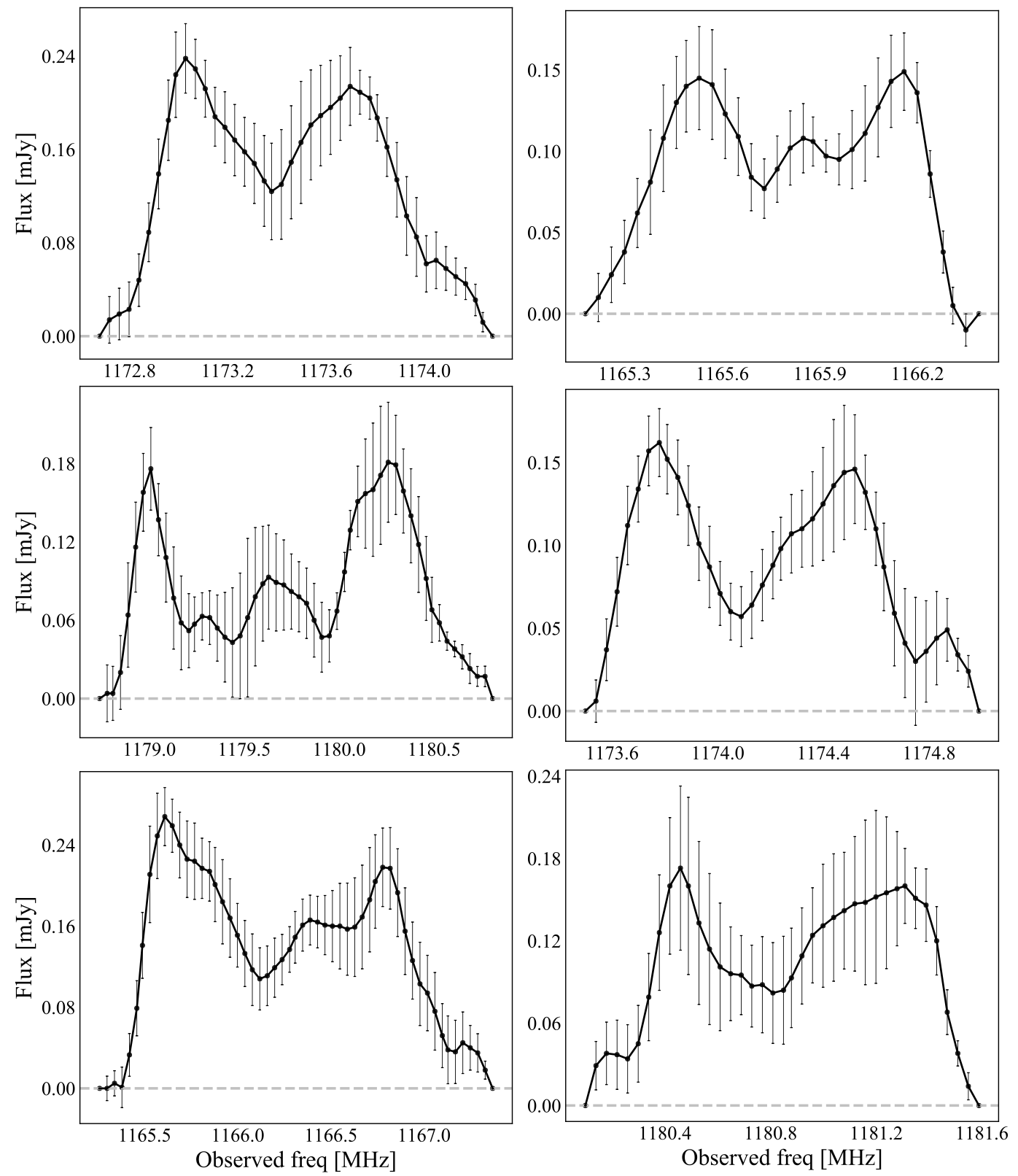}
\caption{Some examples of asymmetric \H\ profiles that show the presence low-level \H\ which is responsible for broadening the observed $\mathrm{W_{20}}$ line widths. The velocity resolution of these profiles is 38 \kms.}
\label{fig:asym_ex}
\end{figure}

It is important to note that while this method for estimating inclinations is widely used for unresolved or partially resolved samples, it still constitutes a significant source of uncertainty. This is because \textit{galfit} estimates inclinations at the effective radius, which may not be representative of the outer disk of galaxies, from which the inclination should ideally be derived. The uncertainty in the inclination may be further enhanced by the presence of an unidentified bulge.

For the UMa and HIGHz samples, we adopted the inclinations from the respective papers \citep[][]{Verheijen01,Catinella15}. For the UMa sample, inclination measurements are based on both optical axis ratios and \H\ kinematics. These inclinations are robust because the UMa galaxies are nearby and also have spatially resolved \H\ kinematics. For the HIGHz sample, these inclinations are based on the $(b/a)_\mathrm{r}$ axis ratios provided in the SDSS database.

\subsubsection{Choice of velocity measures}\label{velmeasure}


 \begin{figure}

    \includegraphics[trim={1.9cm 0cm 2cm 1cm },clip,width=\linewidth]{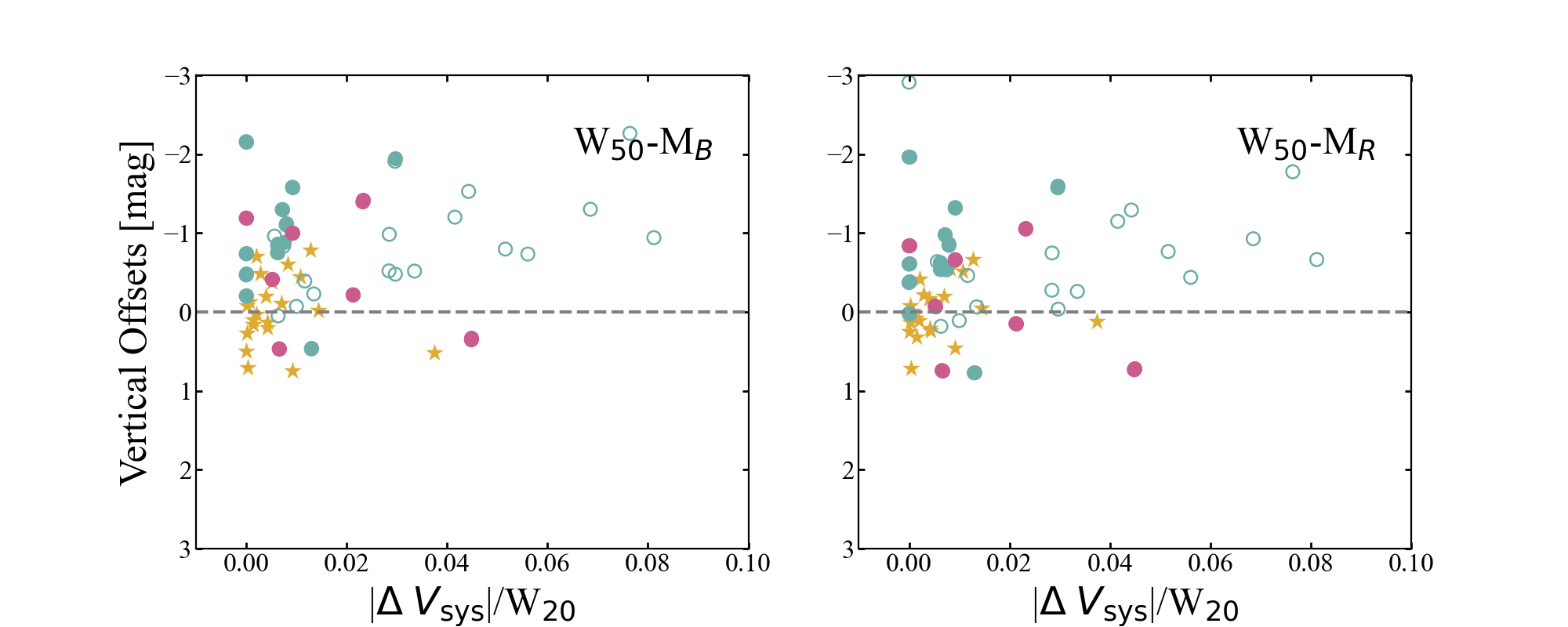}
    \includegraphics[trim={1.9cm 0cm 2cm 1cm },clip,width=\linewidth]{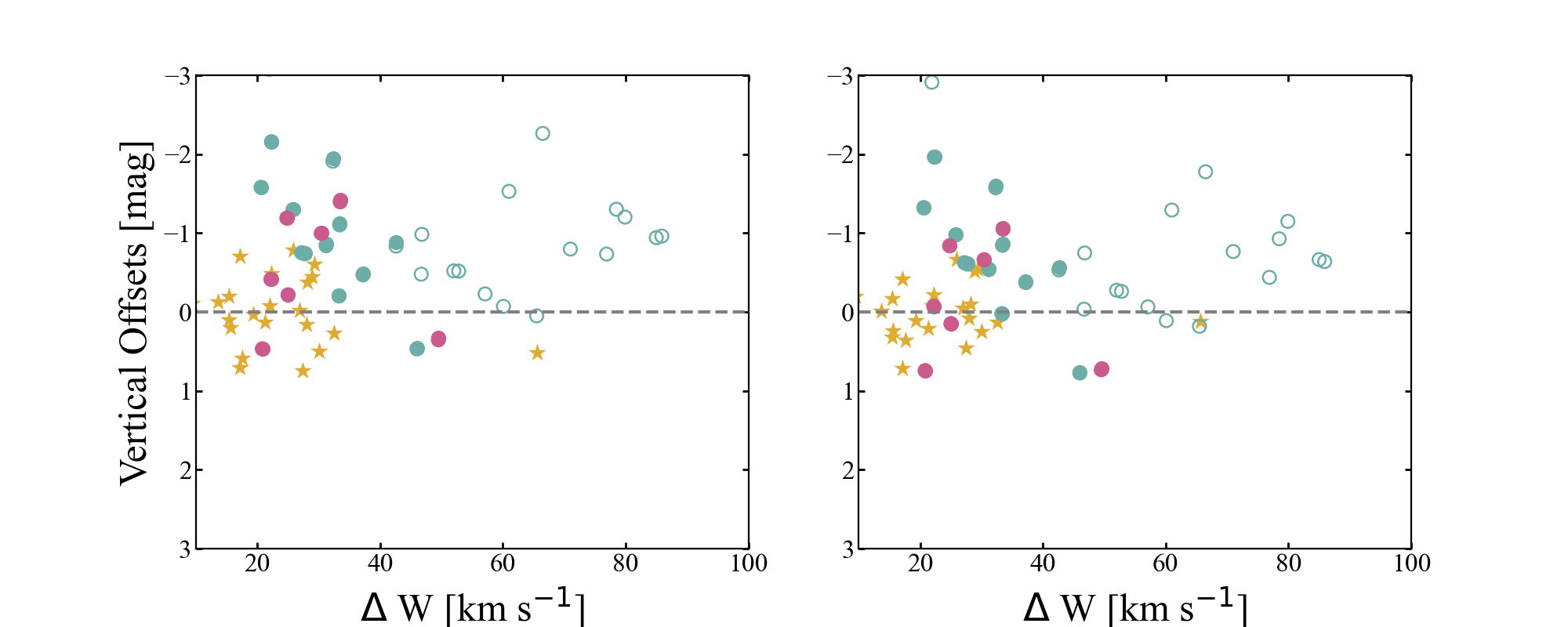}
    \includegraphics[trim={1.9cm 0cm 2cm 1cm },clip,width=\linewidth]{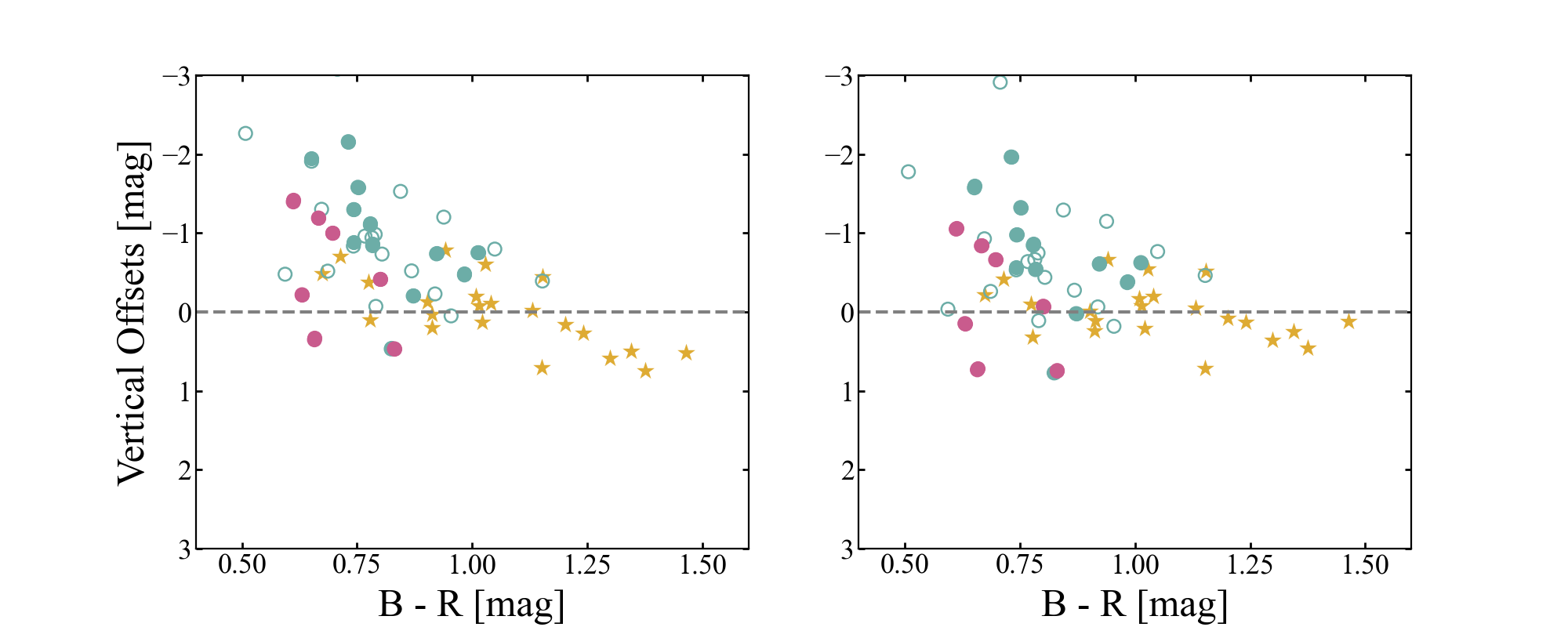}
    \includegraphics[trim={1.9cm 0cm 2cm 1cm },clip,width=\linewidth]{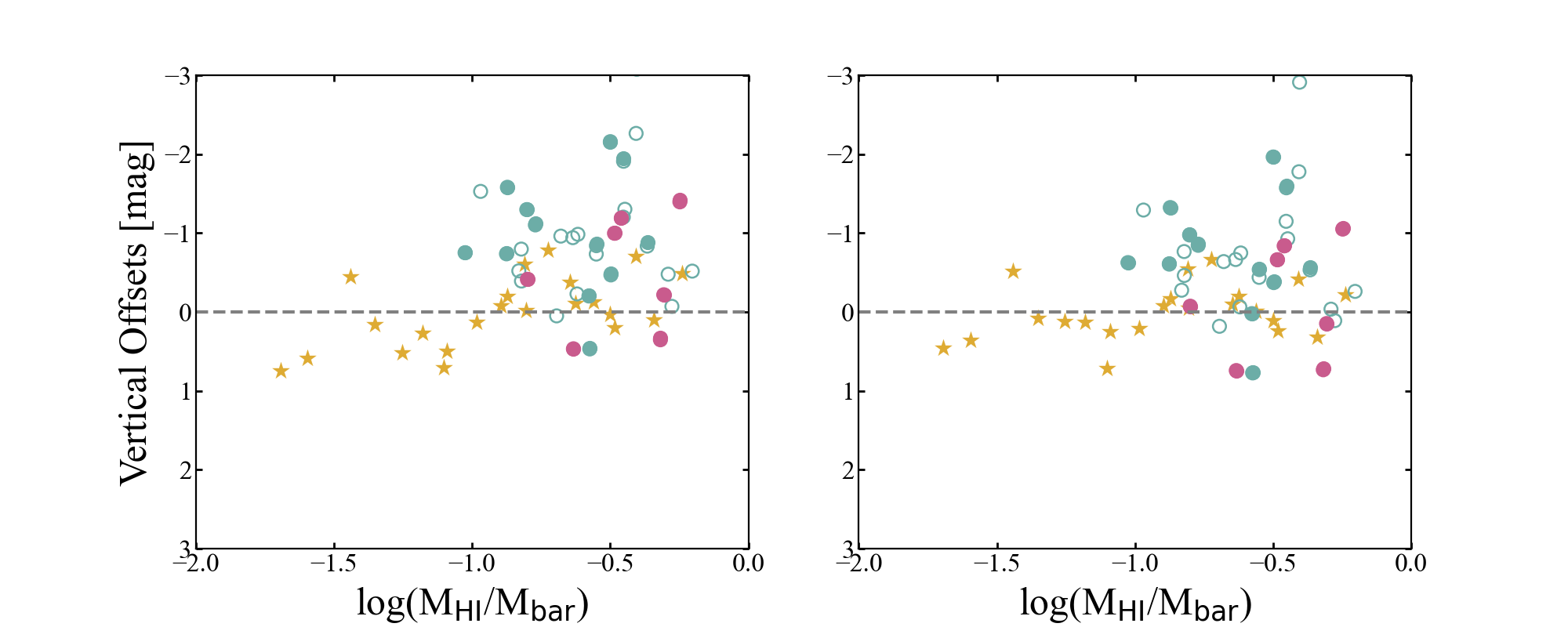} 
    
    \caption{Scatter plot of the vertical offsets (difference in magnitudes) of the individual galaxies from the luminosity-based UMa TFR using \wcf\ as the velocity measure, as a function of quantified asymmetry (first row), quantified steepness (second row), B-R colour (third row) and the ratio of the \H-to-baryonic mass (fourth row). Points lying above the dashed 'zero' line are brighter than the UMa TFR while those below are fainter. The left column corresponds to the TFR in the photometric B-band while the right column is the R-band TFR. All colours and symbols are identical to Fig. \ref{fig:btfr}.} 
    \label{fig:offset_col}
\end{figure}

For single dish and spatially unresolved \H\ studies, the classic approach for estimating the circular velocities of galaxies is based on the corrected widths of the global \H\ profiles \citep{Tully_Fisher77}, where inferred circular velocities are roughly half of the corrected \H\ profile line widths. \citet{Lelli19} found significantly tighter TFRs using \H\ line-widths as compared to other velocity tracers such as the H$\alpha$ and CO emission lines, which probe velocities in the inner star-forming regions of galaxies, while the amplitude of the outer, flat part of \H\ rotation curves provides the tightest TFRs \citep{Verheijen01,Ponomareva17,Lelli19}. This implies that the extended, cold \H\ discs are better tracers of the relevant circular velocity, and that the circular velocity of the dark matter halo is more fundamental to the TFR than the inner circular velocity affected by the distribution of the baryons. However, interferometric observations to measure spatially resolved \H\ rotation curves are often not available. Since our \b\ sample consists mostly of spatially unresolved \H\ sources, we are limited to using the corrected \H\ profile line widths \wct\ and \wcf\ as proxies for the circular velocities of the dark matter halo. Interestingly, while \wct\ and \wcf\ yield the same circular velocity for UMa galaxies, we found a striking systematic difference in \wct\ and \wcf\ for the \b\ galaxies, independent of the average SNR of the \H\ profiles (see Fig. \ref{fig:fracdiff1}). This difference in corrected line widths does not disappear for the galaxies in the \textit{HQS} which still have more asymmetric and shallower \H\ profiles as compared to the UMa galaxies (see fig. \ref{fig:steep}). We conclude that the \b\ galaxies have intrinsically shallower \H\ profile wings compared to the Local Universe counterparts. Furthermore, the \H\ profiles of the \b\ galaxies tend to be more asymmetric than the UMa galaxies. Some of these asymmetric profiles are shown in Fig. \ref{fig:asym_ex}.

\begin{figure}

\includegraphics[trim={0cm 0cm 0.2cm 0cm },clip,width=\linewidth]{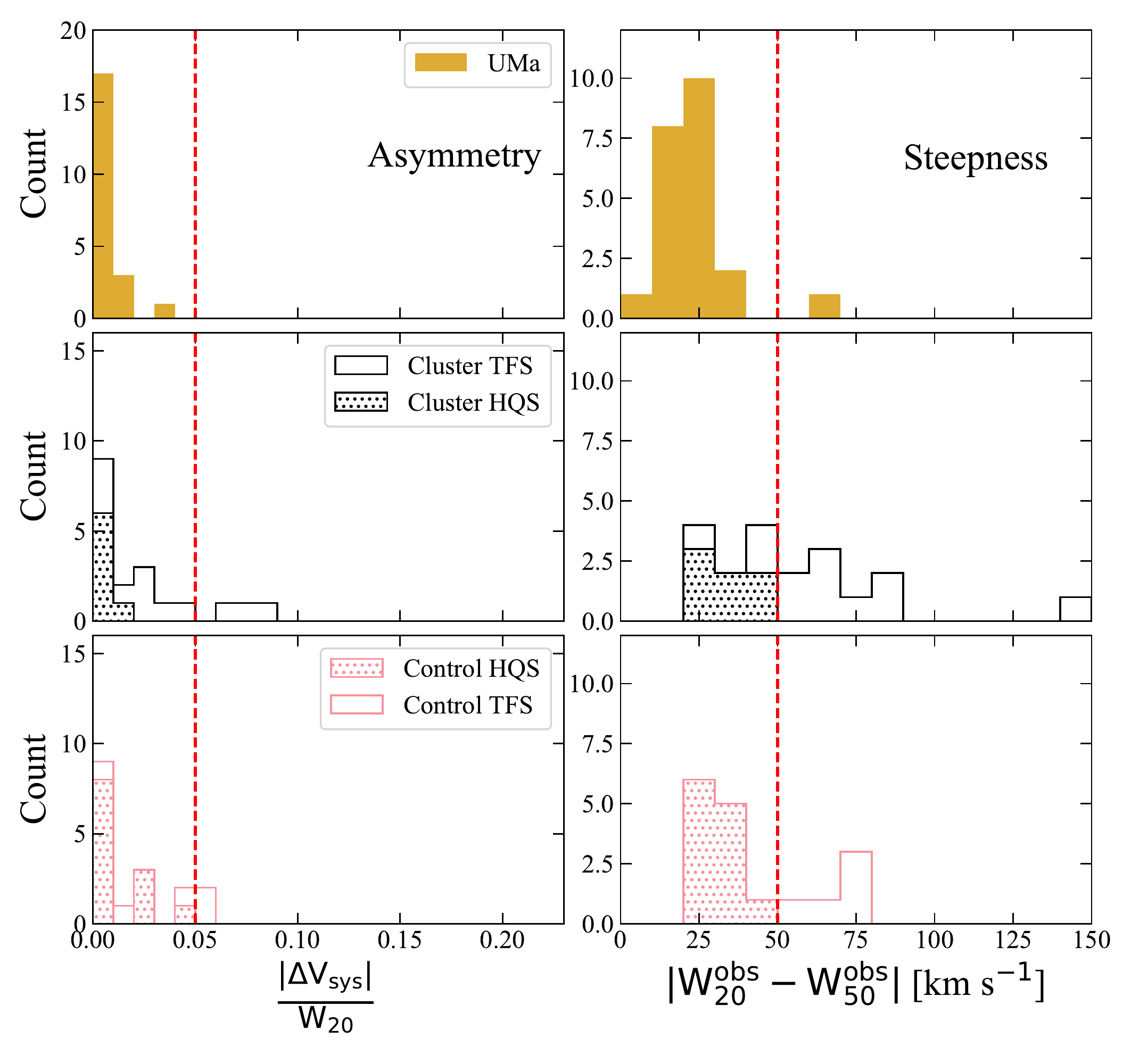} 
    
\caption{ Similar to Fig. \ref{fig:steep} but for the control (pink) and cluster (black) samples respectively: Left: histograms showing the absolute fractional differences in the systemic velocities derived from the two line width measures (asymmetry); Right: histograms showcasing the difference in the observed line widths (steepness).  The orange histograms show the UMa sample. Open histograms correspond to the \textit{TFS} while the hatched histograms show the \textit{HQS}. }
\label{fig:steep_bo}
\end{figure}


These systematic differences in the \wct\ and \wcf\ line widths of the \b\ galaxies raise the question which velocity measure to use, since they result in two different TFR zero point offsets (Fig. \ref{fig:tfr}). The asymmetric and shallow wings of the \b\ \H\ profiles suggest the presence of \H\ gas which is likely not participating in the rotation of these galaxies but appears to broaden the \H\ profiles at low flux levels, thus affecting the \wct\ line widths. The \wcf\ line widths on the other hand, seem less affected by this broadening. Thus, smaller zero point offsets with UMa seen in the \wct\ luminosity-based TFRs (Fig. \ref{fig:tfr}, left) are likely due to an overestimation of the circular velocities derived from the \wct\ line widths, causing the \b\ galaxies to shift towards larger circular velocities and reducing the vertical zero point offset of the \b\ sample. Notably, plotting the vertical offsets of the individual \b\ galaxies from the UMa TFRs as a function of increasing asymmetry (|$\Delta$ V$_{\mathrm{sys}}$|/\wot) and decreasing steepness ($\Delta$W$=$|\wct-\wcf|) in Fig. \ref{fig:offset_col} (top two rows), we do not see any particular trend.

To explore a possible correlation between the environment and the occurrence of asymmetric and shallow \H\ profiles, the \b\ galaxies were divided into a \textit{Cluster} and a \textit{Control} sample (see Sect. \ref{bo}). Histograms similar to Fig. \ref{fig:steep} are shown for these sub-samples in Fig. \ref{fig:steep_bo}. The \textit{Cluster} sample contains relatively more asymmetric and shallow profiles than the \textit{Control} sample, although the statistics are rather poor. \citet{Jaffe11b} found similar disturbances in their optical emission-line profiles of cluster galaxies. Other studies such as \citet{Watts20a} and \citet{Watts20b} found that the environment is not the only driver of \H\ profile asymmetries. Physical processes within galaxies, regardless of environment, may also effect the shape and symmetry of global \H\ profiles. Moreover, they found that the majority of these asymmetric profiles belong to satellite galaxies. For our \b\ sample, investigating the nature and origin of the asymmetry and steepness of the \H\ profiles in the context of evolutionary signatures would require further analysis of the galaxies and a larger sample, which is beyond the scope of this paper. See sect. \ref{envir} for a further discussion on the environmental effects on the TFR and BTFR.

\subsubsection{Choice of corrections for turbulent motion}\label{corrdisc}

In the previous sections, we have already stressed the importance of having consistently selected samples with uniformly applied corrections to the observables. For instance, applying different line width corrections for the different samples could result in significantly different zero points of the TFRs at different redshifts, which could result in erroneous conclusions regarding the evolution of the TF scaling relation.

The correction for line width broadening due to turbulent motion of the \H\ gas are adopted from \citet{Tully85} and optimised by \citet{Verheijen&Sancisi01} based on rotation curve measurements of UMa galaxies. If the \H\ gas in the \b\ galaxies at z=0.2 is significantly more turbulent than in the UMa galaxies at z=0 then this may result in a systematic under-correction for the \b\ galaxies and an artificial shift of the zero point of the TFR. Maybe the shallower wings of the \H\ profiles of the \b\ galaxies hint at a higher turbulence in their \H\ discs, justifying the use of \wcf\ as the correction for turbulent motion has less impact on this line width measure. For the spatially unresolved \b\ galaxies, however, there is no way of knowing whether the corrections for turbulent motions based on the UMa galaxies are adequate.

\subsubsection{Choice of corrections for internal extinction}\label{corrdisc2}

The correction for internal extinction that we adopted, is based on \citet[][referred to as T98]{Tully98} (see Eq. \ref{eq:intext_tully98} in Sect. \ref{photcor}) and depends both on a galaxy's inclination and corrected \H\ line width. Prior to this work, internal extinction corrections were often based on the prescription by \citet[][referred to as TFq]{Tully85}, which depends on inclination only, and is given by:


\begin{equation}
   \mathrm{ A_\lambda^{i,TFq} = -2.5 log \left[f(1 + e^{-\tau_\lambda sec\; i}) + (1 - 2f)\left(\frac{1 - e^{-\tau_\lambda sec \:i}}{\tau_\lambda sec \:i}\right)\right]}
\end{equation}

\noindent
where, for a slab of dust containing a homogeneous mixture of stars of fraction (1-2f), f signifies the fraction of stars above and below this slab, while $\tau_\lambda$ gives the optical depth of the dust layer as a function of wavelength. Here, we use f = 0.1, $\tau_B$ = 0.81 and $\tau_R$ = 0.40 \citep{Verheijen97}. This extinction prescription is applicable for galaxies with 45$^\circ$ $<$ $i$ $<$ 80$^\circ$. For more edge-on galaxies, the TFq prescription assigns reddening corrections corresponding to $i$=80$^\circ$, assuming the extinction to plateau for extremely inclined systems, as the `back' of the disk and bulge becomes visible below the dustlane. Four, two and two galaxies in the UMa, \b\ and HIGHz samples respectively have $i$ $>$ 80$^\circ$, and thus were assigned TFq corrections $\mathrm{A_\lambda^{i,TFq}}$ =  $\mathrm{A_\lambda^{80,TFq}}$. 

\begin{figure}
\includegraphics[trim={12.7cm 0.2cm 0cm 0cm },clip,width=\linewidth]{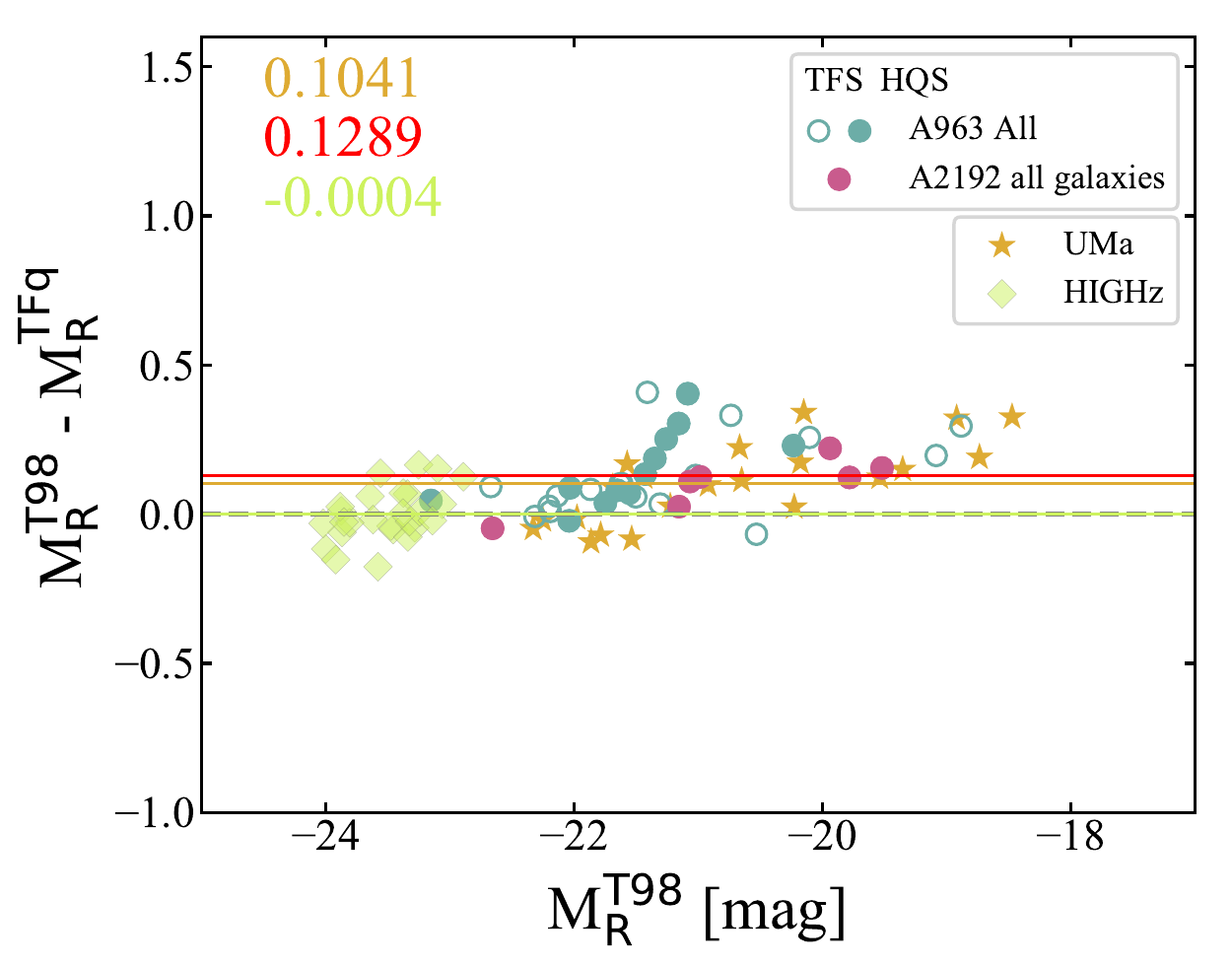}

\caption{Difference in the absolute magnitudes based on two different internal extinction corrections for the R-band. The magnitudes were corrected for Galactic extinction, K-corrections and the internal extinction corrections were adopted from \citet{Tully98} and \citet{Tully85} respectively. The differences are plotted against the absolute magnitudes based on \citet{Tully98}, which is the standard correction used in this chapter. The horizontal lines indicate the average differences for each sample, denoted by their respective colours, with their corresponding values printed at the top-left. Note that the red solid line and corresponding value of the average difference represents the TFS.} Other colours and symbols are identical to Fig. \ref{fig:cmd}. The dashed black line indicates no differences between the T98 and TFq corrections.
\label{fig:tfq_diff}
\end{figure}

Fig. \ref{fig:tfq_diff} shows the differences in the absolute R-band magnitudes of the various samples, based on the two internal extinction corrections (T98 and TFq). The sample average difference between the two prescriptions is $\sim$0.13 mag for the R-band and $\sim$0.35 mag for the B-band (not shown), suggesting large systematic offsets in the TFR zero points if the samples were to be treated with different internal extinction prescriptions. Interestingly, the \b\ and UMa samples show a similar behaviour in their relative differences of $\sim$0.025 in the R-band and $\sim$0.04 in the B-band. This suggests that the distribution of dust inside the \b\ and UMa galaxies is rather similar and that our choice of prescription has an insignificant impact on the relative shifts of the TFR zero points. We note that the T98 and TFq prescriptions result in similar corrections for the HIGHz galaxies. This may not be surprising as the HIGHz galaxies tend to be more face-on and, consequently, receive smaller corrections. Finally, Fig. \ref{fig:tfq_diff} suggests an increasing difference in corrections with fainter magnitude. This is probably due to the fact that lower mass galaxies tend to be less dusty, an effect accommodated by the T98 prescription but not by the TFq prescription. This would also introduce a change in the slope of the TFR if different prescriptions are used for different samples.

\subsubsection{Choice of photometric band}\label{coldisc}

Some studies such as \citet[][z$\sim$0.6]{Flores06} claim to find a larger scatter in the B-band TFR compared to other bands. We find that both the K-corrections and internal extinction corrections are notably larger in the B-band than in the R-band yet, after applying these corrections, the vertical scatters in the \b\ TFRs are similar in the B and R bands (see Fig. \ref{fig:tfr} and Table \ref{tab:fitparams}). This implies that these corrections do not drive the scatter in the \b\ TFRs. After applying the various photometric corrections, however, the zero point offsets with respect to the z=0 UMa TFR do depend on the photometric band, where a stronger evolution with redshift may be inferred from the B-band TFR than from the R-band TFR for all sub-samples.

Furthermore, Fig. \ref{fig:offset_col} (third row) illustrates the vertical offsets in magnitude for each individual \b\ galaxy from the z=0 UMa TFR as a function of B-R colour. We find a mild colour dependence for the UMa galaxies, particularly in the B-band, in the sense that bluer UMa galaxies have a positive offset. This mild trend is continued towards bluer colours by the \b\ galaxies, albeit with a larger scatter. This implies that the scatter in the luminosity based TFRs can be reduced by introducing a colour term as a second parameter. The scatter in the TFRs, however, is not the focus of our investigation. The TFRs in the respective photometric bands are discussed in more detail in Sect. \ref{redshiftevol}.

\noindent\subsubsection{Choice of stellar mass estimator}

Similar to Sects. \ref{corrdisc} and \ref{corrdisc2}, here we briefly discuss the systematics involved in the choice of stellar mass prescription used to estimate baryonic masses. Studies such as \citet{Ponomareva18,Ponomareva21}, whose stellar masses are based on Spectral Energy Distribution (SED) fitting, have shown that the quality of the photometry and the choice of stellar mass estimator can impact the statistical properties of the BTFR. To verify this notion, although we are not able to make full SED fits, we used two different stellar mass prescriptions, Eqs. \ref{eq:mstar1} and \ref{eq:mstar2}, to derive stellar and baryonic masses, and found that they provide consistent results with a difference in the zero points of the BTFRs of $\sim$0.01 dex. The findings of \citet{Ponomareva18}, however, remain a stark warning that using different stellar mass estimators for different samples may result in different zero points and slopes, which may result in wrong conclusions about the evolution of the BTFR.

\bigskip

In conclusion of this subsection \ref{scatter}, we find that some choices regarding the velocity measure, the photometric band and the corrections applied to the data, may affect the TFR and BTFR zero point offsets more than others. For a study such as this, one of the most significant sources of uncertainty is the inclination measurement using disk fitting algorithms such as \textit{galfit}. The effective radii of galaxies derived from optical images with limited spatial resolution do not always guarantee accurate inclination measurements (especially in bulge-dominated systems) and can thus impact the interpretation of the TFR zero points. Other choices such as the correction for turbulent motion, choice of photometric band etc. can also bias TFR results, though we find that in our case the impact of these choices is limited thanks to the rigorous sample selection procedure applied. Most importantly, it is essential that these choices are applied consistently throughout all comparison samples to avoid the influence of systemic biases on the measured TFR offsets.

\subsection{Evolution in the TFR with redshift}\label{redshiftevol}


Conflicting results and ongoing debates in the literature regarding the evolution of the TFR parameters with redshift \citep[e.g.,][]{Vogt97, Rix97, Simard98,Ferreras01,Ziegler02,Bohm04,Ferreras04,Conselice05,Flores06,Bamford06,Weiner06, Kassin07,Puech08} are likely caused by differences in the construction of the comparison samples and inconsistencies in the applied corrections, which, as discussed previously, could introduce systematic offsets that may be mistaken for evolutionary signatures. In our analysis, we have chosen to compare our \b\ TFRs with the local UMa TFR from \citet{Verheijen01} because of the similarities in the available data, full control over the applied corrections and the fact that the UMa galaxies are nearly equidistant. 

\subsubsection{Evolution in the luminosity-based TFR}

In Sect. \ref{velmeasure} we discussed in detail the differences observed in the \wct\ and \wcf\ line widths, and the possible reasons for the  asymmetries and the shallow edges seen in the \H\ profiles of the \b\ galaxies. We limit this discussion to the offsets in the zero points from the \wcf\ TFRs, which imply that galaxies in the past were both brighter and bluer ($\Delta$ZP$_\mathrm{B}$ = 0.72 $\pm$ 0.06 mag and $\Delta$ZP$_\mathrm{R}$ = 0.44  $\pm$ 0.06 mag) than galaxies in the present epoch. This is most likely due to the rest frame B-band luminosity being significantly more sensitive to star formation activity compared to the R-band luminosity. Qualitatively, this is expected, since galaxies at higher redshifts are predicted to have younger stellar populations, and it is known that the cosmic star formation rate density increases towards higher redshifts \citep[e.g.,][]{Dutton11, Madau14}. Such offsets are also reported in the literature. For example, \citet{Bohm16} found that their sample of higher redshift field galaxies (0.05 $<$ z $<$ 1) were either brighter in the B-band for their dynamical masses or had lower dynamical masses for their B-band luminosities. The underlying difference in colour seen in the CMD presented in Fig. \ref{fig:cmd} and discussed in Sect. \ref{compprop} could be the reason behind the brightening of the \b\ TFR with respect to UMa. However, the fact that our \b\ galaxies are not only brighter in the B-band but also bluer suggests that, indeed, galaxies were brighter for their dynamical mass.




\subsubsection{Evolution in the Baryonic TFR}\label{disc_stel_bar}

Turning our attention to the \b\ BTFRs, we find that their zero points are comparable to the z=0 BTFRs, using either stellar mass estimator (Eqs. \ref{eq:mstar1} and \ref{eq:mstar2}). For the W$_{50}^{\rm corr}$ BTFRs constructed with the \textit{TFS}, the offsets in the zero points are as small as +0.08 $\pm$ 0.02 dex and -0.02 $\pm$ 0.03 dex, based on Eqs. \ref{eq:mstar1} and \ref{eq:mstar2} respectively. We therefore conclude that there is no evolution in the zero point of the BTFR with redshift up to z$\approx$0.2. This supports the finding by \citet{Dutton11}, who concluded that the scaling relations using baryonic masses show a weak evolution. 

Interestingly, as shown in Fig. \ref{fig:mratios}, the higher \mhi/\mbar\ ratios of the \b\ galaxies compared to UMa (see Sect. \ref{compprop}) imply that a larger fraction of the baryonic content in the \b\ galaxies is in the form of \H. Figure \ref{fig:mratios} also shows a larger intrinsic spread in \mhi/\mbar\ for the UMa sample, which is not reflected in a larger scatter in the TFR or BTFR.  In particular, the two UMa galaxies (NGC 3729 and NGC 4102) with the lowest \mhi/\mstel\ and \mhi/\mbar\ ratios in Fig. \ref{fig:mratios} (b) and (d) have the lowest \H\ masses and small \H\ discs, yet they follow the TFR and BTFR. This implies that the BTFR is insensitive to the gas fraction of the galaxies at least out to z$\approx$0.2. 

Shown in the bottom four panels of Fig. \ref{fig:offset_col} are the vertical offsets of the UMa and \b\ galaxies from the luminosity-based UMa TFRs as a function of B$-$R colour and \mhi/\mbar. While the BTFR does not evolve with redshift, the fact that the \b\ galaxies are on average brighter, bluer and more gas rich suggests that an enhanced star formation activity in the \b\ galaxies is responsible for the perceived evolution of the luminosity-based TFRs, and not an evolving M$_{\rm bar}$/M$_{\rm halo}$ ratio.

\bigskip

In conclusion to this section \ref{redshiftevol}, we find that while the baryonic content of galaxies in the past has remained largely unchanged, the components making up the baryonic content (\H\ and stellar masses) have changed in the past 2.5 Gyrs. At the same time, the higher gas fraction of the \b\ galaxies seems to be linked to a higher star formation rate, which makes the \b\ galaxies brighter and bluer in the past.

\subsection{Effect of cosmic environment on the TFR and BTFR}\label{envir}

To probe the effect of the environment on the TFR and BTFR parameters, the \b\ galaxies in the \textit{TFS} and \textit{HQS} were divided into \textit{"Cluster"} and \textit{"Control"} sub-samples (see Sect. \ref{bo} for the sample construction and a summary of the zero point offsets). We discussed the correlation between the \H\ profile shapes, and the environment in Sect. \ref{velmeasure}. Here, we discuss the environment-specific TFRs and BTFRs shown in Fig. \ref{fig:bo}. The sub-sample of \textit{Cluster} galaxies shows a more extreme brightening but a similar blueing as compared to the \textit{Control} sample. We hypothesise that the extreme brightening of the \textit{Cluster} galaxies could be the result of enhanced star formation activity induced by the cluster environment, while these galaxies still contain a detectable amount of \H\ gas \citep[e.g.,][]{Mahajan12,Yara3_16,vulcani18c}. The brightening and blueing of the \textit{Control} galaxies, however, is still significant compared to the UMa sample. This indicates that, though the cluster environment increases the zero point offset, field galaxies in the past were still intrinsically brighter and bluer than local galaxies for a given circular velocity.

Focusing instead on the environment-specific BTFRs (Fig. \ref{fig:bo}, bottom panel), we find that the BTFR of the \textit{Control} sample is identical to the UMa BTFR within the uncertainties. The zero point of the \textit{Cluster} BTFR is also consistent with the UMa BTFR at the 2.5$\sigma$ level (see Table \ref{tab:fitparamsbo}). This finding and the fact that the galaxies in the \textit{Cluster} sample have similar \mhi/\mbar\ ratios as our \textit{Control} sample galaxies (see Fig. \ref{fig:mratios}) imply that the cluster environment does not seem to remove the baryonic mass from the dark matter halos of these cluster galaxies. This is not surprising, however, as our \textit{Cluster} sample galaxies are all outside the estimated stripping cone of A963 (see Fig.7 in \citealp{Yara2_15}). Nevertheless, since the \textit{Cluster} galaxies are brighter than the \textit{Control} galaxies, the outer cluster environment may already be enhancing the star formation activity compared to field galaxies. 

Thus, reiterating our conclusion from Sect. \ref{disc_stel_bar}, while the baryonic content of these galaxies do not show an evolution with redshift for a given rotational velocity, the \mhi/\mbar\ ratio seems to have evolved. At the same time, our results also show that the star-formation activity was higher in the past and further enhanced by the cluster environment.



\subsection{Comparison with HIGHz}

Lastly, for a cursory comparison, we also indicated the HIGHz sample in various \wcf\ TFR and BTFR figures. The HIGHz sample consists of optically selected massive, luminous galaxies at redshifts similar to \b\ (see Sect. \ref{compprop} for more details). \citet{Catinella15} found that they follow the local BTFR defined by \citet{Catinella12} and \citet{McGaugh00}, though the galaxies had not been corrected for turbulent motion or internal extinction. In our work, however, corrections applied to the \b\ galaxies were identically applied to the HIGHz galaxies, including the correction for internal extinction. We do find that the HIGHz galaxies seem to be offset from the local BTFR (see Fig. \ref{fig:btfr}), underlining the importance of applying consistent corrections to all samples.

We find that the HIGHz galaxies are more consistent with our luminosity-based TFRs at z=0.2 (Fig. \ref{fig:tfr}) than with the z=0 UMa TFR, although there is a significant offset from the \b\ M$_{\rm R}$-\wcf\ TFR. However, a correlation between luminosity and line width is nearly absent for the HIGHz sample. In the case of the BTFR (right panel of Fig. \ref{fig:btfr}), we find that the HIGHz galaxies seem to have significantly larger baryonic masses for their circular velocities compared to both the \b\ and z=0 BTFR. Whether this behaviour of the HIGHz galaxies is due to a selection bias, and over-estimate of their inclinations, or photometric uncertainties remains uncertain.

\section{Summary and Conclusions}\label{summary}

With this work, we present the first dedicated study of an \H-based TFR at z=0.2 using direct \H\ detections. We have studied the luminosity-based B- and R-band TFRs as well as the BTFR at z=0.2 from a meticulously selected sample of \H\ detected galaxies from the Blind Ultra-Deep \H\ Environmental Survey (\b, \citealp{Gogate20}). In addition, we used deep B- and R-band images, obtained with the INT/WFC. The \b\ TF sample (\textit{TFS}) comprises 36 galaxies, of which 29 are in the survey volume containing A963 and 7 in the volume containing A2192. Of these, 11 and 7 galaxies together make up the \textit{High Quality} sub-sample (\textit{HQS}) based on stringent quantitative thresholds imposed on the shape of the \H\ global profile. Our results were compared with an identically constructed TFR from the Ursa Major association of galaxies in the Local Universe at a distance of 18.6 Mpc \citep[UMa,][]{Verheijen01}. Our main results are as follows: \\

\noindent
(1) The best fit luminosity based and Baryonic TFRs using \wcf\ for the \b\ TF sample with all parameters left free are:

\medskip\noindent
 M$_{\rm B}$ [mag] = ($-$8.2 $\pm$ 0.5) Log (W$_{50}^{\rm corr}$) [km$\:$s$^{-1}$] + (0.1 $\pm$ 1.4)

\medskip\noindent
 M$_{\rm R}$ [mag] = ($-$8.4 $\pm$ 0.5) Log (W$_{50}^{\rm corr}$) [km$\:$s$^{-1}$] + ($-$0.3 $\pm$ 1.4)

\medskip\noindent
 Log(\mbar/M$_\odot$) = (3.4 $\pm$ 0.2) log (W$\mathrm{_{50}^{corr}}$) [km$\:$s$^{-1}$] + (1.9 $\pm$ 0.6)

\medskip\noindent
The stellar masses for the BTFR are computed according to Eq. \ref{eq:mstar1}. However, due to the relatively large scatter of the \b\ sample compared to UMa, our analysis was restricted to the zero points of the \b\ TFR and BTFR only, by fixing their slope to the UMa TFR. The results of these fits are compiled in Table \ref{tab:fitparams}.\\

\noindent
(2) The scatter in the \textit{TFS} and \textit{HQS} TFRs are very similar, suggesting that stricter quantitative selection criteria on the shapes of the \H\ profiles do not reduce the scatter. \\

\noindent
(3) Similarly, differences in the zero points between the \textit{TFS} and \textit{HQS} TFRs are always less than 0.05 magnitudes, implying that stricter selection criteria on the \H\ profile shapes also do not affect the TFR zero points. Furthermore, the \b\ galaxies in the \textit{HQS} still seem to have more asymmetric \H\ profiles with shallower wings than the UMa galaxies. The \H\ profile shapes of the \b\ galaxies suggest the presence of low-level unresolved \H\ gas that likely does not participate in the rotation of the galaxies, possibly from optically unidentified nearby companions. The shallow profile edges may also suggest larger turbulent motions of the \H\ gas in z=0.2 galaxies. This leads to a broadening of the \wct\ line widths while the \wcf\ line widths seem to remain unaffected. Galaxies associated with A963 tend to have more disturbed \H\ profiles. \\

\noindent
(4) The zero point offsets of the z=0.2 \b\ TFRs depend on the choice of photometric band and velocity measure. Offsets in the B-band are always larger than in the R-band. An systematic over-estimation of the circular velocities using \wct\ causes the \b\ galaxies to shift towards larger circular velocities and thus negate the above mentioned vertical offset of the \b\ sample. \\

\noindent
(5) Adopting \wcf\ as the velocity measure, the z=0.2 \b\ TFR is brighter and bluer than the z=0 UMa TFR. We divided the \textit{TFS} into a \textit{Cluster} and a \textit{Control} sub-sample. While the \textit{Control} sample is brighter and bluer than the UMa sample ($\Delta$M$_B$ = 0.61 $\pm$ 0.07 mag and $\Delta$M$_R$ = 0.31 $\pm$ 0.08 mag), the \textit{Cluster} sample is even brighter but not bluer ($\Delta$M$_B$ = 0.90 $\pm$ 0.09 mag and $\Delta$M$_R$ = 0.65 $\pm$ 0.09 mag). \\

\noindent
(6) The Baryonic TFR at z=0.2 is consistent with the local BTFR despite the fact that the \mhi/\mbar\ ratio of the \b\ galaxies is somewhat higher than that of the UMa galaxies. Although the \mhi/\mbar\ ratios are similar for the \textit{Cluster} and \textit{Control} galaxies, the cluster environment seems to enhance the luminosities of the \textit{Cluster} sample galaxies.\\

\noindent
(7) Lastly, it is very important to ensure that the comparison samples trace similar galaxy populations for a proper and unbiased comparison, since the various corrections, even if consistently applied, may affect different galaxy populations differently. \\

\noindent
Finally, we reinforce the fact that \H\ as a kinematical tracer of the circular velocities of galaxies is an important tool to further our knowledge of the evolution of scaling relations. With this study, we have provided a first reference for future, large-scale, \H-based surveys at higher redshifts such as the Deep Investigation of Neutral Gas Origins \citep[DINGO;][]{Meyer09} with the Australian Square Kilometre Array Pathfinder (ASKAP) and the Looking At the Distant Universe with the MeerKAT Array \citep[LADUMA;][]{Holwerda12, Blyth16}, and ultimately, the Square Kilometre Array (SKA-1)\footnote{https://www.skatelescope.org}.

\section*{Acknowledgements}

AG and MV would like to acknowledge the Netherlands Foundation for Scientific Research support through VICI grant 016.130.338 and the Leids Kerkhoven-Bosscha Fonds (LKBF) for travel support. The authors thank the referee for useful inputs that helped sharpen and improve several aspects of the paper. AG thanks P. Bilimogga for useful discussions and for providing the systemic velocities for the Ursa Major sample. AG also thanks N. Choque. and J. Healy for useful discussions on galaxy modelling with \textit{galfit}. JMvdH acknowledges support from the European Research Council under the European Union's Seventh Framework Programme (FP/2007-2013)/ERC Grant Agreement nr. 291531. Y.J. acknowledges financial support from CONICYT PAI (Concurso Nacional de Insercion en la Academia 2017) No. 79170132 and FONDECYT Iniciacion 2018 No. 11180558. The WSRT is operated by the Netherlands Foundation for Research in Astronomy, supported by the Netherlands Foundation for Scientific Research. The full acknowledgement of the Sloan Digital Sky Survey Archive used in this paper can be found at http://www.sdss.org.



\section*{Data availability}
The data underlying this article will be shared on reasonable request to the corresponding author.

\bibliographystyle{mnras}
\bibliography{references}

\clearpage

\renewcommand{\thefootnote}{\fnsymbol{footnote}}

\input{tf_tab_forpaper.tex}

\onecolumn


\setlength\LTcapwidth{15 cm}
\begin{longtable}{c | c  c  c  c  c | c  c}

\caption{\label{tab:fitparams} Parameters and offsets of the various TFRs and BTFRs, shown in Figs. \ref{fig:tfr} and \ref{fig:btfr}.\\
\\
\textit{Column} (1): Describes which TFR has been fit. Given in bold are the headers, consisting of the TF categories, for instance, the M$_{\rm B}$$\mathrm{W_{20}}$, M$_{\rm R}$$\mathrm{W_{50}}$ etc.;
\textit{Column} (2): The number of galaxies belonging to the specified sample;
\textit{Column} (3): The slopes ($\alpha$) with errors obtained from fitting. This is true for the UMa sample. For the others, the slope is fixed to the UMa value;
\textit{Column} (4): The zero points (zp) with errors obtained from fitting;
\textit{Column} (5): The reduced $\mathrm{\chi^2_{red}}$ signifying whether the observed scatter can be explained given the errors;
\textit{Column} (6): The total vertical rms scatter taking into account all the data points;
\textit{Column} (7): Offsets in zero points with respect to the UMa TFR. Offsets with an estimated significance in excess of 5$\sigma$ are highlighted in bold;
\textit{Column} (8): Offset in zero point of the \textit{HQS} TFR with respect to the \textit{TFS} TFR.} \\

\hline \hline

& & & \textit{Parameters} & &  &\multicolumn{2}{c}{zero point offsets} \\ \cline{2-8}
TFR & Count & $\alpha$  & zp  & $\chi^2_{\mathrm{red}}$ & $\sigma$ & Offset UMa & Offset \textit{TFS} \\
(1)  & (2)  & (3) & (4) & (5)  & (6) & (7) & (8)\\

\hline
\endfirsthead
\caption{continued} \\
\hline \hline

& & & \textit{Parameters} & &  & \multicolumn{2}{c}{zero point offsets} \\ \cline{2-8}
TFR & Count & $\alpha$  & zp  & $\chi^2_{\mathrm{red}}$ & $\sigma$ & Offset UMa & Offset \textit{TFS} \\
(1)  & (2)  & (3) & (4) & (5)  & (6) & (7) & (8)\\

\hline 
\endhead
\hline
\endfoot

 & \multicolumn{5}{c}{luminosity based TFRs} & & \\
\hline
\textbf{M$_{\rm B}$$-$W$_{20}$} & & & & & & & \\
\cline{1-1}  
UMa (free)  &  22  &  -7.94  $\pm$ 0.28  &  -0.02  $\pm$  0.69  &  7.62  &  0.45  & $-$ & $-$ \\ 
UMa         &  22  &  -7.94  (fixed)     &  -0.02  $\pm$  0.04  &  7.26  &  0.45  & $-$ & $-$ \\ 
\textit{TFS}         &  36  &  -7.94  (fixed)     &  -0.49  $\pm$  0.04  &  7.02  &  0.69  & \textbf{0.47} $\pm$ \textbf{0.06} & $-$ \\ 
\textit{HQS}         &  19  &  -7.94  (fixed)     &  -0.48  $\pm$  0.05  & 10.23  &  0.68  & \textbf{0.46} $\pm$ \textbf{0.06} & 0.01 $\pm$ 0.06 \\ 
\textbf{M$_{\rm B}$$-$W$_{50}$} & & & & & & & \\
\cline{1-1} 
UMa (free)  &  22  &  -8.25  $\pm$ 0.30  &   0.72  $\pm$  0.72  &  6.56  &  0.44  & $-$ & $-$ \\ 
UMa         &  22  &  -8.25  (fixed)     &   0.72  $\pm$  0.04  &  6.25  &  0.44  & $-$ & $-$ \\ 
\textit{TFS}         &  36  &  -8.25  (fixed)     &   0.00  $\pm$  0.05  &  3.73  &  0.56  & \textbf{0.72} $\pm$ \textbf{0.06} & $-$ \\ 
\textit{HQS}         &  19  &  -8.25  (fixed)     &   0.04  $\pm$  0.06  &  5.55  &  0.56  & \textbf{0.68} $\pm$ \textbf{0.07} & 0.04 $\pm$ 0.08 \\ 
\textbf{M$_{\rm R}$$-$W$_{20}$} & & & & & & & \\
\cline{1-1}  
UMa (free)  &  22  &  -8.29  $\pm$ 0.28  &  -0.20  $\pm$  0.69  &  3.88  &  0.34  & $-$ & $-$ \\ 
UMa         &  22  &  -8.29  (fixed)     &  -0.20  $\pm$  0.04  &  3.69  &  0.34  & $-$ & $-$ \\ 
\textit{TFS}         &  36  &  -8.29  (fixed)     &  -0.39  $\pm$  0.05  &  6.35  &  0.68  & 0.19 $\pm$ 0.06 & $-$ \\ 
\textit{HQS}         &  19  &  -8.29  (fixed)     &  -0.38  $\pm$  0.05  & 10.00  &  0.70  & 0.18 $\pm$ 0.06 & 0.01 $\pm$ 0.07\\ 
\textbf{M$_{\rm R}$$-$W$_{50}$} & & & & & & & \\
\cline{1-1} 
UMa (free)  &  22  &  -8.68  $\pm$ 0.31  &   0.74  $\pm$  0.73  &  3.50  &  0.34  & $-$ & $-$ \\ 
UMa         &  22  &  -8.68  (fixed)     &   0.74  $\pm$  0.04  &  3.33  &  0.34  & $-$ & $-$ \\ 
\textit{TFS}         &  36  &  -8.68  (fixed)     &   0.30  $\pm$  0.05  &  3.40  &  0.56  & \textbf{0.44} $\pm$ \textbf{0.06} & $-$ \\
\textit{HQS}         &  19  &  -8.68  (fixed)     &   0.35  $\pm$  0.06  &  5.20  &  0.57  & \textbf{0.39} $\pm$ \textbf{0.07} & 0.05 $\pm$ 0.08\\ 
\hline
 & \multicolumn{5}{c}{Baryonic TFRs} & & \\
\hline
\textbf{M$_{\rm bar}$$-$W$_{20}$} & & & & & & & \\
\cline{1-1} 
UMa (free)  &  22  &   3.02  $\pm$ 0.10  &   2.80  $\pm$  0.36  &  3.64  &  0.12  & $-$ & $-$ \\ 
UMa         &  22  &   3.02  (fixed)     &   2.80  $\pm$  0.01  &  3.46  &  0.12  & $-$ & $-$ \\ 
\textit{TFS}         &  36  &   3.02  (fixed)     &   2.79  $\pm$  0.02  &  5.28  &  0.23  & 0.01 $\pm$ 0.02 & $-$ \\ 
\textit{HQS}         &  19  &   3.02  (fixed)     &   2.79  $\pm$  0.02  &  8.85  &  0.24  & 0.01 $\pm$ 0.02 & 0.00 $\pm$ 0.03\\ 
\textbf{M$_{\rm bar}$$-$W$_{50}$} & & & & & & & \\
\cline{1-1} 
UMa (free)  &  22  &   3.20  $\pm$ 0.11  &   2.36  $\pm$  0.37  &  3.79  &  0.13  & $-$ & $-$ \\ 
UMa         &  22  &   3.20  (fixed)     &   2.36  $\pm$  0.01  &  3.61  &  0.13  & $-$ & $-$ \\ 
\textit{TFS}         &  36  &   3.20  (fixed)     &   2.44  $\pm$  0.02  &  3.38  &  0.21  & 0.08 $\pm$ 0.02 & $-$ \\ 
\textit{HQS}         &  19  &   3.20  (fixed)     &   2.42  $\pm$  0.02  &  5.35  &  0.21  & 0.06 $\pm$ 0.02 & 0.02 $\pm$ 0.03 \\ 

\end{longtable}


\pagebreak

\LTcapwidth=\textwidth
\setlength\LTcapwidth{15 cm}
\begin{longtable}{c | c  c  c  c  c | c  c}
 
\caption{\label{tab:fitparamsbo} Fit parameters of the environment dependent TFRs and BTFRs, shown in Fig. \ref{fig:bo}. The description of individual columns is identical to Table \ref{tab:fitparams}.}\\

\hline
\hline
& & & \textit{Parameters} & & & \multicolumn{2}{c}{zero point offsets} \\
\cline{2-8}
TFR & Count & $\alpha$  & zp  & $\chi^2_{\mathrm{red}}$ & $\sigma$ & UMa & Cluster sample \\
(1)  & (2)  & (3) & (4) & (5)  & (6) & (7) & (8) \\
\hline

\hline
\endfirsthead
\caption{continued} \\
\hline
\hline

& & & \textit{Parameters} & & & \multicolumn{2}{c}{zero point offsets} \\
\cline{2-8}
TFR & Count & $\alpha$  & zp  & $\chi^2_{\mathrm{red}}$ & $\sigma$ & UMa & Cluster sample \\
(1)  & (2)  & (3) & (4) & (5)  & (6) & (7) & (8)\\
\hline
\hline 
\endhead
\hline
\endfoot

 & \multicolumn{5}{c}{luminosity based TFRs} & & \\
\hline
\textbf{M$_{\rm B}$$-$W$_{20}$} & & & & & & & \\
\cline{1-1} 
UMa                 &  22  &  -7.94  (fixed)  &  -0.02  $\pm$  0.04  &  7.26  &  0.45  &                 $-$                 &        $-$       \\ 
\textit{TFS} Cluster sample  &  19  &  -7.94  (fixed)  &  -0.56  $\pm$  0.06  &  6.58  &  0.68  &  \textbf{0.54} $\pm$ \textbf{0.07}  &        $-$       \\ 
\textit{TFS} Control sample  &  17  &  -7.94  (fixed)  &  -0.42  $\pm$  0.06  &  7.80  &  0.70  &  \textbf{0.40} $\pm$ \textbf{0.07}  &  0.14 $\pm$ 0.09 \\ 
\textit{HQS} Cluster sample  &   7  &  -7.94  (fixed)  &  -0.49  $\pm$  0.07  & 11.28  &  0.61  &  \textbf{0.47} $\pm$ \textbf{0.08}  &        $-$       \\ 
\textit{HQS} Control sample  &  12  &  -7.94  (fixed)  &  -0.47  $\pm$  0.07  & 10.72  &  0.72  &  \textbf{0.45} $\pm$ \textbf{0.08}  &  0.02 $\pm$ 0.10 \\ 
\textbf{M$_{\rm B}$$-$W$_{50}$} & & & & & & & \\
\cline{1-1} 
UMa                 &  22  &  -8.25  (fixed)  &   0.72  $\pm$  0.04  &  6.25  &  0.44  &                 $-$                 &        $-$       \\ 
\textit{TFS} Cluster sample  &  19  &  -8.25  (fixed)  &  -0.18  $\pm$  0.08  &  2.14  &  0.44  &  \textbf{0.90} $\pm$ \textbf{0.09}  &        $-$       \\ 
\textit{TFS} Control sample  &  17  &  -8.25  (fixed)  &   0.11  $\pm$  0.06  &  6.29  &  0.60  &  \textbf{0.61} $\pm$ \textbf{0.07}  &  0.29 $\pm$ 0.10 \\ 
\textit{HQS} Cluster sample  &   7  &  -8.25  (fixed)  &  -0.08  $\pm$  0.10  &  3.73  &  0.39  &  \textbf{0.80} $\pm$ \textbf{0.11}  &        $-$       \\ 
\textit{HQS} Control sample  &  12  &  -8.25  (fixed)  &   0.08  $\pm$  0.07  &  8.30  &  0.61  &  \textbf{0.64} $\pm$ \textbf{0.08}  &  0.16 $\pm$ 0.12 \\ 
\textbf{M$_{\rm R}$$-$W$_{20}$} & & & & & & & \\
\cline{1-1} 
UMa                 &  22  &  -8.29  (fixed)  &  -0.20  $\pm$  0.04  &  3.69  &  0.34  &        $-$        &        $-$       \\ 
\textit{TFS} Cluster sample  &  19  &  -8.29  (fixed)  &  -0.47  $\pm$  0.06  &  5.78  &  0.68  &  0.27 $\pm$ 0.08  &        $-$       \\ 
\textit{TFS} Control sample  &  17  &  -8.29  (fixed)  &  -0.31  $\pm$  0.07  &  7.20  &  0.70  &  0.11 $\pm$ 0.08  &  0.16 $\pm$ 0.09 \\ 
\textit{HQS} Cluster sample  &   7  &  -8.29  (fixed)  &  -0.44  $\pm$  0.08  & 11.43  &  0.61  &  0.24 $\pm$ 0.09  &        $-$       \\ 
\textit{HQS} Control sample  &  12  &  -8.29  (fixed)  &  -0.33  $\pm$  0.07  & 10.03  &  0.72  &  0.13 $\pm$ 0.08  &  0.11 $\pm$ 0.10 \\ 
\textbf{M$_{\rm R}$$-$W$_{50}$} & & & & & & & \\
\cline{1-1} 
UMa                 &  22  &  -8.68  (fixed)  &   0.74  $\pm$  0.04  &  3.33  &  0.34  &        $-$        &         $-$      \\ 
\textit{TFS} Cluster sample  &  19  &  -8.68  (fixed)  &   0.09  $\pm$  0.08  &  1.58  &  0.44  &  \textbf{0.65} $\pm$ \textbf{0.09}  & $-$ \\ 
\textit{TFS} Control sample  &  17  &  -8.68  (fixed)  &   0.43  $\pm$  0.07  &  5.03  &  0.60  &  0.31 $\pm$ 0.08  &  0.34 $\pm$ 0.11 \\ 
\textit{HQS} Cluster sample  &   7  &  -8.68  (fixed)  &   0.18  $\pm$  0.11  &  2.21  &  0.39  &  0.56 $\pm$ 0.12  &        $-$       \\ 
\textit{HQS} Control sample  &  12  &  -8.68  (fixed)  &   0.41  $\pm$  0.07  &  7.02  &  0.61  &  0.33 $\pm$ 0.08  &  0.23 $\pm$ 0.13 \\ 
\hline
 & \multicolumn{5}{c}{Baryonic TFRs} & & \\
\hline
\textbf{M$_{\rm bar}$$-$W$_{20}$} & & & & & & \\
\cline{1-1} 
UMa                 &  22  &  3.02  (fixed)  &  2.80  $\pm$  0.01  &  3.46  &  0.12  &        $-$        &        $-$       \\ 
\textit{TFS} Cluster sample  &  19  &  3.02  (fixed)  &  2.83  $\pm$  0.07  &  5.03  &  0.22  &  0.03 $\pm$ 0.07  &        $-$       \\ 
\textit{TFS} Control sample  &  17  &  3.02  (fixed)  &  2.75  $\pm$  0.07  &  5.55  &  0.22  &  0.05 $\pm$ 0.07  &  0.08 $\pm$ 0.09 \\ 
\textit{HQS} Cluster sample  &   7  &  3.02  (fixed)  &  2.85  $\pm$  0.08  & 11.60  &  0.22  &  0.05 $\pm$ 0.08  &        $-$       \\ 
\textit{HQS} Control sample  &  12  &  3.02  (fixed)  &  2.75  $\pm$  0.07  &  7.51  &  0.23  &  0.05 $\pm$ 0.07  &  0.10 $\pm$ 0.11 \\ 
\textbf{M$_{\rm bar}$$-$W$_{50}$} & & & & & & \\
\cline{1-1} 
UMa                 &  22  &  3.20  (fixed)  &  2.36  $\pm$  0.01  &  3.61  &  0.13  &        $-$        &        $-$       \\ 
\textit{TFS} Cluster sample  &  19  &  3.20  (fixed)  &  2.56  $\pm$  0.08  &  1.41  &  0.15  &  0.20 $\pm$ 0.08  &        $-$       \\ 
\textit{TFS} Control sample  &  17  &  3.20  (fixed)  &  2.36  $\pm$  0.06  &  4.20  &  0.20  &  0.00 $\pm$ 0.06  &  0.20 $\pm$ 0.10 \\ 
\textit{HQS} Cluster sample  &   7  &  3.20  (fixed)  &  2.57  $\pm$  0.10  &  1.98  &  0.15  &  0.21 $\pm$ 0.11  &        $-$       \\ 
\textit{HQS} Control sample  &  12  &  3.20  (fixed)  &  2.36  $\pm$  0.06  &  5.83  &  0.21  &  0.00 $\pm$ 0.06  &  0.22 $\pm$ 0.12 \\

\end{longtable}

\bsp	
\label{lastpage}

\newpage


  




\appendix

\end{document}

%% file: tf_tab_forpaper.tex
\onecolumn

\begin{landscape}
\setlength\LTcapwidth{22 cm}
\begin{longtable}{c  c  c  c  c  c  c  c  c  c  c  c  c}

\caption{\label{tab:A963_HI_table_tf} \H\ properties of the \b\ TF galaxies. The full table is available as supplementary material online.\\ \\
\textit{Column} (1): The running identification number of the galaxies for easy cross-reference with entries in Table \ref{tab:A963_opt_table_tf};
\textit{Column} (2): The catalogue number assigned to these galaxies in Paper 1, for easy cross-reference to the atlas pages.;
\textit{Column} (3): The \H\ ID which contains the Right Ascension and Declination of the \H\ source [J2000].;
\textit{Column} (4): The rest-frame velocity resolution R at which the \H\ profile, total \H\ map and Position-Velocity diagrams were extracted from the data cubes. The velocity resolution is set at 4 $\times$ $\Delta$v (R4), where $\Delta$v is the redshift-dependent rest-frame width of the channel in \kms.;
\textit{Column} (5): The channel-average Signal-to-Noise Ratio (SNR) of the extracted \H\ profiles of each galaxy;
\textit{Column} (6): The galaxy redshift based on the \H\ profile;
\textit{Column} (7): The luminosity distance to the galaxy, calculated using its \H\ redshift and the adopted cosmology;
\textit{Columns} (8) \& (9): The observed (rest frame) $\mathrm{W_{20}}$ and $\mathrm{W_{50}}$ line widths, including their errors;
\textit{Column} (10) \& (11): The \wct\ and \wcf\ line widths corrected for instrumental resolution, turbulent motions and inclination, following the methodology described in Sect. \ref{corr};
\textit{Column} (12): \H\ masses ($\times$ 10$^9$ \Msun) calculated from the integrated flux densities of the extracted \H\ profiles, including their uncertainties. For further details, see Paper 1;
\textit{Column} (13): The volume (A963 or A2192) and sample (\textit{TFS} or \textit{HQS}) that a galaxy belongs to.} \\

\hline \hline

Sr. no. & Cat no. & HI ID  & R & SNR & z$_{HI}$ & D$_{lum}$ & w$_{20}^{obs}$ & w$_{50}^{obs}$ & w$_{20}^{\rm corr}$ & w$_{50}^{\rm corr}$ & M$_{HI}$ &  \\
 &   &  & km s$^{-1}$ &  &  & Mpc & km s$^{-1}$ & km s$^{-1}$ & km s$^{-1}$ &km s$^{-1}$ & $\times$ 10$^9$ M$_{\odot}$ & \multicolumn{1}{c}{Sample}\\
(1)  & (2)  & (3) & (4) & (5)  & (6) & (7) & (8) & (9) & (10) & (11) & (12) &  \multicolumn{1}{c}{(13)}\\

\hline
\endfirsthead
\caption{continued} \\
\hline \hline

Sr. no. & Cat no. & HI ID  & R & SNR & z$_{HI}$ & D$_{lum}$ & w$_{20}^{obs}$ & w$_{50}^{obs}$ & w$_{20}^{\rm corr}$ & w$_{50}^{\rm corr}$ & log M$_{HI}$ & \\
 &   &  & km s$^{-1}$ &  &  & Mpc & km s$^{-1}$ & km s$^{-1}$ & km s$^{-1}$ &km s$^{-1}$ & $\times$ 10$^9$ M$_{\odot}$ & \multicolumn{1}{c}{Sample}\\
(1)  & (2)  & (3) & (4) & (5)  & (6) & (7) & (8) & (9) & (10) & (11) & (12) & \multicolumn{1}{c}{(13)}\\

\hline 
\endhead
\hline
\endfoot

1  & 7   & HIJ101600.44+385211.7 & 39.7 & 2.6 & 0.20861 & 1026 & 126 $\pm$ 10 & 104 $\pm$ 12 & 115 $\pm$ 27  & 117 $\pm$ 31  & 3.7 $\pm$ 0.4  & TFS A963 \\
2  & 10  & HIJ101613.62+390438.4 & 39.1 & 3.8 & 0.18954 & 922  & 372 $\pm$ 15 & 315 $\pm$ 18 & 400 $\pm$ 37  & 359 $\pm$ 42  & 8.4 $\pm$ 0.3  & TFS A963 \\
3  & 27  & HIJ101641.11+391025.1 & 39.8 & 4.6 & 0.21047 & 1036 & 335 $\pm$ 22 & 256 $\pm$ 10 & 363 $\pm$ 54  & 296 $\pm$ 25  & 11.2 $\pm$ 0.4 & TFS A963 \\
4  & 37  & HIJ101701.14+384258.2 & 39.5 & 2.7 & 0.20328 & 996 & 268 $\pm$ 22 & 220 $\pm$ 23 & 295 $\pm$ 55  & 264 $\pm$ 59  & 4.3 $\pm$ 0.3  & TFS A963 \\
5  & 45  & HIJ101705.51+384927.4 & 39.5 & 3.1 & 0.20401 & 1000 & 446 $\pm$ 19 & 394 $\pm$ 13 & 454 $\pm$ 41  & 422 $\pm$ 30  & 8.4 $\pm$ 0.4  & TFS A963 \\

\end{longtable}

\setlength\LTcapwidth{22 cm}
\begin{longtable}{ c  c  c  c  c  c  c  c  c  c  c  c  c  c}

\caption{\label{tab:A963_opt_table_tf} Optical properties of the \b\ galaxies. The full table is available as supplementary material online.\\ \\
\textit{Column} (1): The running identification number of the galaxies for easy cross-reference with entries in Table \ref{tab:A963_HI_table_tf};
\textit{Column} (2): The SDSS ID of the adopted optical counterpart of the \H\ detection, indicating the optical coordinates of the galaxy;
\textit{Column} (3): The minor-to-major axis ratios (b/a) obtained from \textit{galfit};
\textit{Column} (4): The inclination as inferred from the optical axis ratio (b/a), following Eq. (1) and assuming an intrinsic disc thickness q$_0$ of 0.2;
\textit{Columns} (5) \& (6): Total apparent INT B- and R- band magnitudes, respectively, including a small aperture correction as described in Sect. 4.2;
\textit{Columns} (7) \& (8): The k-corrections applied to the Galactic extinction corrected, apparent B- and R- band magnitudes, as described in Sect. 3;
\textit{Columns} (9) \& (10): The internal extinction corrections applied to the Galactic extinction and k-corrected B- and R- band magnitudes, as described in Sect. 3;
\textit{Columns} (11) \& (12): The corrected, absolute B- and R- band magnitudes, respectively;
\textit{Column} (13): Log stellar mass (\Msun) calculated using Eq. \ref{eq:mstar1};
\textit{Column} (14): Log baryonic mass (\Msun) calculated from the \H\ masses in Col. (12) of Table \ref{tab:A963_HI_table_tf}, and the stellar masses in Col. (13) respectively.
} \\

\hline \hline

Sr. no. & SDSS ID  & b/a & incl & m$_B$ & m$_R$ & k$_B$ & k$_R$ & A$^i_B$ & A$^i_R$ & M$_B$ & M$_R$ & log M$_{\ast}$ & log M$_{bar}$\\
 &   &  & deg &  mag & mag & mag & mag & mag & mag & mag & mag & M$_{\odot}$ & M$_{\odot}$ \\
(1)  & (2)  & (3) & (4) & (5)  & (6) & (7) & (8) & (9) & (10) & (11) & (12) & (13)  & (14) \\

\hline
\endfirsthead
\caption{continued} \\
\hline \hline

Sr. no. & SDSS ID  & b/a & incl & m$_B$ & m$_R$ & k$_B$ & k$_R$ & A$^i_B$ & A$^i_R$ M$_B$ & M$_R$ & log M$_{\ast}$ & log M$_{bar}$\\
 &   &  & deg &  mag & mag & mag & mag & mag & mag & mag & mag & M$_{\odot}$ & M$_{\odot}$ \\
(1)  & (2)  & (3) & (4) & (5)  & (6) & (7) & (8) & (9) & (10) & (11) & (12) & (13)  & (14) \\

\hline 
\endhead
\hline
\endfoot

1  & SJ101600.10+385205.5 & 0.646 $\pm$ 0.011 & 51.2  & 21.19 & 20.08 & 0.07 & 0.06 & 0.41 & 0.04 & -19.40  & -20.10  & 9.63          & 9.98      \\
2  & SJ101613.68+390437.8 & 0.568 $\pm$ 0.003 & 57.1  & 20.46 & 18.81 & 0.45 & 0.33 & 0.74 & 0.15 & -20.62 & -21.53 & 10.37         & 10.55     \\
3  & SJ101641.08+391025.7 & 0.594 $\pm$ 0.007 & 55.2 & 19.98 & 18.78 & 0.39 & 0.29 & 0.47 & 0.06 & -21.01 & -21.67 & 10.19         & 10.5      \\
4  & SJ101701.11+384258.9 & 0.642 $\pm$ 0.006 & 51.5  & 20.65 & 19.29 & 0.29 & 0.21 & 0.58 & 0.10  & -20.26 & -21.04 & 10.07         & 10.25     \\
5  & SJ101705.49+384924.8 & 0.468 $\pm$ 0.002 & 64.4 & 20.04 & 18.34 & 0.66 & 0.48 & 0.80  & 0.17 & -21.48 & -22.34 & 10.66         & 10.76     \\

\end{longtable}
\end{landscape}
\twocolumn